\documentclass[a4paper,12pt,twoside]{article}
\usepackage[T1]{fontenc}
\usepackage[latin1]{inputenc}

\usepackage[in,myheadings]{fullpage}
\usepackage{epsfig}
\usepackage{enumerate}
\usepackage[active]{srcltx}
\usepackage{amsmath,amssymb,epsf,epic}
\usepackage{times}
\usepackage{latexsym,amsopn}
\usepackage{amsfonts,amsbsy,amscd,stmaryrd}
\usepackage{theorem}
\usepackage{paralist}
\usepackage{geometry}

\numberwithin{equation}{section}
\newcounter{resultcounter}[section]

\newtheorem{thm}{Theorem}[section]
\newtheorem{prop}[thm]{Proposition}
\newtheorem{lem}[thm]{Lemma}
\newtheorem{defn}[thm]{Definition}
\newtheorem{cor}[thm]{Corollary}
\newtheorem{rem}[thm]{Remark}
\newtheorem{ex}[thm]{Example}
\newtheorem{exo}[thm]{Exercise}
\newtheorem{open}{Open problem}

\def\cA{{\cal A}} \def\cB{{\cal B}} \def\cC{{\cal C}}
\def\cD{{\cal D}} \def\cE{{\cal E}} \def\cF{{\cal F}}
 \def\cH{{\cal H}} 
  \def\cL{{\cal L}}
\def\cM{{\cal M}}  
\def\cP{{\cal P}}  \def\cR{{\cal R}}
\def\cS{{\cal S}}  \def\cU{{\cal U}}

\def\R{{\mathbb R}}
\def\N{{\mathbb N}}
\def\Q{{\mathbb Q}}
\def\C{{\mathbb C}}
\def\Z{{\mathbb Z}}
\def\E{{\mathbb E}}
\def\P{{\mathbb P}}
\def\B{{\mathbb B}}

\def\I{{\mathbb I}}
\def\D{{\mathbb D}}
\def\T{{\mathbb T}}

\def\fa{{\mathfrak A}}
\def\fm{{\mathfrak M}}
\def\fh{{\mathfrak h}}

\def\e{{\rm e}}
\def\i{{\rm i}}
\def\d{{\rm d}}
\def\p{{\rm p}}

\def\tr{{\rm Tr}}
\def\sp{{\rm sp}}

\def\pp{{\rm pp}}

\def\ran{{\rm Ran}}

\def\env{{\rm env}}
\def\RI{{\rm RI}}

\def\tOmega{\tilde\Omega}
\newcommand{\vd}{v_{\mathrm d}}
\def\sinc{{\rm sinc}}
\def\ent{{\rm Ent}}

\def\ss{{\sigma'\sigma}}

\newcommand{\bra}{\langle} 
\newcommand{\ket}{\rangle}

\def\gs{{|0\rangle}}
\def\exs{{|1\rangle}}
\newcommand{\ds}{\displaystyle}
\def\Im{{\rm Im}}

\renewcommand{\th}{\tilde h}
\newcommand{\tL}{\widetilde L}

\newcommand{\scalprod}[2]{\left\langle {#1}, {#2}\right\rangle}
\def\one{{\mathchoice {\rm 1\mskip-4mu l} {\rm 1\mskip-4mu l} {\rm 1\mskip-4.5mu l} {\rm 1\mskip-5mu l}}}
\newcommand{\fer}[1]{(\ref{#1})}

\newcommand{\be}{\begin{equation}}
\newcommand{\ee}{\end{equation}}
\newcommand{\bea}{\begin{eqnarray}}
\newcommand{\eea}{\end{eqnarray}}

\begin{document}

 \title{Repeated interactions in open quantum systems}
\author{
Laurent Bruneau\footnote{Laboratoire AGM, Universit\'e de Cergy-Pontoise, Site Saint-Martin, BP 222, 95302 Cergy-Pontoise, France. Email: laurent.bruneau@u-cergy.fr, http://bruneau.u-cergy.fr/}\ ,
Alain Joye\footnote{ Institut Fourier, UMR 5582, CNRS-Universit\'e Grenoble I, BP 74, 38402 Saint-Martin d'H\`eres, France. Email: Alain.Joye@ujf-grenoble.fr, http://www-fourier.ujf-grenoble.fr/\ $\widetilde{}$\,joye }\ ,  
Marco Merkli\footnote{Department of Mathematics, Memorial University of Newfoundland, Canada. Supported by NSERC under Discovery Grant
205247. Email: merkli@mun.ca, http://www.math.mun.ca/\ $\widetilde{}$\,merkli/}
}

\date{\today}

\maketitle

\tableofcontents

\section{Introduction}\label{sec:intro}

\subsection{Motivations}
 
The study of the dynamics of open quantum systems is a venerable topic due to its relevance in the description of several basic physical mechanisms of interest, such as convergence towards a thermodynamical equilibrium state of onset of heat or particle fluxes between reservoirs at different temperatures or chemical potentials for example. At the same time, it is a very active field of present research in mathematical physics. One of the reasons for this is  to be found in the fact that the description of return to equilibrium or onset of stationary states in open quantum systems appeals explicitly to the description in the large times regime of the unitary dynamics of quantum systems and the effective dispersive effects induced by the intrinsic properties of the reservoirs. Besides non trivial modeling aspects, the mathematical analysis still represents a challenge for many physically relevant models. 

The repeated interactions systems considered in these notes are models of non-equilibrium quantum statistical mechanics whose specificities to be explained below make it possible to determine large times asymptotic properties which are characteristic of stationary systems out of equilibrium. Repeated interactions models are quite relevant in quantum optics, and also appear as approximate quantum dynamics, as will be detailed below.

\medskip
 
In order to put repeated interactions systems in perspective within the wider framework of open quantum systems, we briefly recall the description and characteristics of what we call open quantum systems in these notes and the main features of two popular and fruitful approaches of their dynamical properties.

Open quantum systems often consist of a reference quantum system also called "small system", $\cS$ characterized by its Hilbert space $\fh_\cS$ and Hamiltonian $h_\cS$ put in contact with an infinitely extended quantum environment or reservoir $\cR$. The state of the latter is characterized by macroscopic thermodynamical properties such as temperature or chemical potential and can be formally described as a density matrix in a Hilbert space $\fh_\cR$ which is invariant under the dynamics driven by a Hamiltonian $h_\cR$. The coupling between these two systems is provided by a interaction operator $v$ acting on the Hilbert space of the total system given by the tensor product $\fh_\cS\otimes \fh_\cR$. In order to describe genuine non equilibrium effects, the environment may have more structure and consist, for example, in the union of several reservoirs $\cR=\vee_{j=1, \dots, n}\cR_j$,  the state of each reservoir $\cR_j$ characterized by its own thermodynamical properties, on a formal Hilbert space, and being driven by its Hamiltonian.

 \medskip
 
The general approach of open systems consists in focusing on the dynamics of the reference system $\cS$ under the influence of the environment, giving up the idea to describe precisely the dynamics of the environment. This is in keeping with the {\it a priori} difference in nature of the reference system $\cS$ and of the environment $\cR$: the latter plays the role of an infinite reservoir of energy and/or particles, so that the small system of interest has very little influence on $\cR$. Hence the macroscopic characteristics of the environment will remain constant in time, whereas the dynamics of the reference system will be significantly influenced by the presence of the environment. 
 
One approach of this question, dubbed the Hamiltonian approach, consists in the following procedure. One adopts a microscopic description of both the small system and the environment on their Hilbert space as above, and one considers a decoupled initial state of the form $\rho_\cS\otimes\rho_\cR$, where $\rho_\#$ is a (formal) density matrix on $\fh_\#$, $\#=\cS, \cR$. One then lets this state evolve up to time $t>0$ under the dynamics generated by the coupled microscopic Hamiltonian  $h_\cS + h_\cR+v$ on the space of the full system $\fh_\cS\otimes\fh_\cR$, to get a state of the full system. Then, tracing out the degrees of freedom of the environment, one gets from this procedure the state $\rho_\cS(t)$ of the small system $\cS$ at time $t$, from the initial state $\rho_\cS$, in which the influence of the environment is encoded exactly. The price to pay is that the dynamical process which maps $\rho_\cS$ to $\rho_\cS(t)$ is, in general not a semigroup and is not the solution of a differential equation, but that of a complicated integro-differential equation. In certain asymptotic regimes like the van Hove limit, approximations of $\rho_\cS(t)$ are provided in terms of solutions to an effective evolution equation of Lindblad type.

Another popular approach of quantum open systems, called the Markovian approach, is based on the general assumption that the perturbations of the state of the reservoir generated by the interaction with the small system propagate to infinity fast enough so that for all practical purposes, the memory effects can be neglected.  In other words, without attempting to provide a description of the reservoir, it is assumed that its effect  on the reduced dynamics of states on the space $h_\cS$ of the small system amounts  to a modification of the free generator by a dissipative part which yields a Markovian effective dynamics. The actual  determination of the effective dissipative generator usually takes into account the physical peculiarities of the small system, of  the reservoir and of the interaction operator. Moreover, the generator of the effective dynamics is often chosen to take the form of a Lindbladian in order to produce a CP map. One of the main features of the Markovian framework is that the main dynamical properties of the evolution of states of the small system can be read off the spectral properties of the effective generator. The drawback lies in the deliberate approximation of the effective evolution by a one parameter semigroup.

\medskip

Let us come to repeated interaction systems now. The situation addressed in these notes shares the same general characteristics with the setup loosely described above: a small system $\cS$ interacting with a large environment. However, the environment in repeated interaction quantum systems is structured in a quite different way: it consists in an infinite chain of quantum subsystems $\cE_1, \cE_2, \cE_3, \dots$, each of which is defined on its Hilbert space $\fh_{\cE_j}$ by its Hamiltonian $h_{\cE_j}$, $j\in \N^*$. The formal Hilbert space of this structured environment $\cE_1\vee \cE_2\vee \cE_3\vee \dots$ is denoted by $\fh_\env = \bigotimes_{j\geq 1} \fh_{\cE_j}$ with formal free Hamiltonian $\sum_{j\geq 1}h_{\cE_j}$. The dynamics of the compound system $\cS$ plus chain taking place on $h_\cS\otimes \fh_\env$ is characterized by a time dependent interaction with the following property: the small system interacts with the elements of the chain $\cE_j$, in sequence and one by one, for a duration that may depend on the considered element. In the simplest case which we call ideal, all elements of the chain are identical, {\it i.e.} $\cE_j=\cE$, $\fh_{\cE_j}=\fh_\cE$, $h_{\cE_j}=h_\cE$, and interact with $\cS$ by means of the same coupling operator $v$ on $\cS\otimes \cE$ for the same duration $\tau$. 

The defining features of repeated interaction systems provide their dynamics with the unique property of being at the same time Hamiltonian and Markovian. Indeed, in the ideal case, the dynamics restricted to the small system is shown to be determined by the map $\cL$ which assigns $\rho_\cS(\tau)$ to $\rho_\cS$, as the result of the interaction of $\cS$ with one subsystem $\cE$ for the duration $\tau$. Heuristically, from the point of view of the small system, all subsystems interacting in sequence with $\cS$ are equivalent, so that the result of $n\in\N$ repeated interactions amounts to iterating  $n$ times the map $\cL$ on the initial condition $\rho_\cS$. This expresses the Markovian character of repeated interactions in discrete times. As a consequence, spectral methods will be available to perform the analysis of the exact dynamics restricted to states on the Hilbert space $h_\cS$ of the small systems, which allows us to determine efficiently several thermodynamical properties of repeated interaction systems. Let us note here that in case the dimension of $\fh_\cS$ is finite, the spectral analysis of the map $\cL$ is, in principle, straightforward. However, in case $\fh_\cS$ is infinite dimensional, as is necessary in some of the physical applications described below, it becomes much more delicate and requires rather sophisticated tools.

It will be shown in this notes that this picture is correct, together with generalizations to non-ideal cases, both in deterministic and random setups.

\medskip
 
After this rather abstract descriptions of repeated interaction systems, let us provide some insight on the physical relevance of this kind of model. The physical situation which is perhaps the most tightly linked to the repeated interactions model is that of the one atom maser \cite{FJM, CDG, MWM, WVHW, WBKM}, and some of its subsequent elaborations  \cite{DRBH, G-al, RH, RBH}. In rough terms, the setup is the following. The system of reference $\cS$ consists in the monochromatic electromagnetic field in a cavity, or a finite number of its modes. This field interacts with a beam of atoms coming out of an oven which are sent through the cavity. With a good approximation, one can consider that the atoms enter the cavity one by one and interact with the field for a duration $\tau$ corresponding to their time of flight through the cavity, before leaving it and never returning. Therefore, the infinite sequence of atoms leaving the oven form the chain of subsystems $\cE_j$, $j\geq 1$, which interact one by one in sequence with the reference system $\cS$, for a duration $\tau$. Assuming ideal experimental conditions, the atoms, their incoming states and times of flight can be considered as identical, which amounts to adopting the model we dubbed ideal. Note that even in the ideal case, considering a full mode of the electromagnetic field as the small system amounts to describing $\cS$ as a harmonic oscillator, which leads to an infinite dimensional Hilbert space $\fh_\cS$. Also, in order to take into account more realistic experimental conditions in which the times of flight through the cavity, incoming state of the atoms, etc, vary slightly around some  fixed quantities, we consider below  random versions of the model allowing some of these parameters to fluctuate. 
 
An important aspect of the physics of one atom masers is the control of effects due to losses within the cavity, measured in terms of the quality factor of the cavity. Such effects are often taken into account in an effective way, by the inclusion of {\it ad hoc} dissipative terms in the generator of the dynamics. In order to have a hamiltonian  description of the losses the reference system experiences with the environment, we generalize the model in the following way. We add to the small system $\cS$ and chain $\cE_1\vee \cE_2\vee \cE_3\vee \dots$ considered so far a reservoir $\cR$ with which $\cS$ keeps interacting continuously, whereas there is no interaction between the chain and $\cR$. This extra reservoir models in an effective but Hamiltonian way the effect of the environment which is responsible for losses of the cavity. Moreover, it doesn't spoil the repeated interaction dynamical structure, since it can be considered jointly with the small system, so that $\cS\vee\cR$ becomes formally the new reference system in repeated interaction with the chain. This generalized model allows us to study the effect of losses by means of more sophisticated spectral methods including resonance theory.

\medskip

From a mathematical point of view, repeated interaction systems first appeared in the papers \cite{A, APa}. These works are devoted to the analysis of models of open quantum systems in which the reservoir $\cR$ is modeled by fields of quantum noises living in a Fock space based on a $L^2$ space in a continuous time variable. Introducing interactions with the small system $\cS$, the dynamics of the coupled systems gives rise to quantum stochastic differential equations, in a similar way to what is done in the classical setting, see {\it e.g.} \cite{HP}. Brought back to the Hilbert space $\fh_\cS$  of the small system, these models provide an exact Markovian description of the dynamics of states. Based on \cite{A}, the paper \cite{APa} proposes a version of the model in which the continuous variable of the reservoir space is discretized so that the reservoir becomes effectively a chain of independent quantum subsystems, with which  the small system interacts in sequence, according to the repeated interaction scheme described above. The point of \cite{APa} is to show that within a subtle continuous limit procedure where the discretization step tends to zero in some definite regime of other  interaction parameters, a natural model of open quantum system with a continuous reservoir of quantum noises emerges, providing the reservoir with intrinsic quantum features. Moreover, the corresponding effective Markovian dynamics on $\fh_\cS$  is generated   by an explicit Lindblad generator obtained in terms of the chosen interations between $\cS$ and $\cR$.  A versions of this construction providing the reservoir with thermal properties can be found in \cite{AJ2}.

While we shall not address models of quantum stochastic differential equations below, we will consider various regimes of repeated interaction systems in which the strength of the coupling, the duration of each interaction and number of interactions can be coupled in ways that are reminiscent of the weak coupling or Van Hove limit in  continuous Hamiltonian systems. Although simpler technically, they bear some resemblance with the limiting procedure of \cite{A, APa} alluded to above, and give rise to various Lindblad operators depending on the chosen scaling.


\section{Mathematical description of RI systems}\label{sec:framework}


We now describe more precisely the mathematical setup of repeated interaction systems, and explain how their particular structure allows one to derive a Markovian, discrete-time, dynamics for $\cS$ from the Hamiltonian dynamics of the entire system.

The various elements needed to describe a RI system are:
\begin{enumerate}
 \item the Hilbert space $\fh_\cS$ and Hamiltonian $h_\cS$ describing the small system $\cS$ ``alone'',
 \item Hilbert spaces $\fh_{\cE_n}$ and Hamiltonians $h_{\cE_n}$ describing the various subsytems $\cE_n$,
 \item a sequence of duration times $(\tau_n)_n$ where $\tau_n\geq \tau>0$ for any $n$ and some given $\tau$. The time $\tau_n$ is the amount of time the system $\cS$ spends interacting with the subsytem $\cE_n$,
 \item operators $v_n$ describing the interactions between $\cS$ and the subsystems $\cE_n$.
\end{enumerate}
The Hilbert space describing the RI system is then
$$
\fh := \fh_\cS\otimes\fh_\env,\qquad \fh_\env := \bigotimes_{n\geq 1} \fh_{\cE_n}.
$$
We also denote $t_n:=\tau_1+\cdots+\tau_n$. During the time interval $[t_{n-1},t_n)$, the system $\cS$ interacts with the $n$-th subsystem, i.e. $\cE_n$, and with none of the others. The full evolution of the system is thus described by the Hamiltonian 
\begin{equation}\label{def:totalham}
h(t)=h_\cS+ \sum_{n\geq 1} h_{\cE_n} +\sum_{n\geq 1}\chi_n(t)v_n,
\end{equation}
where $\chi_n$ is the characteristic function of the interval $[t_{n-1},t_n)$. We will use the following notation:
$$
h_n:= h_\cS+h_{\cE_n}+ v_n, \quad \mbox{and} \quad \th_n:= h_n+\sum_{k\neq n} h_{\cE_k}.
$$
Note that $h(t)\equiv \th_n$ when $t\in [t_{n-1},t_n)$. We have also omitted trivial factors $\one$, e.g. $h_\cS$ should be $h_\cS\otimes\one_\env$.

Given any initial state $\rho$ for the system $\cS$ at time $t=0$ (i.e. $\rho$ is a positive trace class operator on $\fh_\cS$ with trace one), and a sequence $(\rho_{\cE_n})_n$ of initial states for the subsystems $\cE_n$, the state of the total repeated interaction system after $n$ interactions is thus given by
$$
\rho^{\rm tot}(n):= \e^{-\i\tau_n \th_n}\cdots\e^{-\i\tau_1 \th_1} \left(
 \rho_\cS\otimes \bigotimes_{k\geq 1} \rho_{\cE_k}\right) \e^{\i\tau_1 \th_1}\cdots\e^{\i\tau_n \th_n}.
$$
We are mainly interested in the system $\cS$ (see however Section \ref{ssec:idealri} for more general observables), i.e. in expectation values of observables of the form 
$$
O=O_\cS\otimes \bigotimes_{k\geq 1} \one_{\cE_k}.
$$
Therefore, we are rather interested in $\rho(n):= \tr_{\fh_\env} (\rho^{\rm tot}(n))$, the reduced density matrix on $\cS$. It is defined as the unique state on $\fh_\cS$ such that, for any observable $O_\cS$ on $\cS$,
$$
\tr_{\fh_\cS} \left(\rho(n) O_\cS\right) = \tr_{\fh} \left( \rho^{\rm tot}(n) \times \left(O_\cS\otimes \bigotimes_{k\geq 1} \one_{\cE_k} \right)\right).
$$
To obtain the state $\rho(n)$ of the system $\cS$ after these $n$ interactions we thus take 
the following partial trace:
\begin{equation}\label{eq:systevol}
\rho(n):=\tr_{\fh_\env}\left[ \e^{-\i\tau_n \th_n}\cdots\e^{-\i\tau_1 \th_1} \left(
 \rho\otimes \bigotimes_{k\geq 1} \rho_{\cE_k}\right) \e^{\i\tau_1 \th_1}\cdots\e^{\i\tau_n \th_n} \right].
\end{equation}

Of course, the above calculation is a little bit formal. Indeed, in order to define a countable tensor product of Hilbert spaces one should specify a stabilizing sequence, i.e. a sequence of vectors $(\psi_n)_n$ where $\psi_n\in\fh_{\cE_n}$. The Hilbert space $\fh_\env$ is then obtained by taking the completion of the vector space of finite linear combinations of the form $\otimes_{n\geq 1}\phi_n$, where $\phi_n\in\fh_{\cE_n}$, $\phi_n=\psi_n$ except for finitely many indices, in the norm induced by the inner product
$$
\bra\otimes_n\varphi_n,\otimes_n\phi_n\ket = \prod_n \, \bra\varphi_n,\phi_n\ket_{\fh_{\cE_n}}.
$$
In general, the infinite tensor product $\bigotimes_{k\geq 1} \rho_{\cE_k}$ then does not make sense. It is however easy to make sense of the formal expression (\ref{eq:systevol}). Indeed, at time $t_n$ only the $n$ first elements of the environment have played a role so that we can replace $\bigotimes_{k\geq 1} \rho_{\cE_k}$ by $\rho_\env^{(n)} := \bigotimes_{k=1}^n \rho_{\cE_k}$ and the partial trace over the environment by the partial trace over the finite tensor product $\fh_\env^{(n)}:=\bigotimes_{k=1}^n \fh_{\cE_k}$, i.e.
\begin{equation}\label{eq:systevol2}
\rho(n)=\tr_{\fh_\env^{(n)}}\left[ \e^{-\i\tau_n \th_n}\cdots\e^{-\i\tau_1 \th_1} \left(
 \rho\otimes \bigotimes_{k=1}^n \rho_{\cE_k}\right) \e^{\i\tau_1 \th_1}\cdots\e^{\i\tau_n \th_n} \right].
\end{equation}

\noindent {\bf Remark. \ } {\em Another possibility would be to define the infinite tensor product ``with respect to the sequence of states $(\rho_{\cE_n})_n$''. For that purpose one should first represent the states $\rho_{\cE_n}$ as vector states with vector $\Psi_n$ (using the GNS representation), and then consider the stabilizing sequence $(\Psi_n)_n$. This then leads to the ``Liouvillian'' description of the RI system which will be presented in details in Section \ref{ssec:riliouv}.}

\medskip

The very particular structure of the repeated interaction systems allows us to rewrite $\rho(n)$ in a much more convenient way. The two main characteristics of these systems are:
\begin{enumerate}
\item The various subsystems of the environment do not interact directly (only via $\cS$), i.e. $[h_{\cE_k},h_{\cE_n}]=0$ for any $k\neq n$,
\item The system $\cS$ interacts only once with each subsystem $\cE_n$, and with only one at a time, i.e. $[h_{\cE_k},h_n]=0$ for any $k\neq n$.
\end{enumerate}
We therefore have the following decomposition which serves to isolate the dynamics of the subsystems which do not interact at a given time
\begin{equation}\label{eq:dynsplit}
\e^{-\i\tau_n \th_n}\cdots\e^{-\i\tau_1 \th_1} = u_n^- \times \e^{-\i\tau_n h_n}\cdots\e^{-\i\tau_1 h_1}\times u_n^+,
\end{equation}
where
$$
u_n^-= \exp\left(-\i\sum_{k=1}^{n-1}(t_n-t_k)h_{\cE_k}\right) \mbox { and } u_n^+= \exp\left(-\i \sum_{k=2}^n t_{k-1} h_{\cE_k} -\i t_n\sum_{k>n} h_{\cE_k}\right)
$$
are respectivley the propagators at time $t_n$ of the subsystems $\cE_k$ after their interaction with $\cS$, and the one before their interaction. Inserting (\ref{eq:dynsplit}) into (\ref{eq:systevol2}) we get
$$
\rho(n)=\tr_{\fh_\env^{(n)}}\left[ \e^{-\i\tau_n h_n}\cdots\e^{-\i\tau_1 h_1} \left(
 \rho\otimes \bigotimes_{k=1}^n \rho_{\cE_k}(t_{k-1})\right) \e^{\i\tau_1 h_1}\cdots\e^{\i\tau_n h_n} \right],
$$
where $\ds \rho_{\cE_k}(t_{k-1})=\e^{-\i t_{k-1} h_{\cE_k}}\rho_{\cE_k} \e^{\i t_{k-1} h_{\cE_k}}$ is the state of the $k$-th subsytem when it begins to interact with $\cS$.

It is now easy to see that the evolution of $\cS$ is Markovian: the state $\rho(n)$ only depends on the state $\rho(n-1)$ and the $n$-th interaction. More precisely, one can write 
\begin{equation}\label{eq:markov}
\rho(n)=\cL_n(\rho(n-1)),
\end{equation}
where
\begin{equation}\label{def:rdm0}
\cL_n(\rho):= \tr_{\fh_{\cE_n}}\left[ \e^{-\i\tau_n h_n} \, \rho\otimes \rho_{\cE_n}(t_{n-1})\ \e^{\i\tau_n h_n} \right].
\end{equation}
This explains why we take as an initial state for $\cS$ a density matrix and not necessarily a pure state. Since $\cS$ interacts with another system, and because of the reduction procedure, after already a single interaction we are led to a density matrix. Of course, this formula simplifies if $\rho_{\cE_k}$ is invariant under the free dynamics of $\cE_k$, e.g. a thermal state,  and from now on we will always assume that this is the case.

\begin{defn} The map $\cL_n$, from $\cB^1(\fh_\cS)$ to itself, is called the \emph{reduced dynamics map} (RDM) at time $n$.
\end{defn}

\noindent Note: $\cB^1(\fh_\cS)$ denotes the space of trace class operators on $\fh_\cS$.\\

The following properties of a reduced dynamics map follow directly from its definition.
\begin{prop}\label{prop:rdmprop} A RDM $\cL$ is a contracting, completely positive and trace preserving map.
\end{prop}
As a corollary of the trace preserving property, $1$ is always an eigenvalue of the dual map $\cL^*$ (for the $\cB^1(\fh_\cS) / \cB(\fh_\cS)$ duality) with eigenstate the identity operator.

The map $\cL_n$ describes the effective evolution of $\cS$ under the influence of the $n$-th subsystem. Using (\ref{eq:markov}), we therefore have for any initial state $\rho$ of the small system $\cS$
\begin{equation}\label{eq:markov2}
\rho(n)=\cL_n\circ\cL_{n-1}\circ\cdots\circ\cL_1(\rho).
\end{equation}

In the particular case of ideal interactions, i.e. $\fh_{\cE_n}\equiv \fh_\cE$, $h_{\cE_n}\equiv h_\cE$,... and if the $\rho_{\cE_n}$ are invariant for the dynamics of $\cE_n$, we then have $\cL_n\equiv\cL$ for all $n$ and (\ref{eq:markov2}) becomes simply
\begin{equation}\label{eq:rimarkov}
\rho(n)=\cL^n(\rho).
\end{equation}
The map $\cL$ is the discrete-time generator of a semi-group of completely positive, trace preserving maps on the state space of $\cS$. In other words, the particular structure of RI systems leads to an effective description of $\cS$ as in the markovian approach, starting from an Hamiltonian description and without any further scaling limit.

The study of the large time behaviour of $\cS$ reduces to the analysis of the RDM $\cL$ defined in (\ref{def:rdm0}). Since $\cL$ is a contraction, to understand the limit $n\to\infty$ of $\rho(n)=\cL^n(\rho)$ we therefore have to understand what are the invariant states, i.e. the eigenspace for the eigenvalue $1$, and also if there are other eigenvalues of modulus $1$. Note that, in the case where $\fh_\cS$ has finite dimension, the fact that $\cL$ is trace preserving implies that $1$ is also an eigenvalue of $\cL$ so that there is always an invariant state when $\cL_n\equiv\cL$ (moreover, because of the contraction property, the remaining part of the spectrum is inside the open unit disk and thus leads to exponential decay). However, if $\fh_\cS$ has infinite dimension, $\cL$ may have an invariant state or not (see Sections \ref{ssec:qedcavity} and \ref{ssec:tightbinding}).

\begin{ex}\label{ex:toy-description-rdm} As a first example, let us consider the simplest non-trivial example of RI system, namely all the subsystems ($\cS$ and the $\cE_k$'s) are 2-level systems (or spin $\frac{1}{2}$). The Hilbert spaces for $\cS$ and the $\cE_m$ are copies of $\C^2$.  Let $E, E_0>0$ be the ``excited'' energy level of $\cS$ and of $\cE$, respectively. Accordingly, the Hamiltonians are given by
\begin{equation*}\label{def:freeham}
h_\cS= \left( \begin{array}{cc} 0 & 0 \\ 0 & E \end{array} \right) \quad {\rm and} \quad
h_\cE= \left( \begin{array}{cc} 0 & 0 \\ 0 & E_0 \end{array} \right).
\end{equation*}
We will denote by $\gs$, resp. $\exs$, the ground state, resp. excited state, of $\cS$ or $\cE$. 
If we denote by $a/a^*$, resp. $b/b^*$, the annihilation/creation operators for $\cS$, resp. $\cE$, i.e.
\begin{equation}\label{def:fermicreation}
a=b=\left( \begin{array}{cc} 0 & 1 \\ 0 & 0 \end{array} \right), \qquad a^*=b^*=\left( \begin{array}{cc} 0 & 0 \\ 1 & 0 \end{array} \right),
\end{equation}
we can then write
$$
h_\cS=Ea^*a, \quad {\rm and} \quad h_\cE=E_0b^*b.
$$
The interaction operator is
\begin{equation*}\label{def:interaction}
v(\lambda)=\frac{\lambda}{2}(a\otimes b^*+a^*\otimes b). 
\end{equation*}
It induces an exchange process between $\cS$ and the subsystem $\cE_k$ it is coupled to: $\cS$ flips from its ground state to its excited state while $\cE_k$ flips the other way around, or vice versa (the parameter $\lambda$ is just a coupling constant). Note that $v$ has the particular feature that it commutes with the total number operator $N^{\rm tot}=a^*a\otimes\one+\one\otimes b^*b$.

It remains to specify the reference states of the subsystems $\cE_k$. They will be thermal states at some inverse temperature $\beta_k$:
$$
\rho_{\cE_k}=\frac{\e^{-\beta_k h_\cE}}{\tr \left(\e^{-\beta_k h_\cE}\right)}= Z_{\beta_k}^{-1}\e^{-\beta_k h_\cE}=:\rho_{\cE_k,\beta_k}.
$$

The calulation of the RDM $\cL$ associated to a subsystem $\cE$ at inverse temperature $\beta$ and interacting with $\cS$ for an amount of time $\tau$ is a straightforward calculation since $h=h_\cS+h_\cE+v(\lambda)$ can be easily diagonalized.
\end{ex}

\begin{exo} Prove that
\begin{equation*}\label{eq:spinspinrdm}
\cL(\rho)=\sum_{\sigma,\sigma'=0,1} V_\ss \rho V_\ss^*
\end{equation*}
where the operators $V_\ss$ are given by
\begin{equation*}\label{eq:vsigma}
\begin{array}{ll}
\ds V_{00}=\frac{1}{\sqrt{Z_{\beta}}}\,\e^{-\i\tau\frac{E+E_0}{2} N}\,C(N)^*,& 
\ds V_{10}=\frac{1}{\sqrt{Z_{\beta}}}\,\e^{-\i\tau\frac{E+E_0}{2} N}\,S(1-N)\,a, \\[16pt]
\ds V_{01}=\frac{\e^{-\beta E_0/2}}{\sqrt{Z_{\beta}}}\,\e^{-\i\tau\frac{E+E_0}{2} N}\,S(N)\,a^*,& 
\ds V_{11}=\frac{\e^{-\beta E_0/2}}{\sqrt{Z_{\beta}}}\,\e^{-\i\tau\frac{E+E_0}{2} N}\,C(1-N),
\end{array}
\end{equation*}
with $N=a^*a$ the number operator for $\cS$, 
$$
C(N)=\cos \left(\frac{\nu\tau}{2} N\right) +\i\frac{\Delta}{\nu}\sin\left(\frac{\nu\tau}{2} N\right), \quad S(N)=\frac{\lambda}{\nu}\sin\left(\frac{\nu\tau}{2} N\right),
$$
and where $\Delta=E-E_0$ and $\nu=\sqrt{\Delta^2+\lambda^2}$.
\end{exo}


\section{Asymptotic state of RIS}\label{sec:asymptoticstate}

In the first two sections, and in order to keep the exposition as simple as possible, we shall stick to the case where the RI system is described using the hamiltonian formalism. Only when we will add an extra reservoir we will have to turn to the Liouvillian description. Throughout this section, we will assume that the small system is finite dimensional, i.e. ${\rm dim}(\fh_\cS)<+\infty$.


\subsection{Ideal case}\label{ssec:idealri}

We start our analysis of RI systems with the simplest ``ideal'' case of identical interactions, i.e. $\cE_n\equiv\cE$, $\tau_n\equiv\tau$, etc. In this case, the reduced dynamics maps $\cL_n$ do not depend on $n$ and we are essentially led to the study of powers of $\cL$. Since $\cL$ is a contraction for the trace norm, its spectrum lies in the complex unit disk. Moreover, since $1$ is an eigenvalue for $\cL^*$ and $\cS$ has finite dimension, it is also an eigenvalue for $\cL$. This means that the system possesses at least one invariant state which is therefore a natural candidate for the limiting state of the system. The results of this Section are the direct translation in the hamiltonian formalism of the ones of \cite{BJM1}.

We first consider observables on the small system $\cS$, i.e. $A=A_\cS\otimes\one_\env$. As we argued in the previous section, the asymptotic behaviour of expectation values for such observables can be reduced to the analysis of $\cL^n$ as $n$ goes to infinity. In all this section we will assume the following ergodicity hypothesis which is a kind of Fermi Golden Rule.\\
\begin{itemize}
\item[\!\!\!\!\! {\bf (E)}]\ \ The spectrum of $\cL$ on the complex unit circle consists in the single eigenvalue $\{1\}$ and this eigenvalue is simple.
\end{itemize}
\medskip

\begin{rem} If Assumption {\bf (E)} holds then the eigenspace corresponding to the eigenvalue $1$ has dimension $1$. It then follows from the fact that $\cL$ is a (completely) positive and trace preserving map that $\cL$ possesses a unique invariant positive and trace one element $\rho_{\cS,+}$, i.e. a unique invariant state. 
\end{rem}

\begin{thm}\label{thm:idealsmall} Suppose Assumption {\bf (E)} is satisfied. Then there exist $C,\gamma>0$ such that for any initial state $\rho \in \cB^1(\fh_\cS)$
$$
\| \cL^n(\rho)-\rho_{\cS,+}\|_1 \leq C\e^{-\gamma n}, \qquad \forall n\in \N,
$$
where $\rho_{\cS,+}$ is the unique invariant state of $\cL$. In other words, for any observable $O_\cS\in\cB(\fh_\cS)$,
\begin{equation}\label{eq:idealstate}
\tr_\fh\left( \rho^{\rm tot}(n) O_\cS\otimes \one_\env\right)=\tr_{\fh_\cS}\left(\rho(n) O_\cS\right)=\rho_{\cS,+}(O_\cS)+O(\e^{-\gamma n}).
\end{equation}
\end{thm}

\noindent Note that the asymptotic state does not depend on the initial state of $\cS$.

\begin{rem} If the ergodic assumption {\bf (E)} is not satisfied then the limit $\ds \lim_{n\to\infty} \cL^n(\rho)$ still exists, in a weaker sense. Namely, if there are eigenvalues different from $1$ on the circle, then the limit exists in the ergodic mean sense, $\ds \frac{1}{N}\sum_{n=0}^{N-1} \cL^n(\rho)=\rho_{\cS,+}+O\left(\frac{1}{N}\right)$. Further, if $1$ is a degenerate eigenvalue of $\cL$ then $\cL$ possesses several invariant states. Hence one can still prove that $\cL^n(\rho)$ has a limit but the latter will depend on the initial state $\rho$.
\end{rem}

\begin{exo}\label{exo:toyideal} Consider the RI system of Example \ref{ex:toy-description-rdm}. 

1) Prove that the eigenvalues of $\cL$ are $1$, $\ds e_\pm=\e^{\pm\i\tau\frac{E+E_0}{2}}\left( \cos\left(\frac{\nu\tau}{2}\right)\pm\i \frac{\Delta}{\nu}\sin\left(\frac{\nu\tau}{2}\right)\right),$ and
\begin{equation}\label{def:e0}
e_0=1-\frac{\lambda^2}{\nu^2}\sin^2\left(\frac{\nu\tau}{2}\right). 
\end{equation}

2) At what condition does the map $\cL$ satisfies {\bf (E)}? In that case, show that the unique invariant state is $\rho_{\cS,+}=\rho_{\cS,\beta^*}$ where $\rho_{\cS,\beta^*}$ is the Gibbs state of $\cS$ at inverse temperature $\beta^*=\frac{E_0}{E}\beta$, and that one can take $\gamma=-\log\left(\sqrt{e_0}\right)$.
\end{exo}


\subsection{Random case}\label{ssec:randomri}

In this section, we turn to a more general situation where the various interactions are not identical. Of course, if one considers arbitrary interactions it is hopeless to expect any convergence (even in the ergodic mean) to some invariant state. As we mentioned in Section \ref{sec:framework}, see (\ref{eq:markov2}), the asymptotic behaviour of the system is essentially described by the product of reduced dynamics operators: $\cL_n\circ \cdots \circ \cL_1$. If the $\cL_n$'s are more or less arbitrary, anything can happen. We shall consider here the case where the interactions are random (but still independent identically distributed). Closely related results can be found in \cite{NPe}. This randomness may have various origins: the interaction time, the reference state of the $\cE_n$'s (via e.g. the temperature), the subsystems $\cE_n$ themselves (and hence the interaction operators),... All these parameters are eventually encoded in the RDM and our assumption will be that the sequence $(\cL_n)_n$ will be independent and identically distributed (i.i.d.).

To motivate this analysis, consider the ``One-Atom Maser'' experiment where a beam of atoms interacts with modes of the quantized electromagnetic field. It is clear that in actual experiments, neither the interaction time $\tau_n$ nor the reference states of the subsystems $\cE_n$ can be exactly the same for all $n$! Typically, the interaction time will be {\it random} (because of the random velocities of the atoms in the beam, see \cite{BRH,FJM}), given e.g. by a Gaussian distribution around some mean value, and the state of the incoming atoms will be random as well, for instance determined by a temperature that fluctuates slightly around a mean temperature (in experiments, the atoms are ejected from an atom oven, then they are cooled down to a wanted temperature before entering the cavity). One could also imagine that the subsystems $\cE_n$ themselves are not all the same (e.g. different kind of atoms, or maybe some impurities).

Another motivation is to consider a non-equilibrium situation. In the general setup of open quantum systems one gets a non-equilibrium situation when the environment is made of several reservoirs, each of them being in an equilibrium state but with different intensive thermodynamic parameters (different temperatures for instance). Then one expects the joint system $\cS+\cR_1+\cdots$ to relax towards a non-equilibrium steady state (NESS). Such states have been constructed in \cite{R,AH,JP2,AP,MMS,CNZ}. Among other things, they carry currents and have non vanishing entropy production rate. These transport properties were investigated in \cite{FMU,CJM,AJPP,N}. The linear response theory (Green-Kubo formula, Onsager reciprocity relations, central limit theorem) was developed in \cite{FMU},\cite{JOP1}-\cite{JOP4},\cite{JPP1}.

In the framework of RI systems, we can create a non-equilibrium situation by imposing the initial state of the subsystems $\cE_n$ to be for example thermal equilibrium states at different temperatures. In other words, we assume that the system $\cS$ interacts with $K$ ``reservoirs'' at a priori different temperatures, i.e. for any $m\in\N$, $\rho_{\cE_{mK+1}}=\rho_{\cE_1,\beta_1}$, $\rho_{\cE_{mK+2}}=\rho_{\cE_2,\beta_2}$, etc... where $\rho_{\cE,\beta}$ is the KMS-state of $\cE$ at inverse temperature $\beta$, see Section \ref{ssec:fluxes}. (One could imagine a ``One-Atom Maser'' where the field in the cavity is coupled to $K$ beams at different temperatures.) However, the particular structure imposed here leads to a lack of symmetry and in particular the system is not at all time reversal invariant (reservoir $2$ always interacts right after reservoir $1$ while the inverse is not true). A direct consequence is that Onsager reciprocity relations do not hold. One way to restore symmetry is then to chose the temperature of the $n$-th subsystem in a random way from the set $\{\beta_1,\ldots,\beta_K\}$, each temperature having probability $\frac{1}{K}$ to occur. The results of this Section come from \cite{BJM3}.

Let $(\Omega_0,\cF,\p)$ be a probability space. To describe the stochastic dynamic process at hand, we introduce the standard probability measure $\d\P$ on $\Omega:=\Omega_0^{\N^*}$,
\begin{equation*}\label{def:probaspace}
\d\P =\Pi_{j\geq 1}\d \p_j, \ \ \ \mbox{where } \ \ \ \d \p_j\equiv \d
\p, \ \  \forall j\in \N^*.
\end{equation*}
We denote by $\omega=(\omega_n)_n$ the elements of $\Omega$. As we already mentioned, we will assume that the various interactions are independent and identically distributed (i.i.d.). This is precisely the meaning of the following randomness assumption 
\bigskip
\begin{itemize}
\item[\!\!\!\!\! {\bf (R1)}]\ The reduced dynamics maps $\cL_n$ are i.i.d. random operators (on $\cB^1(\fh_\cS)$). We write $\cL_n=\cL(\omega_n)$, where $\cL:\Omega\rightarrow \cB\left(\cB^1(\fh_\cS)\right)$ is an operator valued random variable.
\end{itemize}
\bigskip
Throughout this section, we will assume, without further mentioning it, that Assumption {\bf (R1)} is satisfied. Finally let $\cM_{(E)}$ be the set of RDM's which satisfy the ergodic assumption {\bf (E)}. 

To indicate the randomness we shall denote the state ``at time $n$'' (i.e. after $n$ interactions) by
$$
\rho(n,\omega)=\left(\cL(\omega_n)\circ \cdots \circ \cL(\omega_1)\right)(\rho).
$$
The following theorem shows that the RI system relaxes almost surely in the ergodic mean towards a deterministic asymptotic state.
\begin{thm}\label{thm:randomstatesmall} \ Suppose that $\p(\cL(\omega_0)\in\cM_{(E)})>0$. Then,
\begin{enumerate}
 \item $\E(\cL)$ satisfies (E),
 \item there exists a set $\tOmega\subset \Omega$, s.t.  $\P(\tOmega)=1$, and s.t. for any $\omega\in\tOmega$, any initial state $\rho$,
\begin{equation*}\label{eq:randomstate}
\lim_{N\to\infty} \frac{1}{N}\sum_{n=1}^N \rho(n,\omega)=\rho_{\cS,+},
\end{equation*}
where $\rho_{\cS,+}$ is the unique invariant state of $\E(\cL)$.
\end{enumerate}
\end{thm}

The convergence in the ergodic mean is rather generic for system out of equilibrium. Actually, in an equilibrium-like situation one could expect a stronger convergence result. In our setting of random RI systems this appears in the following
\begin{thm}\label{thm:randomstatesmallstronger} \ Suppose that $\p(\cL(\omega_0)\in\cM_{(E)})>0$ and that there exists $\rho_+$ such that $\cL(\omega_0)(\rho_{\cS,+})=\rho_{\cS,+}$ for a.e. $\omega_0$, i.e. there is a deterministic invariant state. Then there exists a set $\tOmega\subset \Omega$, s.t. $\P(\tOmega)=1$ and $\alpha>0$ s.t. for any $\omega\in\tOmega$ there exists $C(\omega)>0$ such that for any $\rho\in\cB^1(\fh_\cS)$
$$
\|\rho(n,\omega)-\rho_{\cS,+}\|_1 \leq C(\omega)\e^{-\alpha n}, \quad \forall n\in\N.
$$
\end{thm}

\begin{exo}\label{exo:toyrandom} Consider the RI system of Example \ref{ex:toy-description-rdm}.

1) Suppose that $\beta_n\equiv\beta$, and that $\tau(\omega_0)$ is a random variable satisfying $\p \left(\nu\tau\notin 2\pi\N \right)\neq 0$. Prove that there exists $\alpha>0$ such that for any initial state $\rho$ and $\P$-a.s. 
$$
\|\rho(n,\omega)-\rho_{\cS,\beta^*}\|\leq C(\omega)\e^{-\alpha n}, \qquad \forall n\in\N,
$$
for some $C(\omega)>0$ and where $\beta^*=\frac{E_0}{E}\beta$. 

2) Suppose that $\beta(\omega_0)$ is a random variable, and that $\tau_n\equiv\tau>0$ satisfies $\nu\tau\notin 2\pi\N$. Prove that, for any initial state $\rho$, $\rho(n,\omega)$ converges $\P$-a.s. in the ergodic mean towards the Gibbs state of $\cS$ at inverse temperature $\beta':=-E^{-1}\log \big(\E[Z_{\cS,\beta(\omega_0)E_0/E}^{-1}]^{-1}-1\big)$. In other words,
$\ds \rho_{\cS,+}=\E\left( \rho_{\cS,\beta^*(\omega_0)} \right)$ where $\beta^*(\omega_0)=\beta(\omega_0)E_0/E$.
\end{exo}


\section{RIS in various limiting regimes}\label{sec:limitingregimes}

This section is devoted to the study the weak coupling limit, or Van Hove limit, and variations thereof, of RIS of the sort described above. The weak coupling limit is a widely used tool to produce effective dynamics in a regime in which the total duration of the evolution scales like the square of the  inverse power of the coupling intensity between the small system and the environment. In the weak coupling limit, the focus is again put on the small system so that the environment only appears in the effective dynamics provided in the limit. When dealing with a discrete time dynamics, this procedure of perturbative nature allows one in general to define a continuous dynamics on the small system that captures some essential features of the original dynamics. The existence of effective dynamics obtained by a weak limit procedure is known for a large class of time-independent Hamiltonian systems, as well as in certain time-dependent situations, see e.g. \cite{Da1}, \cite{Da2}, \cite{DS}, \cite{LS}, \cite{DJ}, \cite{AJP}. The results provided in \cite{AJ1} which we describe here, show that RIS can be added to this list. See also \cite{V} for generalizations. 

As mentioned above, the primary motivation to study weak coupling regimes comes from \cite{AP}. Although the setup considered here is much simpler, some conclusions of \cite{AP} can be reached without going to a continuous limit of Quantum Noises. We refer the reader to volume two of \cite{AJP} for a detailed presentation of the latter. A particular consequence of models in which the environment is described in terms of  Quantum Noises, is that if one traces over the environment degrees of freedoms, a Lindbladian description of  the reduced dynamics of states on $\fh_\cS$ naturally emerges from the analysis, see \cite{AP}. See also \cite{AJ2} for positive temperature results. Similar conclusions are drawn below in the simpler setup of RIS.

\bigskip

The framework is that of the ideal case in which the small quantum system is defined  on a finite dimensional Hilbert space $\fh_\cS$ coupled to an environment made of an infinite chain of identical independent $n+1$-level sub-systems, with $n$ finite, on 
$\fh_\env := \bigotimes_{m\geq 1} \fh_{\cE_m}$. The coupling between the system 
$\cS$ and the environment is provided by identical interactions with each individual sub-system of the chain, for the same duration $\tau>0$. 
Hence, over a macroscopic time interval 
$]0,k\tau]$, $k\in\N^*$, the small system is coupled with elements 
$1,2, ..., k$ of the chain, in sequence with the same interaction of strength $\lambda$. 
The interaction is given by $\lambda$ times an operator acting on $\fh\otimes\fh_{\cE_m}$ of the form
$v_m=\sum_{j=0}^nV_j^*\otimes a_j+V_j\otimes a_j^*$. Here the $a_j^*$'s and $a_j$'s are similar to
creation and annihilation operators relative to the levels of the $k$-th sub-system $\cE_k$ 
and the $V_j$'s are arbitrary operators on $\fh_\cS$. Instead of the evolution of states, we consider equivalently the Heisenberg 
evolution of observables on the small system when the chain is initially 
at equilibrium at positive temperature.
\medskip

The long time results presented in the previous section
allows one to expect that an effective continuous dissipative dynamics on the small system  arises  when the number $k$ of discrete interactions goes to infinity and 
the coupling $\lambda$ with the chain elements is weak. 
\medskip

We start with the familiar weak coupling regime by choosing  $t>0$ fixed and considering $\N\ni k=t/\lambda^2$ 
so that the macroscopic time scale equals $T=\tau t/\lambda^2$. 
We show the existence an effective dynamics driven by a $\tau$ dependent generator
which we determine. 
The supplementary parameter given by the microscopic interaction time $\tau$ 
allows us to explore different asymptotic regimes, as $\tau$ goes to 
zero:
We  extend the analysis
to the whole range $\tau\rightarrow 0$, $\lambda^2\tau\rightarrow 0$
over macroscopic time scales $T=t/(\tau\lambda^2)\rightarrow \infty$.
The analysis of these first two regimes is 
strongly related to regular perturbation theory in the parameter $\lambda^2\tau$.
The divergence of the macroscopic time scale imposes, as usual, some 
renormalization of
the dynamics by the restriction of the uncoupled dynamics. 
Note however that in the second regime,
the interaction strength  $\lambda$ is not required to go to zero and 
can even diverge.
The common feature of the generators of the dynamics of observables obtained in 
these first two regimes is that they commute with the generator 
$i[h_\cS,\cdot ]$ of the uncoupled 
unitary evolution restricted to $\fh_\cS$. In other words, 
the corresponding effective dynamics admits the commutant of $h_0$ as a non trivial 
invariant sub-algebra of observables. This is a well known feature of  the weak 
coupling regime for time-independent Hamiltonians, \cite{Da1}, \cite{LS}, \cite{DJ}.\\

The third regime, tightly linked to the scaling used in \cite{AP}, 
is characterized by $\tau\rightarrow 0$, while the product $\lambda^2\tau$ is kept
constant.  This leads us beyond the perturbative and yields  
a macroscopic time scale $T=t/(\lambda^2\tau)=t$, which is finite. The analysis of this critical case makes 
use of Chernoff's Theorem, rather than perturbative methods.
Within this scaling, one shows that an effective Heisenberg 
dynamics for observables on $\fh_\cS$ arises at any temperature.
It is generated by a general Lindblad operator 
whose dissipative part is explicitly constructed in terms of the
$V_j$'s defining the coupling in the Hamiltonian, whereas its conservative 
part is simply $i[h_\cS,\cdot]$. The Lindbladian generator coincides with the one driving the effective Heisenberg dynamics of observables on $\fh_\cS$ obtained  by means of quantum noises \cite{AP}, \cite{AJ2} at zero and positive temperature.
A particular trait of these generators is that they do not commute with $i[h_\cS,\cdot]$ anymore, 
the generator of the uncoupled evolution restricted to $\fh_\cS$. Hence, there is 
no obvious sub-algebra of observables  left invariant  by the effective dynamics
of observables. 

We remark at his point that the modeling of 
the dynamics of observables (or states) of a small system in contact with a 
reservoir at a certain temperature often starts with a choice of a certain 
Lindblad generator suited to the physical phenomena to be discussed. 
The analysis presented allows to assign to any given Lindblad generator 
a simple model of repeated quantum interactions, with explicit couplings constructed
from the Lindblad generator, whose effective dynamics in the limit 
$\tau\rightarrow 0$, $\lambda=1/\sqrt{\tau}$, is generated by the chosen Lindblad operator.

\medskip

The links between the generators obtained in the three regimes considered are discussed below.

\setcounter{equation}{0}
\subsection{Setup}

The small system, described by the Hilbert
space $\fh_\cS$ of dimension $d+1>1$,  is characterized by a Hamiltonian $h_\cS$. The Hilbert space of the chain elements is $\C^{n+1}$, that of the infinite chain is 
$\fh_\env=\otimes_{j\geq 1}\C^{n+1}$, $n\geq1$ so that the total Hilbert space is $\fh_\cS \otimes \fh_\env$. We introduce some notations that will prove useful below.

The vacuum vector $\Omega\in \fh_\env$ 
is defined as the infinite tensor product of the vacuum vector 
$\omega= \begin{pmatrix}0 &\cdots& 0 & 1\end{pmatrix}^T$ in $\C^{n+1}$, the stabilizing sequence,
\be
\Omega=\omega \otimes \omega \otimes\omega \otimes  \cdots \ \in \  
\C^{n+1}\otimes \C^{n+1}\otimes \C^{n+1}\otimes \cdots .
\ee
We define the $i$th excited vectors by $x_i=\begin{pmatrix}0 & \cdots & 0&1 & 0&\cdots& 0\end{pmatrix}^T$, where the
$1$ sits at the $i$th line, starting from the bottom, $i=1,2, \cdots , d$. Hence
the corresponding excited state at site $j\geq 1$ is given by 
\be
x_i(j)=\omega \otimes \cdots  \otimes\omega \otimes x_i
\otimes\omega \otimes  \cdots ,
\ee
where $x_i$ sits at site $j\geq 1$.
More generally, given a finite set 
\be\label{defs}
S=\{ (k_1, i_1), (k_2,i_2), \cdots, (k_m, i_m) \} \subset 
(\N^*\times \{1,2, \cdots , d\})^m \ \ \mbox{ with all } k_j\mbox{'s distinct},
\ee 
we denote by $X_{S}$ the vector 
given by an infinite tensor product as above, with $i_j$th excited vectors $x_{i_j}(k_j)$
at all sites $k_j\geq 1$, $j=1,\cdots, m$,  and ground state vectors $\omega$ everywhere else. Then, 
$\fh_\env$ is the completion under the norm arising from the inner
product of linear combinations of such vectors.
This construction together with the vacuum $\Omega\equiv X_\emptyset$ yield an orthonormal basis 
 of $\fh_\env$, when $S$ runs over all finite sets of the type above.

Correspondingly, we introduce "creation" and "annihilation" operators associated with the vectors $x_i(j)$ as follows.
Let $a_i$ and $a_i^*$, $i=1,2,\cdots, n$, denote the operators corresponding to 
$\{\omega, x_1, \cdots , x_n\}$ in $\C^{n+1}$, i.e. such that
\bea\label{aadag}
& & a_i x_i=\omega, \ \ a_i \omega =a_i x_j =0, \ \mbox{ if } \ j\neq i,\nonumber\\
& & a_i^* \omega =x_i, \ \ a_i^* x_j =0 \ \mbox{ for any } \ j=1,2,\cdots, n.
\eea
Then, for $j\geq 1$, the operators $a_i(j)$ and $a_i(j)^*$ on 
$\fh_\env$ are defined as acting as $a_i$ and $a_i^*$ on the $j$th copy of 
$\C^{n+1}$ at site $j$, and as the identity everywhere else. 
The Hamiltonian of one sub-system at site $j\geq 1$ can thus be written as
\be
h_{\cE_j}= \sum_{i=1}^n \delta_i a_i(j)^* a_i(j),
\ \mbox{ with } \  \delta_i\in\R.
\ee
\begin{exo} Compare the operators (\ref{aadag}) with the familiar creation and 
annihiliation operators. Show that  for $i$ fixed, they satisfy the canonical anti-commutation rules when restricted to the
two dimensional subspace spanned by $\{\omega, x_i\}$ and are zero on the orthogonal complement 
of this subspace. Show that the spectrum $\sigma(h_{\cE_j})=\{0\}\cup\{\delta_i\}_{i=1,\dots, n}$, for each $j\in \N^*$
\end{exo}

In keeping with this notation, we can introduce a basis of 
eigenvectors of $h_\cS$  in $\fh_\cS$ of the form
\be
\{\omega, x_1, x_2, \cdots, x_d\}, \ \ \mbox{ where } \ d=\dim(\fh_\cS)-1.
\ee
While $d\neq n$ in general, we shall nevertheless write sometimes
 $\omega(0)$ and $\{x_i(0)\}_{i=1,2, \cdots, d}$ to denote these vectors, slightly abusing the notation.
No confusion should arise with vectors of $\fh_\env$ above, since we labelled the sites
of the sub-systems, or spins, by positive integers.

The corresponding  time dependent Hamiltonian $h(t,\lambda)$ on $\fh_\cS\otimes\fh_\env$, (\ref{def:totalham}) is now completely defined
with the interaction given for $t\in [\tau (k-1) ,\tau k[$ by 
\be\label{defint}
v_k=\sum_{i=1}^n V^*_i\otimes a_i(k)+V_i\otimes a_i(k)^*, 
\ee
where the $V_i$'s are bounded operators on $\fh_\cS$.
Notations are made more compact by introducing vectors with operator
valued entries. Let
$a(j)^{\sharp}=\begin{pmatrix} a_1(j)^{\sharp} & a_2(j)^{\sharp} & \cdots & a_n(j)^{\sharp}\end{pmatrix}^T$, 
$V^\sharp=\begin{pmatrix}V_1^\sharp & V_2^{\sharp} & \cdots & V_n^{\sharp}\end{pmatrix}$
where $^{\sharp}$ denotes either nothing or $*$. Then, with the rules of matrix composition, 
we write 
$
V^{\sharp_1}\otimes a(j)^{\sharp_2}=\sum_{i=1}^n V^{\sharp_1}_i\otimes a_i^{\sharp_2}(j),
$
so that the interaction  for $t\in [\tau (k-1) ,\tau k[$ reads 
$
v_k=V^*\otimes a(k)+V\otimes a(k)^*
$.
 Similarly, with $\delta = \begin{pmatrix}\delta_1  & \delta_2  & \cdots & \delta_n \end{pmatrix}$, and
$a(j)^{*}a(j)=\begin{pmatrix}a_1(j)^{*}a_1(j) & a_2(j)^{*}a_2(j) & \cdots & 
a_n(j)^{*}a_n(j)\end{pmatrix}^T $
we have
$
h_{\cE_j}=\delta  a(j)^*a(j).
$

\medskip

Let us denote the corresponding evolution operator
between the time $\tau (k-1)$ and $\tau k$ by $U_k$, 
\be\label{brick}
U_k=e^{-i\tau (h_\cS+ h_{\cE_k}+ \lambda v_k+\sum_{j\neq k}h_{\cE_j})}
\equiv e^{-i\tau (h_k+\sum_{j\neq k}h_{\cE_j})},
\ee
so that the evolution between times $0$ to $\tau n$ is given by 
\be\label{evolAJ}
U(n,0)=U_nU_{n-1} \cdots U_k\cdots U_1.
\ee

Finally, let $P$ be the projection from $\fh_\cS\otimes \fh_\env$ to the subspace $\fh_\cS\otimes \C \Omega$
defined by
\be
P=\one \otimes |\Omega\ket\bra\Omega|,
\ee
such that $P \ \fh_\cS\otimes \fh_\env$ can be identified with 
$\fh_\cS$, the Hilbert space of the small system. This projector is directly linked to 
the partial trace to be performed on $\fh_\env$.

\subsection{Preliminary Estimates}

The limit $\lambda\rightarrow 0$ of the unitary evolution operators (\ref{evolAJ}) 
in a controlled way as $\tau\rightarrow 0$ rests on general perturbative estimates we describe here.

\medskip

Dropping the index $k$ in (\ref{brick}) the generator takes the form
\be
H(\lambda)=H(0)+\lambda W, \ \ \ \mbox{with} \ \ \ 
H(0)=h_\cS+ h_{\cE_k}+ \sum_{j\neq k}h_{\cE_j}
\ \ \mbox{and } \ \ W=V^* a +V a^*,
\ee
and the projector $P$ restricted to $\fh_\cS\otimes \fh_{\cE_k}$ writes 
$
P=\I- a^*a.
$
\\
Assume the following general framework:
\bigskip
\begin{itemize}
\item[\!\!\!\!\!{\bf (H1)}]  Let $P$ be a projector on a Banach space $\cB$ and $H(\lambda)$ be 
an operator of the form
\be
H(\lambda)=H(0)+\lambda W,
\ee
where $H(0)$ and $W$ are bounded and $0\leq\lambda\leq\lambda_0$ for
some $\lambda_0>0$.  Further assume that
\be
[P,H(0)]=0\ \ \ \mbox{and }\ \  W=PWQ+QWP \ \ \mbox{where }\ \ \ Q=\one -P.
\ee
\end{itemize}
\bigskip

\noindent
We consider 
\be\label{lop}
U_\tau(\lambda)=e^{-i\tau H(\lambda)}. 
\ee
For later purposes, we also take care of the dependence in $\tau$ of the 
error terms, as both $\lambda$ and 
$\tau$ tend to zero, independently of each other. The first perturbative result reads
\begin{lem}\label{easypert}
Let {\bf H1} be true. Then, as $\lambda$ and $\tau$ go to zero, 
\bea \label{ff}
&&e^{-i\tau(H(0)+\lambda W)}=e^{-i\tau H(0)}+\lambda F(\tau)+\lambda^2 G(\tau)
+O(\lambda^3\tau^3)\\
\label{pup}
&&Pe^{-i\tau(H(0)+\lambda W)}P=Pe^{-i\tau H(0)}P+\lambda^2 PG(\tau)P+PO(\lambda^4\tau^4)P,
\eea
where 
\bea\label{deff}
F(\tau)&=&\sum_{n\geq 1}\frac{(-i\tau)^n}{n!}
\sum_{m_j\in \N \atop m_1+m_2=n-1}H(0)^{m_1}WH(0)^{m_2}  \nonumber\\
&=&-i e^{-i\tau H(0)}\int_0^\tau ds_1 e^{i s_1 H(0)}We^{-i s_1 H(0)}\\
\label{defg}
G(\tau)&=&\sum_{n\geq 2}\frac{(-i\tau)^n}{n!}\sum_{m_j\in \N \atop m_1+m_2+m_3=n-2}
H(0)^{m_1}WH(0)^{m_2}WH(0)^{m_3}\nonumber\\
&=&- e^{-i\tau H(0)}\int_0^\tau ds_1 \int_0^{s_1} ds_2  e^{i s_1 H(0)}W e^{-i (s_1-s_2) H(0)}
W e^{-i s_2H(0)}.
\eea
\end{lem}
{\bf Remark:} Formula (\ref{ff}) is true without assuming that
$W$ is off-diagonal with respect to $P$ and $Q$.

Specialized to the Hilbert space context, we have some more properties of the expansion of $U_\tau(\lambda)$ for
$\lambda>0$ small, $\tau>0$ 
\begin{cor}\label{pertuu} Assume $\cB$ is a Hilbert space, $H(0)$, $W$ and $P$
are self-adjoint and $\lambda, \tau $ are real.
As $\lambda\rightarrow 0$, the operator
$U_\tau(\lambda)=e^{-i\tau H(\lambda)}$ satisfies
\bea
&&U_\tau(\lambda)=e^{-i\tau H(0)}+\lambda F(\tau)+\lambda^2 G(\tau)+O(\lambda^3\tau^3)\\
&&U_\tau(\lambda)^{-1}=U_\tau(\lambda)^*=U_{-\tau}(\lambda)\nonumber\\
&&\quad \quad \quad \quad =e^{i\tau H(0)}+\lambda F(-\tau)+\lambda^2 G(-\tau)+O(\lambda^3\tau^3)
\eea
\end{cor}
\begin{exo}
Prove Lemma \ref{easypert} by making use of an expansion of the exponential and by the Dyson series in the interaction picture 
$
\Theta(\lambda, \tau)=e^{i\tau H(0)}e^{-i\tau(H(0)+\lambda W)} 
$
given by the convergent expansion
\bea\label{cv}
\Theta(\lambda, \tau)&=&\sum_{n=0}^{\infty} (-i\lambda)^n\int_0^\tau ds_1\int_0^{s_1} 
ds_2 \cdots \int_0^{s_{n-1}}ds_n e^{i s_1 H(0)}W\times \nonumber\\
& & \times e^{-i (s_1-s_2) H(0)}W e^{-i (s_2-s_3) H(0)}
\cdots e^{-i s_{n-1}-s_n) H(0)}We^{-i s_n H(0)}.
\eea
Then prove Corollary \ref{pertuu}, making use of the fact that $H(\lambda)$ is self-adjoint. 
\end{exo}

The technical basis underlying all weak limit results to follow
is contained in Proposition \ref{weakprop}, essentially due to Davies \cite{Da1}. In words, this proposition relates high powers of an approximate isometry to hight powers of an exponential, to leading order.  

\begin{prop}\label{weakprop}  Let $V(x)$, $x\in [0,x_0)$ and $R$ be 
bounded operators on a Banach  
space $\cB$ such that,  in the operator norm, 
$V(x)=V(0)+ x R + O(x^2)$, where $V(0)$ is an isometry admitting
the spectral decomposition $V(0)=\sum_{j=0}^r e^{-iE_j}P_j$ and 
let $h=\sum_{j=0}^r E_jP_j$. Then, 
for any $0\leq t\leq t_0$, if $x\rightarrow 0$ in such a way that $t/x\in\N$, 
\be
V(0)^{-t/x}V(x)^{t/x}=e^{t{e^{ih}R}^\#}+O(x), \ \ \ \mbox{ in norm,}
\ee
where $K^\#=\sum_{j=0}^r P_jKP_j$, for any $K\in {\cal L}(\cB)$.
\end{prop}
\begin{exo}{ \ \\}
i) Show that the projectors $P_j$  are of norm one, using Von Neumann's ergodic theorem.\\
ii) Prove the alternative expression
$ K^\#=
\lim_{T\rightarrow\infty}\frac{1}{T}\int_0^Te^{i s h}Ke^{-i s h}\, ds
$ for any bounded operator $K$, 
and deduce the estimate  $\|K^\#\|\leq \|K\|$.\\
 iii) Check that the operator in the exponent can be rewritten as
\be
e^{ih}R^\#=R^\#e^{ih}=(e^{ih}R)^\#=(Re^{ih})^\#. 
\ee
\end{exo}
\begin{rem}
All hypotheses are made on the isometry $V(0)$, not on the operator 
$h$.\\
\end{rem}

Let us now further assume:
\bigskip
\begin{itemize}
\item[\!\!\!\!\!{\bf (H2)}]  The restriction $H_P(0)$ of $H(0)$ to $P\cB$ is diagonalizable
and reads 
\be
H_P(0)=\sum_{j=0}^{r} E_j P_j, \ \ \mbox{ with } \ \ \dim(P_j)\leq \infty, \ \ 
r \, \mbox{ finite }.
\ee
Moreover, the operator $Pe^{-i\tau H(0)}=Pe^{-i\tau H_P(0)}$
is an isometry on $P\cB$.\end{itemize}
\bigskip

Note that this implies $Pe^{-i\tau H_P(0)}$ is invertible and
\be
P=\sum_{j=0}^{r}P_j, \ \ \ E_j \in \R \ \ \ \forall 
j=0,\cdots, r,\ \ \ \mbox{ and }\ \ \
Pe^{-i\tau H(0)}=\sum_{j=0}^{r} e^{-i\tau E_j} P_j,
\ee
where the projectors $P_j$ are eigenprojectors of $Pe^{-i\tau H(0)}$ iff the
$e^{-i\tau E_j}$'s are distinct.
In case $\cB$ is a finite dimensional Hilbert space and $H(0)$ is self adjoint, 
{\bf H2} is automatically true.

The perturbation formulae above together with Proposition \ref{weakprop} yield a general statement 
in a Banach space framework under assumptions {\bf H1} and {\bf H2}, 
taking into account both parameters $\lambda$ and $\tau$:

\begin{thm}\label{theo31}
Suppose Hypotheses {\bf H1} and {\bf H2} hold true and further assume the spectral projectors
$P_j$, $j=0,\cdots, r$, of $e^{-i\tau H_P(0)}$ coincide with those of $H_P(0)$ 
on $P\cB$. Set  $K^\#=\sum_{j=0}^r P_j K P_j$, for $K\in {\cal L}(\cB)$ and $G_P(\tau)=PG(\tau)P$.\\

A) Then, for any 
$0<t_0<\infty$, there exists $0<c<\infty$ such that for any $0\leq t \leq t_0$,
the following estimate holds in the limit $\lambda^2\tau \rightarrow 0$, $\lambda^2\tau^2\rightarrow 0$, 
and $t/(\lambda\tau)^2\in \N$:
\be
\left\|e^{iH(0)t/(\lambda^2 \tau)}\left[Pe^{-i\tau (H(0)+\lambda W)}
P\right]^{t/(\lambda \tau)^2}-e^{t \, e^{i\tau H(0)}G_P(\tau)^{\#}/\tau^2}\right\|
\leq c(\lambda^2\tau^2 +\lambda^2\tau).
\ee

B)  Then, for any 
$0<t_0<\infty$, there exists $0<c<\infty$ such that for any $0\leq t \leq t_0$,
the following estimate holds in the limit $\lambda^2\tau \rightarrow 0$, $\tau\rightarrow 0$, 
and $t/(\lambda\tau)^2\in \N$:
\be
\left\|e^{iH(0)t/(\lambda^2 \tau)}\left[Pe^{-i\tau (H(0)+\lambda W)}
P\right]^{t/(\lambda \tau)^2}-e^{-t(W^2)^{\#}/2}\right\|
\leq c(\tau +\lambda^2\tau).
\ee

\end{thm}
\begin{rem}
 If $\tau$ is small enough, the spectral projectors of $e^{-i\tau H_P(0)}$ and $H_P(0)$ 
on $P\cB$ coincide.
\end{rem}

\subsection{Heisenberg representation for non-zero temperature}\label{sechei}

We now come back to our model and consider the evolution of observables, instead of density matrices. The two points of view are equivalent, of course.  The Markovian nature of the model allows us to express the evolution of observables $B$ of the small system at positive temperature 
after $k$ repeated interactions as the action of the $k$-th power of an operator ${\cal U}_\beta(\lambda, \tau)$ on $\fh_\cS$, see Proposition
\ref{bkuk}. Then, we apply Proposition \ref{weakprop} to ${\cal U}_\beta(\lambda, \tau)$ to get weak limit types results stated as Theorem \ref{mtb} and Chernoff's Theorem to go beyond the perturbative regime \ref{thlin}.

We consider the equilibrium state $\bigotimes_{k=1}^N\rho_{\cE, \beta}$ of a chain 
of $N$ spins at inverse temperature $\beta$ where
\be
\rho_{\cE, \beta}=
\frac{e^{-\beta \delta a^*a}}{{\cal Z}(\beta)}.
\ee

If $\rho$ is any state on $\C^{d+1}$, the initial state of the
small system plus spin chain is 
$\rho\otimes \bigotimes_{k=1}^N\rho_{\cE, \beta}$. The Heisenberg
evolution of observables of the form $B\otimes \one_{\fh_\env}$, where 
$B\in M_{d+1}(\C)$, can be conveniently described as follows, for $k\leq N$:
\be\label{bkt}
B_\beta(k,\lambda,\tau)=\tr_{\fh_\env}((\one\otimes\bigotimes_{k=1}^N\rho_{\cE, \beta}) \,\, U(k,0)^{-1}
(B\otimes \one_{\fh_\env})U(k, 0))\in  {\cal L}(\fh_\cS),
\ee
where, for any $A\in {\cal L}(\fh_\cS\otimes \fh_\env)$ and $x_0=\omega$,
\be\label{partra}
\tr_{\fh_\env}(A)=\left(\sum_{S}\bra x_i\otimes x_S |\, A 
\, x_j\otimes x_S\ket 
\right)_{i,j\in\{0,\cdots, d\}}\in  {\cal L}(\fh_\cS)
\ee
 denotes the partial trace taken on the spin variables only.
 \begin{rem} The operator (\ref{bkt}) is actually the dual expression of the reduced dynamics map defined in equation (\ref{def:rdm0}).
 \end{rem}
Hence, the expectation in the state $\rho$ of the observable 
$B$ after $k$ interactions
over a time interval of length $k\tau$ with the chain at 
inverse temperature $\beta$ is given by
\be
\bra B(k, \beta)\ket_\rho=\tr_{\C^{d+1}}(\rho B_\beta(k,\lambda,\tau)).
\ee

Recall that
\be
U(k,0)^{-1}(B\otimes \one_{\fh_\env})U(k, 0)=
U_1^* U_{2}^*\cdots U_k^* (B\otimes \one_{\fh_\env})U_k U_{k-1}\cdots U_1,
\ee
where $U_j$ is non-trivial on $\C^{d+1}\otimes \C_j^{n+1}$ only. 

The partial trace operator $\tr_{\fh_\env}
((\one\otimes \bigotimes_{k=1}^N\rho_{\cE, \beta})\,\, A)$, where $A$ is an operator
on $\C^{d+1}\otimes \Pi_{j=1}^N\C_j^{n+1}$ can made more explicit.

\begin{lem}
Let us denote the matrix elements of $A$ as follows
\be
A^{i,j}_{S, S'}=\bra x_i\otimes X_S |A \,\, 
x_j\otimes X_{S'}\ket,
\ee
where $i,j$ belong to  $\{0,\cdots, d\}$, and
$S$, $S'$ run over subsets of $\left\{\{1, \cdots, N\}\times \{1,\cdots,n\}
\right\}^m$, for $m=0,\cdots, N$ as in (\ref{defs}). Then
\be\label{4.9}
\tr_{\cal H}
((\one\otimes\bigotimes_{k=1}^N\rho_{\cE, \beta})\,\, A)_{i,j}=
\sum_S\frac{e^{-\beta\sum_{l=1}^{n}\delta_l |S|_l}}
{(1+\sum_{l=1}^{n}e^{-\beta\delta_l})^N}A_{S,S}^{i,j}
\ee
where, for
\be
S=\{ (k_1, i_1), (k_2,i_2), \cdots, (k_m, i_m) \} \subset 
(\N\times \{1,2, \cdots , n\})^m 
\ee 
 with all  $1\leq k_j \leq N$ distinct and $m=0,\cdots, N$,
\be
|S|_l=\#\{k_r\ \  \mbox{s.t.}\ \ i_r=l\}.
\ee
\end{lem}
\begin{exo} Prove the formula (\ref{4.9}), making use of the identity
\be
\bigotimes_{k=1}^N\rho_{\cE, \beta}X_S=\frac{\Pi_{r=1}^me^{-\beta \delta_{i_r}}}{(1+\sum_j
e^{-\delta_j\beta})^N}X_S=
\frac{e^{-\beta \sum_{l=1}^n\delta_l |S|_l}}{(1+\sum_j
e^{-\delta_j\beta})^N}X_S.
\ee
\end{exo}

We now further compute the action of $U(k,0)$ given by the product of
$U_j's$. 
Let us denote the vectors $\omega\otimes X_S$ and $x_j\otimes X_S$ by 
$n_0\otimes |n_1, n_2, \cdots , n_N\ket
\equiv n_0\otimes |\vec n \ket $, where $n_0\in\{0,1,\cdots d\}$,
and  $n_j\in \{0,1, \cdots n\}$, for any $j=1,
\cdots N$, with $\omega \simeq  0$ and $x_k \simeq k$ and 
$X_{\{(1,n_1), \cdots, (N,n_N)\}}\simeq |\vec n \ket$.

Recall (\ref{brick})
\be
U_j=e^{-i\tau (h_j+\sum_{k\neq j}h_{\cE_k})}=e^{-i\tau h_j}e^{-i\tau \sum_{k\neq j}h_{\cE_k}},
\ee
where $e^{-i\tau\sum_{k\neq j}h_{\cE_k}}$ is diagonal. More precisely, with 
the convention $\delta_0=0$,
\be
e^{-i\tau\sum_{k\neq j}h_{\cE_k}}n_0\otimes |n_1, n_2, \cdots , n_N\ket=
e^{-i\tau\sum_{k=1\atop k\neq j}^N\delta_{n_k}}n_0\otimes |n_1, n_2, \cdots , n_N\ket.
\ee
Let us denote  the $k$-independent matrix
elements of $e^{-i\tau h_k}|_{\C^{d+1}\otimes \C_k^{n+1}}
$ by 
\be\label{notbloc}
U^{n,n'}_{m,m'}=\bra n\otimes m |e^{-i\tau h_k} \, n'\otimes m'\ket.
\ee
\begin{exo}  Iterating the formula
\be
U_1 \,\, n_0\otimes |n_1,\cdots, n_N\ket=\sum_{m_0^1=0,1, \cdots , d \atop 
m_1=0,1, \cdots, n}
e^{-i\tau \sum_{j>1}\delta_{n_j}}
U_{m_1,n_1}^{m_0^1,n_0}\,\, m_0^1\otimes |m_1, n_2, n_3, \cdots, n_N\ket.
\ee
show that 
for any
$N\geq k$
\bea\label{matel} 
&& U_kU_{k-1}\cdots U_2U_1 \,\, n_0\otimes |n_1,\cdots, n_N\ket=\sum_{\vec{m_0}\in \{0,\cdots, d\}^{k} \atop \vec m\in \{0,\cdots, n\}^{k} }
e^{-i\tau \varphi(\vec m,\vec n)}\times\\
&&\times U_{m_k,n_k}^{m_0^k,m_0^{k-1}}
\cdots U_{m_2,n_2}^{m_0^2,m_0^1} U_{m_1,n_1}^{m_0^1,n_0}\,\,  
m_0^k\otimes |m_1, m_2,\cdots, m_k, n_{k+1}, \cdots, n_N\ket,\nonumber
\eea
where 
\be\label{phase}
\varphi(\vec m,\vec n)= \sum_{j=1}^{k} \left(\sum_{j < l \leq N}\delta_{n_l}+
\sum_{l < j}
\delta_{m_l}\right)
\ee
\end{exo}

As already noted, the tensor product structure of the initial state allows us to consider $k$ spins of the chain only: 
For any $N\geq k$,
\bea
&&\tr_{\fh_\env}(\one\otimes\bigotimes_{k=1}^N\rho_{\cE, \beta} \,\, U_1^* U_{2}^*\cdots U_k^*
(B\otimes \one_{\fh_\env})U_k U_{k-1}\cdots U_1)=\\ \nonumber
&&\hspace{4cm}\tr_{\fh_\env}(\one\otimes\bigotimes_{k=1}^k\rho_{\cE, \beta}\,\, 
U_1^* U_{2}^*\cdots U_k^*
(B\otimes \one_{\fh_\env})U_k U_{k-1}\cdots U_1).
\eea
\vspace{.3cm}

We now adopt the following block matrix notation on
$\fh_\cS\otimes \C^{n+1}_k$ 
as a block matrix with respect to the ordered basis of 
$\fh_\cS\otimes \C^{n+1}_k$
\be\label{ordb}
\begin{matrix}\{\omega \otimes \omega, x_1 \otimes \omega, \cdots, 
 x_d \otimes \omega, \cr \phantom{\{x }
\omega \otimes x_1, x_1\otimes x_1, \cdots  x_d \otimes 
x_1,\cr \vdots \cr \phantom{\{xx }
\omega \otimes x_n, x_1\otimes x_n, \cdots  x_d \otimes 
x_n\}\end{matrix}
\ee
to proceed with $h_k=h$. We get
\be\label{4.20}
U=e^{-i\tau h}=\begin{pmatrix}U_{0,0} & U_{0,1}& \cdots & U_{0,n}\cr 
U_{1,0} & U_{1,1}& \cdots & U_{1,n}\cr \vdots &\vdots &\ddots & \vdots
\cr U_{n,0} & U_{n,1}& \cdots & U_{n,n} \end{pmatrix}
\ee where, see (\ref{notbloc}),
\be
U_{m,m'}=\begin{pmatrix}U_{m,m'}^{0,0} & U_{m,m'}^{0,1}& \cdots & U_{m,m'}^{0,d}\cr 
U_{m,m'}^{1,0} & U_{m,m'}^{1,1}& \cdots & U_{m,m'}^{1,d}\cr 
\vdots &\vdots &\ddots & \vdots
\cr U_{m,m'}^{d,0} & U_{m,m'}^{d,1}& \cdots & U_{m,m'}^{d,d} \end{pmatrix}.
\ee
Also, in terms of the notations of the previous section,
\be
U=\begin{pmatrix}PUP & PUQ \cr QUP& QUQ\end{pmatrix},
\ee 
we have the identifications
\bea\label{identif}
& &PUP\simeq U_{0,0}, \ \ \ \ \  QUQ\simeq \begin{pmatrix} U_{1,1}& \cdots & U_{1,n}\cr 
\vdots &\ddots & \vdots
\cr  U_{n,1}& \cdots & U_{n,n} \end{pmatrix},\nonumber \\ 
& &PUQ\simeq \begin{pmatrix} U_{0,1}& \cdots & U_{0,n}\end{pmatrix}, \ \ \ \ 
QUP\simeq \begin{pmatrix} U_{1, 0}& \cdots & U_{n, 0}\end{pmatrix}^T.
\eea

Let us finally denote the inverse of $U=(U^{n,n'}_{m,m'})$
by 
\be
V=({V}^{n,n'}_{m,m'})=U^{-1}=({U^{-1}}^{n,n'}_{m,m'})\in M_{(1+d)(1+n)}(\C),
\ee
so that we have for any $m$ and $n$
\be
U_{n,m}^*=V_{m,n}\in M_{1+d}(\C).
\ee
\begin{exo} Show with these notations by means of (\ref{matel}) and 
(\ref{phase}) that 
the  matrix elements of 
$U(k,0)^{-1}\,\, (B\otimes \I_{\cal H}) \,\, U(k, 0)$
in the orthonormal basis $\{n_0\otimes |n_1,\cdots,n_k\ket\}=\{n_0\otimes |\vec{n}\ket\}$ read
\bea
& &\bra \tilde{n_0}\otimes \vec{\tilde{n}} |(U_k\cdots U_1)^* B\otimes 
\I_{\cal H}\,(U_k\cdots U_1)
\,\,n_0\otimes\vec{n}\ket=\\
& &\quad \quad  e^{-i\tau(\varphi(0,\vec n)-\varphi(0 ,\vec{\tilde{n}}))}
\sum_{\vec m\in\{0,\cdots, n\}^k} (V_{\tilde n_1, m_1}\cdots V_{\tilde n_k,m_k}BU_{m_k,n_k}\cdots U_{m_1, n_1})^{\tilde n_0, n_0}.
\nonumber
\eea
\end{exo}

We are thus lead with (\ref{bkt}) to study the matrix in $M_{d+1}(\C)$
\be
B_\beta(k,\lambda,\tau)=\sum_{\vec{n}=(n_1,\cdots, n_k)\atop \vec m=(m_1,\cdots m_k)}
\frac{e^{-\beta\sum_{l=0}^{n}\delta_l|\vec n|_l}}{(1+\sum_{j=1}^n e^{-\delta_{j}\beta})^k}V_{ n_1, m_1}\cdots 
V_{n_k,m_k}BU_{m_k,n_k}\cdots U_{m_1, n_1}
\ee 
in various  limiting cases as $\lambda$ and/or $\tau$ go to zero, with the notation 
\be
|\vec n|_l=\sharp\{n_r \ \ \mbox{s.t.}\ \ n_r=l\}= |S|_l.
\ee 

Consider the Hilbert space $M_{d+1}(\C)$ equipped with the scalar product $\bra A | B\ket =\tr (A^* B)$, 
for any $A, B\in M_{d+1}(\C)$, and the following linear operators on this Hilbert space
\be\label{calu}
{\cal U}_{m,m'}(A):=V_{m',m}\, A\, U_{m,m'}, \,\,\,\,\,\, (m,m')\in 
\{0,1,\cdots, n\}^2.
\ee
One has with respect to the above scalar product,
\be
{\cal U}_{m,m'}^*(\cdot)=(V_{m',m}\, \cdot \, U_{m,m'})^*=  U_{m,m'}\, 
\cdot \, V_{m',m},
\ee
 and the composition of such operators will be denoted as follows
\be
{\cal U}_{m',n'}\, {\cal U}_{m,n}(A)=V_{n', m'} V_{n,m}\, A\, U_{m,n} U_{m',n'}.
\ee

The Markovian nature of the evolution of observables reads as follows:
\begin{prop}\label{bkuk}
In terms of the operators defined above, we can write 
\bea
B_\beta(k,\lambda,\tau)=&\frac{1}{(1+\sum_{j=1}^ne^{-\delta_j\beta})^k}&
\left({\cal U}_{0,0}+e^{-\beta\delta_1}{\cal U}_{0,1}+\cdots +
e^{-\beta\delta_n}{\cal U}_{0,n}\right.\nonumber\\
&&+{\cal U}_{1,0}+e^{-\beta\delta_1}{\cal U}_{1,1}+\cdots +
e^{-\beta\delta_n}{\cal U}_{1,n}\nonumber\\
&&+\left.
{\cal U}_{n,0} +e^{-\beta\delta_1}{\cal U}_{n,1}+
\cdots + e^{-\beta\delta_n}{\cal U}_{n,n} \right)^k(B)\nonumber\\
&&\equiv {\cal U}_\beta(\lambda,\tau)^k(B).
\eea
\end{prop}
\begin{exo} Prove this proposition.
\end{exo}

\subsection{Weak Limit in the Heisenberg Picture}

The $\lambda$-dependence in $B_\beta(k,\lambda,\tau)$ comes from the definition 
\be\label{deful}
U=U_\tau(\lambda)=e^{-i\tau(H(0)+\lambda W)},
\ee
which implies that the ${\cal U}_{n,m}$'s depend on $\lambda$
as well, in an analytic fashion, and will be denoted 
${\cal U}_{n,m}(\lambda)$.
Expliciting the $\lambda$ dependence in $B_\beta(k,\lambda,\tau)$, the weak limit 
corresponds to taking $k=t/\lambda^2$ and computing the behavior of
$B_\beta( t/ \lambda^2,\lambda,\tau)$, as $\lambda\rightarrow 0$ (keeping $\tau$ fixed). 
We shall use the same strategy as in the 
previous Section and Lemma \ref{pertuu} to identify the weak limit by
means of perturbation theory. We shall also eventually consider the
possibility of letting $\tau\rightarrow 0$, therefore we explicit the behavior in
$\tau$ of the expansions below. \\

Consequently, with (\ref{identif}) and Corollary \ref{pertuu}, we get
\begin{lem}\label{4.5} Let $U$ be given by (\ref{deful}), with $H(0)$, $W$ 
self adjoint and satisfying {\bf H1}, and further assume $H(0)$ is 
diagonal with respect to the basis (\ref{ordb}). 
If ${\cal U}_{m,m'}(\lambda)$ is defined by (\ref{calu}) 
As $\lambda \rightarrow 0$, we get the expansions
\bea
&&{\cal U}_{0,0}(\lambda)={\cal U}_{0,0}(0)+\lambda^2{\cal U}_{0,0}^{(2)}
+O(\lambda^4\tau^4)\\
&&{\cal U}_{m,m'}(\lambda)={\cal U}_{m,m'}(0)+\lambda^2{\cal U}_{m,m'}^{(2)}
+O(\lambda^4\tau^4), \ \ \ m, m' \geq 1 \\
&&{\cal U}_{0,m}(\lambda)=\lambda^2{\cal U}_{0,m}^{(1)}+O(\lambda^4\tau^4), 
\,\,\,\,\, m\geq 1\\
&&{\cal U}_{m,0}(\lambda)=\lambda^2{\cal U}_{m,0}^{(1)}+O(\lambda^4\tau^4), 
\,\,\,\,\, m\geq 1
\eea
where, for all $0\leq m, m' \leq n$
\bea
&&{\cal U}_{m,m'}(0)(B)=\delta_{m,m'}e^{i\tau H_{m,m}(0)}\, 
B \,e^{-i\tau H_{m,m}(0)}, \ \ \  \\
&&{\cal U}_{m,m'}^{(2)}(B)=\delta_{m,m'}(G_{m,m}(-\tau)Be^{-i\tau H_{m,m}(0)}+
e^{i\tau H_{m,m}(0)}BG_{m,m}(\tau)),
\eea
and, for all $1\leq m$,
\bea
&&{\cal U}_{0,m}^{(1)}(B)=F_{m,0}(-\tau)BF_{0,m}(\tau),
\\
&&{\cal U}_{m,0}^{(1)}(B)=F_{0,m}(-\tau)BF_{m,0}(\tau).
\eea
\end{lem}

This Lemma allows us to perform the analysis of
the operator defined in Proposition \ref{bkuk}
\be\label{ubeta}
{\cal U}_\beta(\lambda,\tau)={\cal Z}(\beta)^{-1}
\sum_{0\leq m \leq n\atop 0\leq l\leq n}{\cal U}_{l,m}(\lambda)
e^{-\delta_{m}\beta}, \ \ \ \mbox{as}\ \ \ \lambda \rightarrow 0,
\ee 
with the convention $\delta_0=0$ and  
${\cal Z}(\beta)=\sum_{j=0}^ne^{-\delta_j\beta}$. 
Recall that
\be
B_\beta(k,\lambda,\tau)={\cal U}_\beta(\lambda,\tau)^k(B).
\ee
Moreover, using the fact, 
\be
H_{m,m}(0)=H_{0,0}(0)+\delta_m \simeq h_\cS+\delta_m,
\ee
we get for all $0\leq m\leq n$
\be
{\cal U}_{m,m}(0)(B)={\cal U}_{0,0}(0)(B)\simeq e^{i\tau h_\cS} B e^{-i\tau h_\cS}
=e^{i\tau [h_0,\cdot]}(B).
\ee
We have thus shown the 
\begin{lem} Assume the hypotheses of Lemma \ref{4.5}. Then  
\bea \label{tbeta}
{\cal U}_\beta(\lambda, \tau)&=&{\cal U}_{0,0}(0)
+\frac{\lambda^2}{ {\cal Z}(\beta)}\left[\sum_{m=1}^n\left\{e^{-\beta\delta_m}
\left({\cal U}_{0,m}^{(1)}+{\cal U}_{m,m}^{(2)}\right)
+{\cal U}_{m,0}^{(1)}
\right\} +{\cal U}_{0,0}^{(2)}\right]+O(\lambda^4\tau^4)\nonumber\\
 &\equiv&{\cal U}_{0,0}(0)
+\lambda^2 {\cal Z}(\beta)^{-1}T_\beta +O(\lambda^4\tau^4),
\eea
with $T_\beta=T_\beta(\tau)=O(\tau^2).$
\end{lem}

\vspace{.3cm}

The weak limit can thus be obtained from Proposition {\ref{weakprop}} to get the following
\begin{thm}\label{mtb} Let ${\cal U}_\beta(\lambda,\tau)$ be given by (\ref{ubeta}), and
${\cal U}_{0,0}(0)$,  $T_\beta$ by (\ref{tbeta}). 
Let $\{e^{i\tau \Delta_l}\}_{l=1,\cdots, r}$
be the set of distinct eigenvalues of ${\cal U}_{0,0}(0)$
and denote by $P_l$ the corresponding orthogonal projectors.
Then
\bea
&&\lim_{\lambda\rightarrow 0\atop t/\lambda^2\in\N }{\cal U}_{0,0}(0)^{-t/\lambda^2}
B_\beta( t/\lambda^2,\lambda,\tau)=\\
&&\quad \quad\quad\quad\quad \lim_{\lambda\rightarrow 0\atop t/\lambda^2\in\N}
{\cal U}_{0,0}(0)^{-t/\lambda^2}
{\cal U}_\beta(\lambda,\tau)^{t/\lambda^2}(B)=e^{t\Gamma^w_\beta}(B),\nonumber
\eea
were
\be
\Gamma^w_\beta(B)=\frac{1}{{\cal Z}(\beta)}\left({\cal  U}_{0,0}(0)^{-1}\ T_\beta
\right)^\#(B),
\ee
with $\#$ corresponding to the set of projectors
$\{P_l\}_{l=1,\cdots, r }$.
\end{thm}
\begin{rem}{\ \\}
i) To make the generator $\Gamma^w_\beta$ completely explicit, one needs 
to investigate the properties of $T_\beta$, i.e. of the operators $V_j$ 
defining the coupling, within the eigenspaces of ${\cal U}_{0,0}(0)$. 
\\
ii) The degeneracy of the eigenvalue $1$ of ${\cal U}_{0,0}(0)$ is
responsible for the existence of a non-trivial invariant sub-algebra of
observables which is the commutant of $h_0$.
\end{rem}

The result is then generalized to the regime
$\lambda^2\tau\rightarrow 0$, $\tau\rightarrow 0$, by switching to the macroscopic time scale 
$T=t/(\lambda^2\tau)\rightarrow \infty$. First compute
\bea
&&\Gamma_\beta(B)=\lim_{\tau \rightarrow 0} \frac{{\cal  U}_{0,0}(0)^{-1}\ T_\beta}{{\cal Z}(\beta)\tau^2}(B)
=-\frac{1}{2{\cal Z}(\beta)}({W^2}_{0,0}B+B{W^2}_{0,0})+\\
&&\frac{1}{{\cal Z}(\beta)}\sum_{m=1}^n\left\{ e^{-\delta_m\beta}\left(   
W_{m,0}BW_{0,m}-\frac{1}{2}({W^2}_{m,m}B+B{W^2}_{m,m})\right)+W_{0,m}BW_{m,0}\right\}.
\nonumber
\eea
Then, the formulas for $m\geq 1$
\be
W_{0,m}=V_m^*, \ \ W_{m,0}=V_m, \ \ W^2_{m,m}=V_mV_m^*, \ \ W^2_{0,0}=
\sum_{j=1}^nV_j^*V_j,
\ee  
allow one to express the operators $W_{mm'}$ in terms of $V_m$, which yields
\bea\label{4.59}
\Gamma_\beta(B)&=&\frac{1}{{\cal Z}(\beta)}\sum_{m=1}^ne^{-\beta\delta_m}
\left(V_{m}BV_{m}^*-
\frac{1}{2}(V_{m}V_{m}^*B+BV_{m}V_{m}^*)\right)\nonumber\\
& &\qquad\qquad\qquad\qquad + V_{m}^*BV_{m}-\frac{1}{2}(V_{m}^*V_{m}B+BV_{m}^*V_{m}).
\eea
\begin{rem}
The operator (\ref{4.59}) has the form of the dissipative part of a Lindblad generator. 
\end{rem}

\begin{cor}\label{tobewritten}
Assume the hypotheses of Theorem \ref{mtb}. Then
with $t/(\tau\lambda)^2=k\in\N$, 
\bea
&&\lim_{\tau\rightarrow 0, \lambda^2\tau \rightarrow 0 \atop 
t/(\tau\lambda)^2\in\N}{\cal U}_{0,0}(0)^{-t/(\tau\lambda)^2}
B_\beta( t/(\tau\lambda)^2,\lambda,\tau))=\\
&&\quad\quad \lim_{\tau\rightarrow 0,  \lambda^2\tau \rightarrow 0\atop 
t/(\tau\lambda)^2\in\N}
{\cal U}_{0,0}(0)^{-t/(\tau\lambda)^2}
{\cal U}_\beta(\lambda,\tau)^{t/\lambda^2}(B)=e^{t{\Gamma_\beta}^\#}(B),\nonumber
\eea
were $\Gamma_\beta(B)$ is defined in (\ref{4.59}).
\end{cor}
\begin{rem}{\ \\}
The proof is obtained essentially along the line of the proof Theorem \ref{theo31}
taking into account the following facts: The operator ${\cal U}_{0,0}(0)=e^{i\tau [h_0,\cdot ]}$ 
is unitary on $M_{d+1}(\C)$ with spectral projectors that are independent of $\tau$ 
as $\tau\rightarrow 0$ and eigenvalues of the form $e^{i\tau \Delta_j}$. Introducing the perturbation parameter $x=(\lambda\tau)^2$, (\ref{tbeta}) states that uniformly in $\tau$,
\be
{\cal U}_\beta(\lambda,\tau)={\cal U}_{0,0}(0)+xT_\beta(\tau)/(\tau^2{\cal Z(\beta)})
+O(x^2),
\ee
where $T_\beta(\tau)/\tau^2\rightarrow \Gamma_\beta$ as $\tau\rightarrow 0$. \\
For generalizations to infinite dimensional Hilbert spaces $\fh_\cS$ and $\fh_\env$, see \cite{V}. 
\\
Explicit formulea for $\Gamma_\beta^w$ and $\Gamma_\beta$ are provided in \cite{AJ1} for the case $d=n=1$.
\end{rem}

\subsection{Beyond the perturbative regime: $\lambda^2\tau=1$}\label{ssec:beyondperturb}

We consider here the regime $\lambda^2\tau=1$, and $\tau\rightarrow 0$. 
It can be viewed as a regime where the weak limit scaling holds at the microscopic level, while, at the  
macroscopic level,  $T=t/(\tau\lambda^2)$ is kept finite.

The determination of the dynamics in this regime amounts to computing the limit
\be
\lim_{\tau\rightarrow 0 \atop t/\tau\in\N} {\cal U}_\beta(1/\sqrt{\tau},\tau)^{t/\tau}(B).
\ee

The technical tool used in this case will be Chernoff's Theorem, see e.g. \cite{BR}.
\begin{thm}
Let $S(\tau)$ defined on a Banach space $\cB$ be such that
$S(0)=\one$, and $\|S(\tau)\|\leq 1$, for all $\tau\geq 0$.
If, 
$\lim_{\tau\rightarrow 0} \tau^{-1}(S(\tau)-\one)=\Gamma$ in the strong sense
 exists in ${\cal L}(\cB)$
and generates a contraction semi-group, then
\be
s-\lim S(t/n)^n=e^{t\Gamma}.
\ee
\end{thm}

\begin{thm}\label{thlin} Assume hypothesis H0 where $\fh_\cS$ is a separable
Hilbert space and the operators $h_\cS$, the $V_j$'s and $B$ are bounded on $\fh_\cS$. Let 
$B_\beta(t/\tau,1/\sqrt\tau,\tau)$ be defined by (\ref{bkt}), 
${\cal U}_\beta(\lambda,\tau)$ is defined by proposition \ref{bkuk} and the 
Remark following it. Then
\be
s-\lim_{\tau\rightarrow 0 \atop t/\tau\in\N} {\cal U}_\beta(1/\sqrt{\tau},\tau)^{t/\tau}(B)=
e^{t(i[h_0,\cdot]
+\Gamma_\beta(\cdot))}(B)
\ee
with a Lindblad generator $i[h_0,\cdot]+\Gamma_\beta(\cdot)$ where $\Gamma_\beta$ is given by
\be\label{lind}
\Gamma_\beta(B)=\sum_{j=1}^{2m}L_jBL_j^*-\frac{1}{2}\left(L_{j}L_{j}^*B+BL_{j}L_{j}^*\right)
\ee
with
\be
L_j=\frac{e^{-\beta\delta_j/2}}{\sqrt{{\cal Z}(\beta)}}V_j, \ 1\leq j\leq m \ \ \
\mbox{ and } \ \ \ L_j=\frac{1}{\sqrt{{\cal Z}(\beta)}}V_j^*, \ m+1\leq j\leq 2m.
\ee
\end{thm}
\begin{rem}{\ \\}
i) In \cite{AP}, Section IV.2,  the scaling is the same with 
a supplementary structure 
allowing one to make the suitably renormalized spins 
forming the chain merge in the limit
$\tau\rightarrow 0$ to yield a heat bath represented by a Fock space
of quantum noises. 
When restricted to $\fh_\cS$, the effective dynamics of observables at zero temperature
corresponds to a contraction semigroup generated by  
\be
\Gamma_{\infty}(\cdot )=i[h_0,\cdot]+\sum_{m=1}^n\left(V_{m}^* \cdot V_{m}-
\frac{1}{2}(V_{m}^*V_{m} \cdot + \cdot V_{m}^*V_{m})\right),
\ee
which coincides with Theorem \ref{thlin} at $\beta=\infty$. A similar comparison holds with \cite{AJ2} which deals with the finite temperature case. \\
ii) The generator $\Gamma^{\beta}$ coincides with the generator 
(\ref{4.59}) obtained in Corollary \ref{tobewritten} in the scaling 
$\lambda^2\tau\rightarrow 0$, $\tau\rightarrow 0$, modulo the $\#$ operation, 
which appears as a trade mark of the perturbative regime.
\end{rem}

\begin{proof} The proof of Theorem \ref{thlin} consists in checking the hypotheses of Chernoff's Theorem.
First recall the formula (see (\ref{4.9}))
\bea
 {\cal U}_\beta(\lambda,\tau)(B)&=&\tr_{\fh_\env}\left((\one\otimes \rho_{\cE,\beta})
U^{-1}(1,0)(B\otimes \one) U(1,0)\right)\\
&=&\sum_{q=0}^n\frac{e^{-\beta \delta_q}}{{\cal Z}(\beta)}\B(\tau)_{qq},\nonumber
\eea
where 
$\B(\tau)_{qq}=(U^{-1}(1,0)(B\otimes \one) U(1,0))_{qq}=P_q U^{-1}(1,0)(B\otimes \one) U(1,0) P_q$
according to the block notation (\ref{4.20}), with the corresponding orthogonal projectors $P_q$. 
Identifying $P_q\C^{(n+1)(d+1)}$ with $\fh_\cS=\C^{d+1}$, we deduce from the above formula that
${\cal U}_\beta(\lambda,\tau)$ is a contraction for any value of the parameters:
\bea
 \|{\cal U}_\beta(\lambda,\tau)(B)\|_{\fh_\cS}&\leq& 
\sum_{q=0}^n\frac{e^{-\beta \delta_q}}{{\cal Z}(\beta)}\|\B(\tau)_{qq}\|_{\fh_\cS}\\
&\leq& \sum_{q=0}^n\frac{e^{-\beta \delta_q}}{{\cal Z}(\beta)}\|P_qU^{-1}(1,0)(B\otimes \one) U(1,0)
 P_{q}\|_{\C^{(n+1)(d+1)}}\nonumber \\
&\leq & \sum_{q=0}^n\frac{e^{-\beta \delta_q}}{{\cal Z}(\beta)}\|(B\otimes \one)
\|_{\C^{(n+1)(d+1)}}=\|B\|_{\fh_\cS}.\nonumber
\eea
Moreover, ${\cal U}_\beta(1/\sqrt{\tau},\tau)|_{\tau=0}=\one$, so we are left with the
computation of the derivative w.r.t. $\tau$ at the origin. This involves the control 
of the operator $U_\tau(\lambda)$ (\ref{lop}) as $\tau\rightarrow 0$ and 
$\lambda=1/\sqrt{\tau}\rightarrow \infty$. 
The expansion of $U_\tau(\lambda)$ in powers of $\lambda$ is convergent, with
$\tau$ dependent coefficients we control sufficiently well. Indeed,  (\ref{cv})
yields
\be
U_\tau(\lambda)=e^{-i\tau H(0)}\Theta(\lambda, \tau)=\sum_{n\geq 0}
e^{-i\tau H(0)}\Theta_n(\lambda, \tau),
\ee
where $\Theta_n$ contains $n$ operators $W$ and satisfies
$
\|\Theta_n(\lambda, \tau)\|=O((\tau\lambda)^n/n!).
$
Using the fact that $(\lambda \tau)^n=\tau^{n/2}\rightarrow 0$ and that $W$ is 
off-diagonal with respect to $P$ and $Q$, we get that the replacement 
of $\lambda$ by $1/\sqrt{\tau}$ doesn't spoil the estimates as $\tau\rightarrow 0$ given
in Proposition \ref{bkuk} and  Lemma  \ref{4.5}. Together with the computation (\ref{4.59}) one gets
\bea
{\cal U}_\beta(1/\sqrt{\tau},\tau)(B)&=&e^{i\tau h_0}Be^{-i\tau h_0}+({\cal Z(\beta)}\tau)^{-1}
T_\beta(\tau)(B)+O(\tau^2)\nonumber\\
&\equiv& e^{i\tau h_0}Be^{-i\tau h_0}+\tau \Gamma_\beta(B) +O(\tau^2).
\eea
Hence, the derivative at the origin exists and is given by
\be
{\cal U}_\beta(1/\sqrt{\tau},\tau)'(B)|_{\tau=0}=i[h_0,B]+\Gamma_\beta(B),
\ee
where $\Gamma_\beta(B)$ is the dissipative part of a Lindblad operator 
which has the form given in the statement.
Hence,
\be
i[h_0,B]+\Gamma_\beta(B)
\ee
generates a completely positive semigroup of contractions.
Therefore, Chernoff's theorem to eventually yields the result. 
\end{proof}

\begin{rem}{\ \\}
We finally recall some consequences
of these results about the evolution of states, within the duality 
provided by  the scalar
product $\bra A | B\ket = \tr (A^* B)$.
If $\Gamma$ is the generator of the dynamics of observables, $B$ is
an observable and $\rho$ is a state, then for any $t\in\R$,
\be
\tr (\rho e^{t\Gamma}(B))=\tr (e^{t \Gamma_*}(\rho) B)
\ee
where the generator of the dynamics of the states is $\Gamma_*$
such that for all states $\rho$ and observables $B$,
\be
\tr ((\Gamma_*(\rho))^* B)=\bra \rho | \Gamma(B)\ket=
\bra \Gamma ^* \rho |B\ket.
\ee
The link between asymptotic states in time described by Theorem \ref{thm:idealsmall} and asymptotic states  of weak limit effective dynamics as $t\rightarrow \infty$ is further explored in \cite{V}. Under certain genericity hypotheses, it is proven there that  the former type of asymptotic states converges to the latter, as $\lambda\rightarrow 0$. 
\end{rem}


\section{Application to concrete models}\label{sec:applic}

In this section we present two concrete models of the repeated interaction type. These models show how repeated interaction systems can be used to adress some physically relevant situations.


\subsection{One atom maser}\label{ssec:qedcavity}

We first consider a specific model describing the ``One-Atom Maser'' experiment where $\cS$ is the quantized  electromagnetic field in a cavity through which a beam of atoms, the subsystems $\cE_n$, is shot. Such systems play a fundamental role in the experimental and theoretical investigations of basic matter-radiation processes. They are also of practical importance in quantum optics and quantum state engineering \cite{MWM,WVHW,WBKM,RH,VAS}. We consider here only the ideal case, i.e. the question of thermal relaxation: is it possible to thermalize a mode of a QED cavity by means of $2$-level atoms if the latter are initially at thermal equilibrium? One particular feature here is that the Hilbert space of the small system $\cS$ is \emph{not} finite dimensional. There are very few models of open quantum systems in the literature with an infinite small system and for which return to equilibrium is proven. The RI structure of the model allows us to provide such a model. Moreover, one usually makes use of perturbation theory in the coupling constant to obtain information on the spectrum of the relevant operator. Here, we do not make use of any perturbation theory, i.e. our results do not restrict to small coupling constants. The results described here come from \cite{BP}.


\subsubsection{Description of the model and the RDM}

We consider the situation where the atoms of the beam are prepared in a stationary mixture of two states with energies $\tilde{E}<E_0$, and without loss of generality we set $\tilde{E}=0$. We assume the cavity to be nearly resonant with the transitions between these two states. Neglecting the non-resonant modes of the cavity, we can describe its quantized electromagnetic field by a single harmonic oscillator of frequency $E\simeq E_0$.

The Hilbert space for a single atom is $\fh_\cE=\C^2\simeq\Gamma_-(\C)$, the Fermionic Fock space over $\C$. The Hamiltonian of a single atom is thus
$$
h_\cE= E_0b^* b,
$$
where $b^*$, $b$ denote the creation/annihilation operators on $\fh_\cE$, see (\ref{def:fermicreation}). The Hilbert space of the cavity field is $\fh_\cS:=\ell^2(\mathbb N)=\Gamma_+(\C)$, the Bosonic Fock space over $\C$. Its Hamiltonian is
$$
h_\cS = E a^* a\equiv E N,
$$
where $a^*$, $a$ are the creation/annihilation operators on $\fh_\cS$ satisfying the commutation relation $[a,a^*]=\one$ and $N$ is the number operator on $\Gamma_+(\C)$.

In the dipole approximation, an atom interacts with the cavity field through its electric dipole moment.  The full dipole coupling is given by $\frac{\lambda}{2}(a+a^*)\otimes(b+b^*)$, acting on $\fh_\cS \otimes \fh_\cE$, where $\lambda\in\R$ is a coupling constant. Neglecting the counter rotating term $a\otimes b+a^*\otimes b^*$ in this coupling (this is the so called {\em rotating wave approximation}) leads to the well known Jaynes-Cummings Hamiltonian
\begin{equation}\label{eq:jaynescummham}
h= h_\cS\otimes \one_\cE+\one_\cS\otimes h_\cE+\lambda v, \qquad v=\frac{1}{2}(a^*\otimes b+a\otimes b^*),
\end{equation}
for the coupled system $\cS+\cE$ (see e.g. \cite{Ba,CDG,Du}). (Example \ref{ex:toy-description-rdm} in Section \ref{sec:framework} is very similar. One simply replaces the bosonic Fock space $h_\cS=\Gamma_+(\C)$ by the fermionic one $\Gamma_-(\C)=\C^2$.) The rotating wave approximation, and thus the dynamics generated by the Jaynes-Cummings Hamiltonian, is known to be in good agreement with experimental data as long as the detuning parameter $\Delta\equiv E-E_0$ satisfies $|\Delta|\ll\min(E_0,E)$ and the coupling is small $|\lambda|\ll E_0$. To our knowledge, there is however no mathematically precise statement about this approximation.

Finally, the initial state of the atoms will be the equilibrium state at inverse temperature $\beta$, i.e. $\rho_{\cE,\beta}\equiv {\e^{-\beta h_\cE}}/{\tr \ \e^{-\beta h_\cE}}$.

\medskip

As for the toy model of Example \ref{ex:toy-description-rdm}, using the fact that $h$ commutes with the total number operator $N^{\rm tot}=a^*a\otimes\one_\cE+\one_\cS\otimes b^*b$, we can calculate explicitly $\e^{-\i\tau h}$ and hence the RDM $\cL_\beta$ associated to this RI system. One gets
\begin{equation}\label{eq:cavityrdm}
\cL_\beta(\rho)=\sum_{\sigma,\sigma'=0,1} V_\ss \rho V_\ss^*,
\end{equation}
where the operators $V_\ss$ are given by
\begin{equation}\label{eq:cavityvsigma}
\begin{array}{ll}
\ds V_{00}=\frac{1}{\sqrt{Z_{\cE,\beta}}}\,\e^{-\i\tau E N}\,C(N),& 
\ds V_{10}=\frac{1}{\sqrt{Z_{\cE,\beta}}}\,\e^{-\i\tau E N}\,S(N+1)\,a, \\[16pt]
\ds V_{01}=\frac{\e^{-\beta E_0/2}}{\sqrt{Z_{\cE,\beta}}}\,\e^{-\i\tau E N}\,S(N)\,a^*,& 
\ds V_{11}=\frac{\e^{-\beta E_0/2}}{\sqrt{Z_{\cE,\beta}}}\,\e^{-\i\tau E N}\,C(N+1)^*,
\end{array}
\end{equation}
with
\begin{equation*}
C(N)\equiv\cos(\pi\sqrt{\xi N+\eta})+\i\eta^{1/2}\, \frac{\sin(\pi\sqrt{\xi N+\eta})}{\sqrt{\xi N+\eta}},\qquad S(N)\equiv\xi^{1/2}\,\frac{\sin(\pi\sqrt{\xi N+\eta})}{\sqrt{\xi N+\eta}},
\end{equation*}
and where
\begin{equation}\label{def:etaxi}
\eta\equiv\left(\frac{\Delta\tau}{2\pi}\right)^2,\qquad \xi\equiv\left(\frac{\lambda\tau}{2\pi}\right)^2,
\end{equation}
are the dimensionless detuning parameter and coupling constant.


\subsubsection{Spectral analysis of the RDM}\label{ssec:cavityspectrum}

We know from the general results on RI systems that $\cL_\beta$ is a contraction on $\cB^1(\fh_\cS)$, and that the state $\rho(n)$ of $\cS$ evolves according to the discrete semigroup $\cL_\beta^n$, i.e. $\rho(n)=\cL_\beta^n(\rho)$. To understand the asymptotic behavior of $\rho(n)$, we shall thus study the spectral properties of $\cL_\beta$. In particular, we will be interested in its peripheral eigenvalues $\e^{\i\theta}$, for $\theta\in\R$, and especially in the eigenvalue $1$ (the corresponding eigenstate(s) will give the candidates for the asymptotic state(s)). 

To understand the difficulty in the spectral analysis, assume that the atom-field coupling is turned off.  The reduced dynamics is then nothing but the free evolution of $\cS$, i.e. $\cL_\beta(\rho)=\e^{-i\tau h_\cS}\rho\,\e^{i\tau h_\cS}$.
It is easy to see that the spectrum of $\cL_\beta$ is then pure point:
$$
\sp(\cL_\beta)=\sp_\pp(\cL_\beta)=\{\e^{\i\tau E d}\,|\,d\in\Z\}.
$$
This spectrum is finite if $\tau E\in2\pi\Q$ and densely fills the unit circle in the opposite case. In any case, all the eigenvalues (and in particular $1$) are infinitely degenerate. This explains why perturbation theory in $\lambda$ fails for this model. Note also that since $\fh_\cS$ has infinite dimension, it is not automatic that $\cL_\beta$ has $1$ as an eigenvalue (we only know that it is in the spectrum since it is in the one of $\cL_\beta^*$).

To describe the spectral results, we need to introduce a notion of resonance. An essential feature of the dynamics generated by the Jaynes-Cummings Hamiltonian $h$ are Rabi oscillations. In the presence of $n$ photons, the probability for the atom to make a transition from its ground state to its excited state is a periodic function of time : 
$$
P(t)=\left|\bra n-1,+|\,\e^{-itH}\,|n,-\ket \right|^2=\left(1-\frac{\Delta^2}{\nu_n^2}\right)\sin^2\left(\frac{\nu_n t}{2}\right),
$$
where the circular frequency is $\nu_n=\sqrt{\lambda^2 n+\Delta^2}$. (In our units, $\lambda$ is thus the one photon Rabi-frequency of the atom in a perfectly tuned cavity.) If the interaction time $\tau$ is a multiple of the Rabi-oscillation period for $n$ photons, then no transition will be possible from the $n$-photon state to the previous one. Such a {\em resonance} occurs when, for some integer $k$,
$$
\tau = k \frac{2\pi}{\sqrt{\lambda^2 n+\Delta^2}} \quad \Longleftrightarrow \quad \xi n+\eta =k^2,
$$
where $\eta$ and $\xi$ are defined in (\ref{def:etaxi}). We therefore introduce the following set
\begin{defn}\label{def:resonant} $R(\eta,\xi):=\{ n\in\N \, |\, \xi n+\eta=k^2 \mbox{ for some } k\in\N\}$. An element $n\in R(\xi,\eta)$ is called a Rabi resonance.
\end{defn}

The Hilbert space $\fh_\cS$ thus has a decomposition $\fh_S=\bigoplus_{k=1}^r\fh_\cS^{(k)},$ where $r-1$ is the number of Rabi resonances, $\fh_\cS^{(k)}\equiv\ell^2(I_k)$ and $\{I_k\,|\,k=1,\ldots, r\}$ is the partition of $\N$ induced by the resonances. Following \cite{BP}, we call $\fh_\cS^{(k)}$ the $k$-th Rabi sector and denote by $P_k$ the corresponding orthogonal projection.

It is easy to show that, according to the arithmetic properties of $\xi$ and $\eta$ (rational or not), the set $R(\eta,\xi)$ possesses either no, one or infinitely many elements (\cite{BP}, Lemma 3.2). We shall say accordingly that the system is non-resonant, simply resonant or fully resonant. A fully resonant system will be called degenerate if there exist $n\in\{0\}\cup R(\eta,\xi)$ and $m\in R(\eta,\xi)$ such that $n<m$ and $n+1,m+1\in R(\eta,\xi)$, i.e. there are two pairs of consecutive Rabi resonances. (Such degenerate systems exist, if e.g. $\xi=840$ and $\eta=1$ then $(1,2)$ and $(52,53)$ are pairs of consecutive resonances. We refer to \cite{BP} for more details on degenerate systems.)

\medskip

\noindent The main ingredients for the spectral analysis of $\cL_\beta$ are:

\noindent 1) {\em The gauge symmetry.}

For any $\theta\in\R$, $\cL(\e^{-i\theta N}\rho\, \e^{i\theta N})=\e^{-i\theta N}\cL(\rho)\, \e^{i\theta N},$ which follows from $\ds [h,N^{\rm tot}]=[h_\cE,\rho_{\cE,\beta}]=0$. As a consequence, $\cL_\beta$ leaves invariant the subspaces
\begin{equation*}\label{JdDef}
\cB^{1,(d)}(\fh_\cS)\equiv\{X\in \cB^1(\fh_\cS)\,|\,\e^{-\i\theta N}X\e^{\i\theta N}
 =\e^{\i\theta d}X \ \text{for all}\ \theta\in\R\},
\end{equation*}
and hence admits a decomposition
\begin{equation*}\label{LDecomp}
\cL_\beta=\bigoplus_{d\in\Z}\cL_\beta^{(d)},
\end{equation*}
so that one can analyze separately the $\cL_\beta^{(d)}$.

\noindent 2) {\em How $\cL_\beta$ acts on diagonal states, i.e. on $\cB^{1,(0)}(\fh_\cS)$.}

Because of the gauge symmetry, if $\rho$ is an invariant state so is its ``diagonal part'' $\rho_0=\sum_n \bra n|\rho n\ket |n\ket\bra n|\in\cB^{1,(0)}$. It is thus important to understand the diagonal invariant states.

If we denote by $x_n$ the diagonal elements of $X\in\cB^{1,(0)}(\fh_\cS)$, we can identify $\cB^{1,(0)}(\fh_\cS)$ with $\ell^1(\N)$.
Introducing the number operator $(Nx)_n\equiv nx_n$ and the finite difference operators
$$
(\nabla x)_n\equiv \left\{ \begin{array}{ll}x_0&\text{for}\ n=0,\\ x_n-x_{n-1}& \text{for}\ n\ge1,
\end{array} \right. \qquad (\nabla^* x)_n\equiv x_n-x_{n+1}\ \ \text{for}\ n\ge0,
$$
a simple algebra from (\ref{eq:cavityrdm})-(\ref{eq:cavityvsigma}) leads to
\begin{equation}\label{Lbetanot}
\cL_\beta^{(0)}=\one-\nabla^* D(N)\e^{-\beta E_0 N}\nabla\e^{\beta E_0 N},
\end{equation}
where
\begin{equation}\label{Ddef}
D(N):=\frac{1}{1+\e^{-\beta E_0}}\,\sin^2(\pi\sqrt{\xi N+\eta})\, \frac{\xi N}{\xi N+\eta}.
\end{equation}
In particular, the diagonal invariant states $\rho$ are solutions of $D(N)\e^{-\beta E_0 N}\nabla\e^{\beta E_0 N}\rho=0$. Hence they satisfy $(\e^{-\beta E_0 N}\nabla\e^{\beta E_0 N}\rho)_n=0 \ \Leftrightarrow \rho_n=\e^{-\beta E_0}\rho_{n-1}$ unless $D(n)=0$, i.e. $n$ is a Rabi resonance. We therefore have three situations:
\begin{itemize}
\item If the system is non-resonant, it follows from (\ref{Ddef}) that $D(n)=0$ if and only if $n=0$ and hence there is a unique diagonal invariant state $\ds \frac{\e^{-\beta E_0 N}}{\tr\ \e^{-\beta E_0 N}}=\rho_{\cS,\beta^*}$ where $\beta^*=\beta\frac{E_0}{E}$ if $\beta>0$ (this is the same renormalization as for the toy model of Section \ref{sec:framework}) and none if $\beta\leq0$.
\item If the system is simply resonant there exists $n_1\in\N^*$ such that $D(n)=0$ if and only if $n=0$ or $n=n_1$. The eigenvalue equation then splits into two decoupled systems
\begin{align*}
\rho_{n}&=\e^{-\beta E_0}\rho_{n-1},  \ \  n\in I_1\equiv\{1,\ldots,n_1-1\},\\
\rho_{n}&=\e^{-\beta E_0}\rho_{n-1},  \ \  n\in I_2\equiv\{n_1+1,\ldots\}.
\end{align*}
The first one yields the invariant state $\ds \frac{\e^{-\beta E_0 N}P_1}{\tr\ (\e^{-\beta E_0 N}P_1)}=\rho_{\cS,\beta^*}^{(1)},$ for any $\beta\in\R$. The second system gives another invariant state
$\ds \frac{\e^{-\beta E_0 N}P_2}{\tr\ (\e^{-\beta E_0 N}P_2)}=\rho_{\cS,\beta^*}^{(2)},$ provided $\beta>0$.
\item If the system is fully resonant, $D(n)$ has an infinite sequence $n_0=0<n_1<n_2<\cdots$ of zeros. The eigenvalue equation now splits into an infinite number of finite dimensional systems
$$
\rho_{n}=\e^{-\beta E_0}\rho_{n-1},  \ \  n\in I_k\equiv\{n_{k-1}+1,\ldots,n_{k}-1\},
$$
where $k=1,2,\ldots$ For any $\beta\in\R$, we thus have an infinite number of invariant states
$\ds \frac{\e^{-\beta E_0 N}P_k}{\tr\ (\e^{-\beta E_0 N}P_k)}=\rho_{\cS,\beta^*}^{(k)},$
one for each Rabi sector.
\end{itemize}

\noindent 3) {\em The following Perron-Frobenius type Theorem due to Schrader.}

\begin{thm}\label{thm:schrader}([Sch], Theorem 4.1) Let $\phi$ be a $2$-positive map on $\cB^1(\cH)$ 
such that its spectral radius $r(\phi)=\|\phi\|$. If $\lambda$ is a peripheral eigenvalue of $\phi$
with eigenvector $X$, i.e. $\phi(X)=\lambda X$, $X\not=0$, $|\lambda|=r(\phi)$,
then $|X|=\sqrt{X^*X}$ is an eigenvector of $\phi$ to the eigenvalue $r(\phi)$: $\phi(|X|)=r(\phi)|X|$.
\end{thm}
\noindent Since the RDM $\cL_\beta$ is completely positive trace preserving map we can apply Theorem \ref{thm:schrader} to it. Hence, if $\e^{\i\theta}$ is a peripheral eigenvalue of $\cL_\beta^{(d)}$ for some $d$, with eigenvector $X$, then $|X|\in \cB^{1,(0)}(\fh_\cS)$ is an invariant state of $\cL_\beta^{(0)}$, which we already know by 2).

\bigskip

Putting all these ingredients together we have a full description of the peripheral eigenvalues of $\cL_\beta$.
\begin{lem}\label{lem:periphev}[BP] 1. The only peripheral eigenvalue of $\cL_\beta^{(0)}$ is $1$.

2. If the system is not degenerate, then the only peripheral eigenvalue of $\cL_\beta$ is $1$ and the corresponding eigenvectors are diagonal.

3. If the system is degenerate we note $N(\eta,\xi):=\{n\in \{0\}\cup R(\eta,\xi)\,|\,n+1\in R(\eta,\xi)\}$ and $\cD(\eta,\xi):=\{d=n-m\,|\,n,m\in N(\eta,\xi), n\not=m\}$. In this case the set of peripheral eigenvalues of $\cL_\beta$ is given by $\{1\}\cup\{\e^{\i(\tau\omega+\xi\pi )d}\,|\,d\in\cD(\eta,\xi)\}.$
\end{lem}


\subsubsection{Convergence results}

Thermal relaxation is an ergodic property of the map $\cL_\beta$ and of its invariant states. For any density matrix $\rho$, we denote the orthogonal projection on the closure of $\ran\,\rho$ by $s(\rho)$, the
support of $\rho$. We also write $\mu\ll\rho$ whenever $s(\mu)\le s(\rho)$.

A state $\rho$ is ergodic, respectively mixing, for the semigroup generated by $\cL_\beta$ whenever
\begin{equation}\label{ergodef}
\lim_{N\to\infty}\frac{1}{N}\,\sum_{n=1}^N\left(\cL_\beta^n(\mu)\right)(A)=\rho(A),
\end{equation}
respectively
\begin{equation}\label{mixdef}
\lim_{n\to\infty}\left(\cL_\beta^n(\mu)\right)(A)=\rho(A),
\end{equation}
holds for all states $\mu\ll\rho$ and all $A\in\cB(\fh_\cS)$. $\rho$ is exponentially mixing if the convergence in (\ref{mixdef}) is exponential, i.e. if
$$
\left|\left(\cL_\beta^n(\mu)\right)(A)-\rho(A)\right|\le C_{A,\mu}\,\e^{-\alpha n},
$$
for some constant $C_{A,\mu}$ which may depend on $A$ and $\mu$ and some $\alpha >0$ independent of $A$ and $\mu$. A mixing state is ergodic and an ergodic state is clearly invariant. 

A state $\rho$ is faithful iff $\rho>0$, that is $s(\rho)=Id$. Thus, if $\rho$ is a faithful ergodic (resp. mixing) state the convergence (\ref{ergodef}) (resp. (\ref{mixdef})) holds for every state $\mu$ and one has global relaxation. In this case, $\rho$ is easily seen to be the only ergodic state of $\cL_\beta$. Conversely, one can show that if $\cL_\beta$ has a unique faithful invariant state, this state is ergodic:
\begin{thm}\label{thm:relax}[BP]\ Let $\phi$ be a completely positive trace preserving map on $\cB^1(\cH)$. If $\phi$ has a faithful invariant state $\rho_\mathrm{inv}$ and $1$ is a simple eigenvalue of $\phi$ then $\rho_\mathrm{inv}$ is ergodic.
\end{thm}

Using Lemma \ref{lem:periphev}, we have the following theorem which is the main result of this section.

\begin{thm}\label{thm:cavityinvstate}[BP]\  1. If the system is non-resonant then $\cL_\beta$ has no invariant state for $\beta\le0$ and has the unique ergodic state
$$
\rho_{\cS,\beta^*}=\frac{\e^{-\beta^* h_\cS}}{\tr \ \e^{-\beta^* h_\cS}}
$$
for $\beta>0$. In the latter case any initial state relaxes in the mean to the thermal equilibrium state at inverse temperature $\beta^*=\beta\frac{E_0}{E}$.

2. If the system is simply resonant then $\cL_\beta$ has the unique ergodic state $\rho_{\cS,\beta^*}^{(1)}$ if $\beta\le0$ and two ergodic states $\rho_{\cS,\beta^*}^{(1)}$, $\rho_{\cS,\beta^*}^{(2)}$ if $\beta>0$. In the latter case, for any initial state $\rho$, one has
\begin{equation*}\label{RelaxSimple}
\lim_{N\to\infty}\frac{1}N\sum_{n=1}^N\left(\cL_\beta^n(\rho)\right)(A) =\rho(P_1)\,\rho_{\cS,\beta^*}^{(1)}(A) +\rho(P_2)\,\rho_{\cS,\beta^*}^{(2)}(A),
\end{equation*}
for all $A\in\cB(\fh_\cS)$ (recall $P_k$ is the projection onto the $k$-th Rabi sector).

3. If the system is fully resonant then for any $\beta\in\R$, $\cL_\beta$ has infinitely many ergodic states $\rho_{\cS,\beta^*}^{(k)}$, $k=1,2,\ldots$ Moreover, if the system is non-degenerate,
\begin{equation}\label{RelaxFull}
\lim_{N\to\infty}\frac{1}N\sum_{n=1}^N\left(\cL_\beta^n(\rho)\right)(A)
=\sum_{k=1}^\infty\rho(P_k)\,\rho_{\cS,\beta^*}^{(k)}(A),
\end{equation}
holds for any initial state $\rho$ and all $A\in\cB(\fh_\cS)$.

4. If the system is non-degenerate, any invariant state is diagonal and can be represented as a convex linear combination of ergodic states.

5. Whenever the state $\rho_{\cS,\beta^*}^{(k)}$ is ergodic, it is also exponentially mixing if the corresponding Rabi sector $\fh_\cS^{(k)}$ is finite dimensional.
\end{thm}

\noindent {\bf Remarks.\ } i) In the non-degenerate cases, this result implies some weak form of decoherence in the energy eigenbasis of the cavity field: the time averaged off-diagonal part of the state $\cL_\beta^n(\rho)$ decays with time.

ii) If the system is degenerate, (\ref{RelaxFull}) and the conclusions of Assertion 4. still hold provided a further non-resonance condition is satisfied.  Namely, if $\e^{\i(\tau E+\xi\pi)d}\not=1$ for all $d\in\cD$ (see Lemma \ref{lem:periphev}), then all eigenvectors of $\cL_\beta$ to the eigenvalue $1$ are diagonal.

iii) Numerical experiments support the conjecture that all the ergodic states are mixing. However, our analysis does not provide a proof of this conjecture if $\fh_\cS^{(k)}$ is infinite dimensional.

\begin{open} Prove that all ergodic states are mixing. 
\end{open}

Actually, due to the presence of an infinite number of metastable states in the non-resonant and simply resonant cases, see \cite{BP} Section 4.5.4., one expects slow, i.e. non-exponential, relaxation.

\begin{open} How slow is the relaxation in infinite dimensional sectors? 
\end{open}

\begin{open} For this model, only the ideal situation has been considered in \cite{BP}. One can also consider the situation where some randomness is included, in particular when the interaction time is random. Besides the mathematical interest of an infinite dimensional example of a random repeated interaction system, this also has some physical relevance. Experimentally one observes exponential convergence to the stationary state. Is the slow convergence mentioned above, and due to metastable states, only an artefact of the assumption that all the interaction times are identical? In other words, do random times enhance mixing? The non-exponential mixing has its origin in the quasi-resonances, the location of which is very sensitive to the various parameters and in particular the interaction time.
\end{open}


\subsection{Electron in tight binding band}\label{ssec:tightbinding}

In the second model we consider, the system $\cS$ describes a spinless electron in the single band tight-binding approximation and subject to an homogeneous static electric field. For the electron alone, Bloch oscillations prevent a current from being set up in the system (see (\ref{eq:xfreeevol})). It is furthermore expected that if the electron is in contact with a thermal environment, the resulting scattering mechanisms will suppress the Bloch oscillations and lead to a steady current. In the model considered here, the environment is described by a chain of two-level atoms with which the electron interacts in the RI scheme. We show that a dc current is indeed created due to the interaction of the particle with its environment. In addition to drifting in the direction of the applied field, the electron diffuses around its mean position. The results concerning this model come from \cite{BDP}.
 

\subsubsection{Description of the model}

\noindent {\em The small system.}

The system $\cS$ consists in a spinless particle on the one-dimensional lattice $\Z$ and submitted to a constant external force $F\geq 0$. The quantum Hilbert space and Hamiltonian of the particle are
\begin{equation*} \label{def:particle}
\fh_\cS=\ell^2(\Z), \qquad h_\cS=-\Delta -FX,
\end{equation*}
where $\Delta$ is the discrete nearest neighbor Laplacian and $X$ the lattice position operator
$$
-\Delta =\sum_{x\in\Z} \bigl(2\,|x\rangle\langle x|-|x\!+\!1\rangle\langle x| - | x\rangle\langle x\!+\!1|\bigr), \quad X=\sum_{x\in\Z}x \,|x\rangle\langle x|.
$$
We shall also identify $\fh_\cS$ with  $L^2(\mathbb T^1,\d\xi)$ via the discrete Fourier transform, so that
$$
-\Delta=2(1-\cos\xi),\qquad X=\i\partial_\xi.
$$
Here $\mathbb T  ^1\simeq [0,2\pi[$ is the first Brillouin zone and $\xi$  the crystal momentum.
If $\ds T=\sum_{x\in\Z} |x\!+\!1\rangle\langle x|=\e^{-\i\xi}$ denotes the translation operator, one easily shows that:
\begin{enumerate}
 \item When $F=0$, $h_\cS$ has a single band of absolutely continuous spectrum, $\sp(h_\cS)=[0,4]$, and the motion of the particle is described by
$$
T(t) = \e^{\i t h_\cS}T\e^{-\i t h_\cS}=T, \quad X(t)=\e^{\i th_\cS}X\e^{-\i th_\cS}=X+\i(T-T^*)t,
$$
showing its ballistic nature.
 \item When $F\not=0$, $h_\cS$ has discrete spectrum, $\sp(h_\cS)=2-F\Z$. This is the well-known Wannier-Stark ladder. In the position representation, the normalized eigenvector $\psi_k$ to the eigenvalue $E_k=2-Fk$ is given by
\begin{equation}\label{eq:psik}\
\psi_k(x)=J_{k-x}\left(\frac{2}{F} \right),
\end{equation}
where the $J_\nu$ are Bessel functions. From their asymptotic behavior for large $\nu$ (see e.g. Formula (10.19.1) in \cite{OLBC}) we infer that
$$
\psi_k(x)\sim\frac1{\sqrt{2\pi|k-x|}}\left(\frac\e{F|k-x|}\right)^{|k-x|} \quad\text{for }|k-x|\to\infty,
$$
which shows that  $\psi_k(x)$ is sharply localized around $x=k$. The motion of the particle, described by
\begin{equation}\label{eq:xfreeevol}
\left. \begin{array}{rcl}
\ds T(t) & = & \e^{\i th_\cS}T\e^{-\i th_\cS}=\e^{-\i tF}T,\\[10pt]
X(t) & = & \ds \e^{\i t h_\cS}X\e^{-\i t h_\cS} = X+\frac{4}{F}\sin\left(\frac{Ft}{2} \right)\sin\left(\xi+\frac{Ft}{2}\right),
\end{array} \right\}
\end{equation}
is now confined by Bloch oscillations.
\end{enumerate}

\medskip

\noindent {\em The environment.}

As in Section \ref{ssec:qedcavity}, it consists of 2-level atoms, each of which has a quantum Hilbert space $\fh_\cE=\C^2$ which we identify with $\Gamma_-(\C)$, the fermionic Fock space over $\C$, and a Hamiltonian given by
$$
h_\cE=Eb^*b,
$$
where $E\geq 0$ is the Bohr frequency of the atom and $b^*$, $b$ are the usual Fermi creation and annihilation operators.

The initial state of the two-level atoms will be their equilibrium state at inverse temperature $\beta$
described by the density matrix
\begin{equation*}\label{def:thermalstate}
\rho_\beta=Z_\beta^{-1}\e^{-\beta h_\cE}, \quad  Z_\beta= \tr (\e^{-\beta h_\cE})=1+\e^{-\beta E}.
 \end{equation*}

\medskip

\noindent {\em The interaction.}

The interaction between the particle and the two-level atom is chosen so that its effect is to give a right or left kick to the particle, depending on whether the atom is in its ground state or in its excited state. More precisely, we set
\begin{equation*}\label{def:tightinteraction}
v=\sum_{x\in\Z}\bigl( |x\!+\!1\ket\bra x| \otimes b^* + |x\ket\bra x\!+\!1| \otimes b\bigr)=T\otimes b^*+T^*\otimes b.
\end{equation*}
To understand this interaction, note that when $F>0$, the translation operator $T$ can also be thought of as a lowering operator for the particle. Indeed, from (\ref{eq:psik}) one finds
\begin{equation}\label{eq:Tpsi}
T\psi_k=\psi_{k+1}.
\end{equation}
Similarly, $T^*$ acts as a raising operator. As a result, $v$ describes an exchange of energy
between the two-level system and the particle. The model considered here is thus very similar to the one studied in Section \ref{ssec:qedcavity} except that the spectrum of $h_\cS$, contrary to the spectrum of the mode of the electromagnetic field, is not  bounded from below. As a result, the system we treat here has no invariant state (see the end of Section \ref{ssec:particlerdm}).


\subsubsection{Interaction with a single atom. The RDM $\cL_\beta$}\label{ssec:particlerdm}

As for the Jaynes-Cummings Hamiltonian, $h=h_\cS+h_\cE+\lambda v$ can easily be diagonalized by exploiting the fact that it commutes with the ``number operator''
\begin{equation}\label{def:Nop}
N=\frac{h_\cS-2}F+\frac{h_\cE}E.
\end{equation}
Introducing the unitary operator
$$
U=(Tb^* b+bb^*)\cos\theta-(Tb^*-b)\sin\theta,
$$
where $\theta$ is chosen such that
$$
\cos(2\theta)=\frac{E-F}{\omega_0}, \quad \sin(2\theta)=\frac{2\lambda}{\omega_0}, \quad \mbox{ and } \quad \omega_0=\sqrt{(E-F)^2+4\lambda^2},
$$
one gets the following explicit formula for the propagator,
\begin{equation}\label{eq:unitarygroup}
\e^{\i th}=U\e^{\i t(E-F)/2}\e^{\i t\omega_0(b^* b-1/2)}\e^{\i th_\cS}U^*.
\end{equation}
It follows then that
\begin{align*}
\e^{\i th}X\,\e^{-\i th}&=\e^{\i th_\cS}X\e^{-\i th_\cS}\nonumber\\
&+\left(\frac{4\lambda^2}{\omega_0^2}(bb^*-b^* b) +\frac{2\lambda(E-F)}{\omega_0^2}(Tb^*+T^*b)\right) \sin^2\left(\frac{\omega_0t}{2}\right) \label{eq:xcoupledevol}\\
&-\i\frac{\lambda}{\omega_0}(Tb^*-T^* b)\sin(\omega_0t).\nonumber
\end{align*}
We conclude that the coupling to a single atom does not substantially alter the long term behavior of the particle: it turns the periodic Bloch oscillations (\ref{eq:xfreeevol}) of frequency $\omega_{\rm Bloch}=F$ into quasi-periodic motion with the two frequencies $\omega_{\rm Bloch}$ and $\omega_0$. In particular, when $F\not=0$, the motion remains bounded. As we will see, the situation is very different for repeated interactions with a sequence of atoms.

The following result describes the RDM $\cL_\beta$ of this system. It follows directly from (\ref{eq:unitarygroup}).
\begin{lem} \label{lem:reduceddyn} 
For any $\rho\in\cB^1(\fh_\cS)$, one has $\cL_\beta(\rho)=\cU\circ\widetilde\cL_\beta(\rho)=\widetilde\cL_\beta\circ \cU(\rho)$ with
\begin{equation}\label{eq:reduceddyn}
\cU(\rho)=\e^{-\i\tau h_\cS}\rho\,\e^{\i\tau h_\cS}, \qquad \widetilde\cL_\beta(\rho)=p_-T^*\rho T + p_0 \rho + p_+ T\rho T^*,
\end{equation}
where
\begin{equation*}
p_-=\frac{\e^{-\beta E}}{1+\e^{-\beta E}}\,p,\quad p_0=1-p, \quad p_+=\frac{1}{1+\e^{-\beta E}}\,p,
\end{equation*}
with $p=\frac{4\lambda^2}{\omega_0^2}\sin^2\left(\frac{\omega_0\tau}{2}\right)$.
\end{lem}
If $\rho$ describes the state of the particle, then $T^*\rho T$ (respectively $T\rho T^*$) represents the same state translated by one lattice spacing to the left (respectively right). Note moreover that 
$$
p_-+p_0+p_+=1,
$$
so that the reduced dynamics consists of a free evolution with the Hamiltonian $h_\cS$, followed by a random translation by $\pm 1$ or $0$, and with probabilities $p_{\pm}$ or $p_0$.  Note that the dynamics is trivial if $p=0$, i.e. if $\omega_0\tau=2\pi m$ with $m\in\Z$. In that case there is no translation and the particle evolves according to $h_\cS$. This can be seen directly on (\ref{eq:unitarygroup}) by noticing that in such a case $U h_\cS U^*=h_\cS+Fb^* b$. It follows that the propagator factorizes
$$
\e^{\i\tau h}=(-1)^m \e^{\i\tau(E-F)/2} \, \e^{\i\tau h_\cS} \otimes\e^{\i\tau Fb^*b},
$$
and, up to an inessential phase factor and a renormalization of the atomic Bohr frequency, the particle
and the two-level system evolve as if they were not coupled. This resembles the ``Rabi oscillation'' 
phenomenon which appears in the Jaynes-Cummings model (see Section \ref{ssec:cavityspectrum}). In the following we will avoid this resonance and assume $p\not=0$. 

We can now see that the system has no stationary state as we already mentioned, i.e. there exists no density matrix $\rho$ on $\fh_\cS$ such that $\cL_\beta(\rho)=\rho$. Indeed, it follows from the gauge invariance $\cL_\beta(\e^{-\i th_\cS}\rho\,\e^{\i th_\cS})=\e^{-\i th_\cS}\cL_\beta(\rho)\,\e^{\i th_\cS}$ that the subspaces $\cB^{1,(d)}(\fh_\cS)$, $d\in\Z$, defined by
\begin{eqnarray*}
\cB^{1,(d)}(\fh_\cS) & = & \{\rho \in \cB^1(\fh_\cS) \,|\, \e^{-\i th_\cS}\rho\,\e^{\i th_\cS}=\e^{\i td}\rho\text{ for all }t\in\R\} \\
 & = & \{\rho \in{\cal B}^1(\fh_\cS)\,|\, \rho=\sum_{k\in\Z} \rho_k |\psi_k \ket\bra \psi_{k+d}| \},
\end{eqnarray*}
are globally invariant under $\cL_\beta$. Hence, if a state $\rho$ is stationary, so is its diagonal part 
$\rho_0=\sum_k p_{k} |\psi_k\ket\bra \psi_k|$, where $p_k=\bra\psi_k|\rho\psi_k\ket$. From (\ref{eq:reduceddyn}) one gets
$$
p_{k-1}-Z_\beta p_k+\e^{-\beta E} p_{k+1}=0,
$$
which implies that $p_k=a+b\,\e^{\beta Ek}$ for some constants $a,b\in\R$. But this contradicts the fact that $1=\tr\,\rho=\sum_kp_k$.


\subsubsection{Asymptotic behaviour of the particle}\label{ssec:particleasympt}

Since the system has no invariant state, our concern here is not the large time behaviour of the state of the particle but of expectation values of some functions of the position observable $X$. Given an observable $B$ on $\fh_\cS$, we write $\bra B\ket_n= \tr (B \cL_\beta^n(\rho))$, for its expectation value at time $t=n\tau$. As already announced, the following theorem shows that the repeated interactions make the motion of the particle diffusive. The motion is characterized by a drift velocity
\begin{equation*}\label{def:vd}
\vd=\vd(E,F)=\frac{p}{\tau} \tanh\left(\frac{\beta E}{2} \right),
\end{equation*}
and a diffusion constant
\begin{equation*}\label{def:dd}
D =D(E,F) =\frac{p}{2\tau}\left( 1-p\tanh^2\left(\frac{\beta E}{2}\right)\right).
\end{equation*}
More precisely, the following holds.

\begin{thm}\label{thm:diffusion}[BDP]\ Assume that $F>0$, $\lambda\not=0$ and $\omega_0\tau\notin2\pi\Z$ so that $p\in\,]0,1]$. Let the density matrix $\rho\in\cB^1(\fh_\cS)$ describe the initial state of the particle and denote by $\mu_n$ the spectral measures of the position observable $X$ on the states $\rho(n)$,
\begin{equation*}\label{eq:mun}
\mu_n(f) =\int f(x)\,\d\mu_n(x) = \langle f(X)\rangle_n.
\end{equation*}
\begin{enumerate}[\rm 1.]
\item The Central Limit Theorem (CLT) holds: For any bounded continuous function $f$ on $\R$,
\begin{equation*}
\lim_{n\to\infty}\int f\left(\frac{x-\vd n\tau}{\sqrt{2Dn\tau}}\right)\,\d\mu_n(x) =\int f(x)\,\e^{-x^2/2}\,\frac{\d x}{\sqrt{2\pi}}.
\end{equation*}
\item If $\tr\left(X^2\rho\right)<+\infty$, then
\begin{equation*}
\lim_{n\to\infty} \frac{\langle X\rangle_n}{n\tau}= \vd,\quad \lim_{n\to\infty} \frac{\bra (X-\vd n\tau)^2\ket_n}{n\tau}=2D.
\end{equation*}
\item If $\tr\left(\e^{\alpha |X|}\rho\right)<+\infty$ for all $\alpha>0$, then a Large Deviation Principle (LDP) holds in the sense that, for any interval $J\subset \R$,
\begin{equation*}\label{eq:gellis}
\lim_{n\to\infty} \frac{1}{n}\log \mu_n(nJ)= - \inf_{x\in J} I(x),
\end{equation*}
where $I(x)$ is the Legendre-Fenchel transform of
\begin{equation}\label{def:ealpha}
e(\alpha)=\log \left( (1-p)+p\frac{\cosh(\frac12\beta E+\alpha)}{\cosh\frac12\beta E}\right),
\end{equation}
i.e. $\ds I(x)=\sup_{\alpha\in\R}\alpha x-e(\alpha).$
\end{enumerate}
\end{thm}

\bigskip
Note that when $E=F$, the mobility
$
\ds \mu=\lim_{F\to 0} \frac{v_\d}{F}=\frac{\beta\sin^2(\lambda\tau)}{2\tau},
$
and the diffusion constant
$
\ds D=\mu\beta^{-1}\left(1-\sin^2(\lambda\tau)\tanh^2\left(\frac{\beta F}{2}\right)\right),
$
satisfy the Einstein relation
$$
\lim_{F\to 0} D=\mu\beta^{-1}=\mu k_B T.
$$

\begin{open} When $E\neq F$ the Einstein relation holds only in the limit $E\to 0$ and not $F\to 0$ (also in the definition of the mobility), meaning that it is actually the interaction with the atoms which drive the electron and not the applied electric field. The chosen coupling is in a sense too strong. Consider a similar model but with a different interaction, e.g. with scattering in the electron momentum instead of the electron position so that there is a sort of momentum conservation.
\end{open}

\medskip
The rate function in Part 3 is explicitly given by
$$
I(x)=\left\{\begin{array}{ll} \ds -x\left(\frac{\beta E}{2}+\log\left(\frac{R(x)-x}{a(1-x)}\right)\right)
-\log\left(\frac{(1-p)(R(x)+1)}{1-x^2}\right) & \text{for } x\in[-1,1],\\[10pt]
+\infty&\text{otherwise}, \end{array}\right.
$$
where
$$
a=\frac{p}{(1-p)\cosh(\beta E/2)},\qquad R(x)=\sqrt{x^2+a^2(1-x^2)}.
$$
It is strictly convex on $[-1,1]$ and satisfies $I(\vd\tau)=0$ and $I(x)>0$ for $x\not= \vd\tau$.

Note that the drift velocity and diffusion constant do not depend on the initial state of the particle. The CLT gives us the probability to find the particle at time $n\tau$ in a region of size $O(\sqrt n)$ around the mean value $\vd n\tau$, whereas the LDP gives information on this probability for a region of size $O(n)$. To put it differently, it yields information on the probability that the particle's mean speed falls asymptotically in an interval of size $O(1)$. Loosely speaking, it says that
$$
\mu_n(\{n(\vd+\delta v)\tau\})\simeq \e^{-nI((\vd+\delta v)\tau)}.
$$

The peculiar symmetry $e(-\beta E-\alpha)=e(\alpha)$ immediately leads to the relation $I(x)=-\beta Ex+ I(-x)$ which tells us that
$$
\lim_{\delta v\downarrow0}\lim_{n\to\infty}\frac1{n\tau}\log
\frac{\mu_n(n[-v-\delta v,-v+\delta v]\tau)}{\mu_n(n[v-\delta v,v+\delta v]\tau)}=-\beta E v,
$$
i.e. that negative mean velocities are exponentially less likely than positive ones. One can recognize here a form of fluctuation theorem. We refer to \cite{BDP} for more details on the origin of this peculiar symmetry.

\begin{open} Consider a similar model but when the electron moves in the continuum. The idea is that now the interactions would slow down the motion instead of creating a current. Can we have similar results as in the discrete case, i.e. drift+diffusion, not instead of a bounded motion but of a uniformly accelerated one?
\end{open}

\begin{open} Look at the various limiting regime à la Attal et al. of this model, in particular the continuous limit, see Section \ref{ssec:beyondperturb}, and analyze the properties of the corresponding process on $\R$.
\end{open}


 \subsection{Link with Quantum Walks}\label{subsec:qw}

The previous model allows us to make a direct link between RIS and another popular kind of discrete quantum dynamical systems, namely Quantum Walks, or QW for short. There are many different versions of QW, suited to the many applications they have in computer science, quantum physics or even probability theory. For a description of the many aspects of QW, the reader is referred to reviews \cite{VA}, \cite{Sa}, \cite{Ko}, \cite{Ke}.

In one of their simplest forms, QW describe the dynamics of a particle or walker on a lattice, say $\Z^d$, $d\geq 1$, which carries an internal degree of freedom, a spin, that lives on $\C^{2d}$. The total Hilbert space of this system is thus $\fh= \C^{2d}\otimes l^2(\Z^d)$ and the discrete dynamics of the walker is {\em defined} by the repeated action of a sequence of unitary operators on $\fh$. The peculiarity of the model is that these unitary operators are not generated by a physical Hamiltonian, but are instead constructed on the basis of an analogy with classical random walks on $\Z^d$. A typical example of dynamics taken from \cite{J}, see also \cite{AVWW}, is the following. Denote the canonical basis of $\C^{2d}$ by $\{|\tau \ket\}_{\tau\in I}$, with $I= \{\pm1, \pm 2, \cdots, \pm d\}$ and define a {\em jump function} as a map
\be\label{jump}
r: I\rightarrow \Z^d.
\ee 
A symmetric choice of jump function often considered is $r(\tau)=\mbox{sign}(\tau)e_{|\tau|}$, where $e_j$, $j\in \{1,2,\dots, d\}$ denotes the canonical basis of $\Z^d$. We will assume that $r(\tau)\neq 0$ for any $\tau$.
Let $C\in U(2d)$ by a unitary matrix on the spin Hilbert space and let $P_\tau$, $\tau\in I$ be the orthogonal projectors of the canonical basis vectors of $\C^{2d}$. We are now ready to construct the dynamics of the walker. 
We first define a shift operator on $\fh$ by 
\be\label{shift}
S=\sum_{x\in \Z^d \atop \tau\in I}P_\tau\otimes |x+r(\tau)\ket\bra x|.
\ee
This shift is unitary, as the direct sum over $\tau$ of shifts on $l^2(\Z^d)$ and makes the walker's spin component along $|\tau\ket$ jump on the lattice by a step of length $r(\tau)$ . Second, we define the unitary operator $C\otimes \one$ on $\fh$, which updates the spin variable of the walker, without making it move on the lattice. The one step unitary evolution operator of the walker is then defined as
\be\label{qw}
U(C)=S (C\otimes \one)=\sum_{x\in \Z^d \atop \tau\in I}P_\tau C\otimes |x+r(\tau)\ket\bra x|, 
\ee
where $C$ can be considered as a parameter.
Correspondingly, the one step evolution of a state in $\cB^1(\fh)$
\be\label{state}
\rho=\sum_{x\in \Z^d \atop y\in \Z^d}\rho(x,y)\otimes |x\ket\bra y|, \ \ \ \mbox{with} \ \ \ \
\rho(x,y)\in M_{2d}(\C)
\ee  is defined by
\be
\cU_C(\rho)=U(C) \rho U(C)^*= \sum_{(x, y) \in \Z^d\times \Z^d \atop (\tau, \sigma) \in I\times I}P_\tau C\rho(x-r(\tau),y-r(\sigma))C^*P_\sigma\otimes |x\ket\bra y|.
\ee
By iteration, we get a state on $\fh$ at time $n$ given by $\cU_C^n(\rho)$. \medskip

Thus, tracing over the spin states, we get a time dependent density matrix on $l^2(\Z^d)\equiv \fh_\cS$ which describes the expectation values of observables on the lattice. Hence there exists a proximity between QW with RIS, although there is no proper environment in the construction of QW. Moreover, the spin state which indirectly makes the walker move on the lattice undergoes a dynamics in which all interferences are taken into account, by contrast with RIS in which the environment induces a loss of coherence. This is transparent if one looks at the spectral properties of $U(C)^n$: Because the definition (\ref{qw}) of $U(C)$ is invariant under the translations on the lattice, the Fourier image of $U(C)$ acts as a multiplication operator on $L^2(\T^d)$. It is not difficult to check that for generic jump functions $r$ and unitary matrices $C$, the spectrum of $U(C)$ is absolutely continuous, which implies ballistic transport on the lattice. In order to mimic the effect of an environment, one can consider different matrices $C\in U(2d)$ at each time step, in particular, one can pick them at random, according to a certain law on $U(2d)$.

\begin{exo} Given a sequence $C_1, C_2, \cdots, C_n$ of unitary matrices on $\C^{2d}$, show that the evolution $\rho_n=\cU_{C_n}\circ \cdots \circ\cU_{C_2}\circ\cU_{C_1}(\rho_0)$ at time $n$ of a compactly supported state $\rho_0$ of the form (\ref{state}) takes the form
\be\label{staten}
\rho_n(x,y)=\sum_{(k,k')\in\Z^{d}\times\Z^d} J_k(n)\rho_0(x-k,y-k')J_{k'}^*(n),
\ee
where 
\be
J_k(n)=\sum_{\tau_1, \tau_2, \dots, \tau_n\in {I_\pm}^n\atop \sum_{s=1}^nr(\tau_s)=k} P_{\tau_n}C_{n}P_{\tau_{n-1}}C_{n-1}\cdots P_{\tau_1}C_{1}\in M_{2d}(\C)
\ee
and $J_k(n)=0$, if $\sum_{s=1}^nr(\tau_s)\neq k$.
\end{exo}

The following Theorem on the random case taken from \cite{J} gives a flavor of the type of results we can get in the framework of unitary random quantum walks. It is to be compared with Theorem \ref{thm:diffusion}. \\

We deal with a quantum walk with random update of the internal degrees of freedom at each time step. Let $C(\omega)$ be a random unitary matrix on $\C^{2d}$ with probability space $(\Omega,{\cal \sigma},d\mu)$, where $d\mu$ is a probability measure. We consider the random evolution operator obtained from sequences of i.i.d. coin matrices on $(\Omega^{\N^*},{\cal F},d\P)$, where ${\cal F}$ is the $\sigma$-algebra generated by cylinders and $d\P = \otimes_{k\in\N^*} d\mu$, by
\be\label{rdqs}
U_{\overline \omega}(n,0)=U_n(\overline \omega) U_{n-1}(\overline \omega)\cdots U_1(\overline \omega), \ \ \mbox{where} \ \ U_k(\overline \omega)=S\ (C(\omega_k)\otimes \I),
\ee
and $\overline \omega=(\omega_{1}, \omega_2, \omega_3, \dots)\in \Omega^{\N^*}$. The evolution operator at time $n$ is now given by a product of i.i.d. unitary operators on $\fh$, and similarly for the evolution of density matrices $\cU_{\overline \omega}(n)(\cdot)=\cU_{C(\omega_n)}\circ \cdots \circ\cU_{C(\omega_2)}\circ\cU_{C(\omega_1)(\cdot)}$.\\

Consider for simplicity the initial condition $\rho_0=|\psi_0\ket\bra\psi_0|\otimes|0\ket\bra 0|$ and define for $p=1,2$ 
\be
\bra X^p\ket_{\psi_0}(n)=\E_{\overline \omega}(\tr_\fh(\cU_{\overline \omega}(n)(\rho_0)\one\otimes X^p)),
\ee
where $X$ is the position operator on $\fh_\cS=l^2(\Z^d)$. Similarly, one defines the corresponding characteristic function $\Phi_n^{\psi_0}$ by
\be\label{charfun}
\Phi_n^{\psi_0}(y)=\E_{\overline \omega}(\tr_\fh(\cU_{\overline \omega}(n)(\rho_0)\one\otimes e^{iXy})), \ \ \ \mbox{for all} \ \ y\in [0,2\pi)^d\equiv \T^d.
\ee

We need a spectral assumption on a certain matrix defined as follows:
Let $ d(y)=\sum_{\tau\in I}e^{iyr(\tau)}|\tau\ket\bra\tau|\in U(2d)$ for $y\in \T^d$ and define 
\be
\cM(y,y')=(d(y)\otimes d(y'))\E(C(\omega)\otimes \overline C(\omega))\in M_{4d^2}(\C).
\ee
\begin{thm} \label{cf} Let $\overline r=\frac{1}{2d}\sum_{\tau\in I}r(\tau)\in \R^d$
and assume that for all $v\in \T^d$,
\be
\sigma(\cM(v,-v))\cap \{|z|=1\}=\{1\} \ \ \mbox{and the eigenvalue 1 is simple}.
\ee
Then, there exists an analytic map $\T^d\ni v\rightarrow \D(v)\in M^+_{d}(\R)$, the set of non-negative  matrices, such that
 uniformly in $y$ in compact sets of $\C^d$ and in $t$ in compact sets of $\R_+^*$,
\bea
&&\lim_{n\rightarrow\infty}\Phi^{\psi_0}_{[tn]}(y/ n)=e^{i{t} y  \overline{r}}\\
&&\lim_{n\rightarrow\infty}e^{-i[tn]\frac{\overline{r}y}{\sqrt{n}}}\Phi^{\psi_0}_{[tn]}(y/\sqrt n)=\int_{\T^d} e^{-\frac{t}{2}\bra y | \D(v) y\ket}\, \frac{d{ v}}{(2\pi)^d},
\eea
where the right hand side admits an analytic continuation in $(t,y)\in \C\times \C^2$.

In particular, for any $(i,j)\in \{1, 2, \dots, d\}^2$, 
\bea
&& \lim_{n\rightarrow\infty}\frac{\bra X_i\ket_{\psi_0}(n)}{n}=\overline{r}_i \\
&&\lim_{n\rightarrow\infty}\frac{\bra (X-n\overline{r})_i (X-n\overline{r})_j \ket_{\psi_0}(n)}{n}=\int_{\T^d} \D_{i\,j}(v)\, \frac{d{ v}}{(2\pi)^d}.
\eea
\end{thm} 
\begin{rem}
The matrix $\D(v)$ is determined explicitly by the spectral data of $\cM(v, v')$.\\
In case $\D(v)=\D$ is independent of $v\in \T^d$, a central limit theorem holds in the limit $n\rightarrow \infty$ for the centered rescaled random variable associated with the characteristic function (\ref{charfun})\\
Moderate deviations results can also be proven  under further hypotheses on the map $\T^d\ni v\rightarrow \D(v)\in M^+_{d}(\R)$, see \cite{J}. 
\end{rem}

Another way to randomize a QW consists in associating a different matrix $C$
to each site of $\Z^d$ in (\ref{qw}) so that
 \be\label{qwsd}
U(\cC)=S (\cC\otimes \one)=\sum_{x\in \Z^d \atop \tau\in I}P_\tau C_x\otimes |x+r(\tau)\ket\bra x|, 
\ee
where $\cC=\{C_x\}_{x\in\Z^d}$, $C_x\in U(2d)$. If the $C_x$'s are given by i.i.d. random matrices, we get a random unitary operator, which bears similarities with the self-adjoint Anderson model of solid state physics. Under certain hypotheses, the dynamics is radically different and since it can be shown that there exists regimes in which dynamical localization takes place, which forbids the walker to propagate on the lattice. For more details, see e.g. the review \cite{J2}.


\section{Thermodynamic properties}\label{sec:thermodynamic}
 
Contrary to the usual context of open systems, in RI systems the total Hamiltonian is (piecewise constant) time-dependent as we can see from (\ref{def:totalham}). Hence the energy of the full system is not necessarily constant. It is constant during each interaction (where the total Hamiltonian is constant) but we may have energy changes when one switches from an interaction to the next one. In other words, the switch from one interaction to the other may require some external work. In this section we show how to define an ``external work'' observable and study the expectation value of the mean external work in the large time limit. Then, in the case where ``several reservoirs'' are present, e.g. RI systems with several beams at different temperature, we also study the energy fluxes. Finally, we consider entropy production in RI systems and relate it to the external work and the fluxes.


\subsection{External work in RI systems}

\subsubsection{General setup}

As already mentioned, since the total Hamiltonian of a repeated interaction system is time-dependent, the total energy is not necessarily constant: to switch from an interaction to the other may require some external work. Since the total system is infinite (there are infinitely many subsystems $\cE_n$), the total energy makes no sense. However, energy variation does. Formally, the total energy at time $t$ is simply
$$
u(t,0)^*h(t)u(t,0),
$$
where $u(t,0)$ is the propagator between time $0$ and $t$, i.e. if $t\in[t_n,t_{n+1})$ then
$$
u(t,0)=\e^{-\i (t-t_n) \th_{n+1}} \e^{-\i\tau_n \th_n}\cdots \e^{-\i\tau_1 \th_1}.
$$
The change of energy between time $t$ and time $t'$ is therefore
\begin{equation*}\label{eq:energyvar}
\Delta E(t',t)= u(t',0)^* \th(t') u(t',0) - u(t,0)^* \th(t) u(t,0).
\end{equation*}
Now, for $t_{n-1}\leq t< t_n\leq t'<t_{n+1}$, it is easy to see that
$$
\Delta E(t',t)=u(t_n,0)^* ( v_{n+1}- v_n) u(t_n,0)=:w(n).
$$
The observable $w(n)$ is the work observable at time $t_n$. If $\cS$ is initially in the state $\rho$, one therefore has
\begin{eqnarray}\label{eq:hamworkexpect}
\delta E(n)\!\! & :=\!\!  & \tr_\fh \big(\rho\otimes\bigotimes_{k\geq 1}\rho_{\cE_k}\times w(n)\big) = \tr_\fh \big( \rho^{\rm tot}(n) \times (v_{n+1}-v_n)\big)\\
 & =\!\! & \tr_{\fh_\cS\otimes\fh_{\cE_{n+1}}}\left[\rho(n)\otimes\rho_{\cE_{n+1}}\ v_{n+1}\right]  -\tr_{\fh_\cS\otimes\fh_{\cE_n}} \left[\rho(n-1)\otimes\rho_{\cE_n} \ \e^{\i\tau_n h_n}v_n\e^{-\i\tau_n h_n}\right].\nonumber
\end{eqnarray}
The mean work per unit time in the large time limit, i.e. the power delivered to the system, finally is (if it exists),
\begin{equation*}\label{eq:meanworkham}
\Delta W:= \lim_{n\to\infty} \frac{1}{t_n}\Delta E(n),
\end{equation*}
where $\ds \Delta E(n)= \sum_{k=1}^n \delta E(k)$ is the total work between time $0$ and time $t_n$.

A simple algebraic computation shows that the total work can also be written as
\begin{eqnarray}\label{eq:totalworkham}
\Delta E(n) & = & \sum_{k=1}^n \tr_{\fh_\cS\otimes\fh_{\cE_k}} \left[ \rho(k-1)\otimes \rho_{\cE_k} \left(\e^{i\tau_k h_k} h_{\cE_k} \e^{-i\tau_k h_k} - h_{\cE_k} \right) \right]\\
 & & + \tr_{\fh_\cS} \left[ \left(\rho(n)-\rho(0)\right) h_\cS\right]\nonumber \\
 & &  + \tr_{\fh_\cS\otimes\fh_{\cE_{n+1}}} \left[ \rho(n)\otimes \rho_{\cE_{n+1}}\, v_{n+1}\right] - \tr_{\fh_\cS\otimes\fh_{\cE_1}} \left[ \rho(0)\otimes \rho_{\cE_1}\, v_1\right].\nonumber
\end{eqnarray}
The first term in the right-hand side is then the amount of energy transfered to the chain, and the second term is the amount of energy gained by $\cS$. Of course, except when the elements $\cE_k$ are finite systems, the free hamiltonians $h_{\cE_k}$ are typically unbounded operators. One may however write the energy difference observable appearing in (\ref{eq:totalworkham}) as
\begin{eqnarray*}
\e^{i\tau_k h_k} h_{\cE_k} \e^{-i\tau_k h_k} - h_{\cE_k} & = & \int_0^{\tau_k} \e^{i s h_k} \Phi_k \e^{-is h_k} \, \d s,
\end{eqnarray*}
where $\Phi_k= \frac{\d}{\d t} \e^{ith_k} h_{\cE_k} \e^{-ith_k} \lceil_{t=0} =  [iv_k,h_{\cE_k}]$ is the energy flux observable corresponding to the $k$-th subsystem.


\subsubsection{The ideal situation}

In the ideal case, the expectation value $\delta E(n)$ of the work observable at time $n$ simplifies as
\begin{equation}\label{eq:idealwork}
\delta E(n) =  \tr_{\fh_\cS\otimes\fh_{\cE}}\left[\rho(n)\otimes\rho_{\cE}\times v\right]  -\tr_{\fh_\cS\otimes\fh_{\cE}} \left[\rho(n-1)\otimes\rho_{\cE} \times \e^{\i\tau h}v\,\e^{-\i\tau h}\right].
\end{equation}
Combining this with Theorem \ref{thm:idealsmall} we immediately get
\begin{prop} Suppose Assumption {\bf (E)} is satisfied. Then for any initial state $\rho\in \cB^1(\fh_\cS)$,
$$
\delta E(n)=\tr_{\fh_\cS\otimes\fh_{\cE}} \left[\rho_{\cS,+}\otimes\rho_{\cE} \times \left(v-\e^{\i\tau h}v\,\e^{-\i\tau h}\right)\right]+O\left(\e^{-\gamma n}\right).
$$
As a consequence, the mean work per unit time exists and equals
\begin{equation}\label{idealmeanwork}
\Delta W = \frac{1}{\tau} \tr_{\fh_\cS\otimes\fh_{\cE}} \left[\rho_{\cS,+}\otimes\rho_{\cE} \times \left(v-\e^{\i\tau h}v\,\e^{-\i\tau h}\right)\right].
\end{equation}
\end{prop}

\begin{rem} Of course in the invariant state $\rho_{\cS,+}$ the external work is constant as one can see from (\ref{eq:idealwork}).
\end{rem}

\begin{exo} Prove that
\begin{equation}\label{idealmeanwork2}
\Delta W = \frac{1}{\tau} \tr_{\fh_\cS\otimes\fh_{\cE}} \left[\rho_{\cS,+}\otimes\rho_{\cE} \times \int_0^\tau \e^{ish} \Phi \, \e^{-ish} \d s\right],
\end{equation}
where $\Phi=[iv,h_\cE]$. Hint : use (\ref{eq:totalworkham}).
\end{exo}

\begin{exo} Consider the RI system of Exercise \ref{exo:toyideal}. Prove that $\Delta W=0$.
\end{exo}

\begin{rem} The fact that the mean work vanishes is specific to this particular example. If one changes the interaction, e.g. taking the full dipole interaction $(a+a^*)\otimes (b+b^*)$, this is not true anymore. Namely, one can then calculate
$$
\Delta W=\frac{\lambda^2\tau^2 E}{2}\tanh\left(\frac{\beta E_0}{2}\right)\times\frac{\sinc^2\left(\frac{\nu\tau}{2}\right)\sinc^2\left(\frac{\mu\tau}{2}\right)}{\sinc^2\left(\frac{\nu\tau}{2}\right)+\sinc^2\left(\frac{\mu\tau}{2}\right)},
$$
where $\mu=\sqrt{(E+E_0)^2+\lambda^2}$ and $\sinc(x)=\frac{\sin(x)}{x}$. To switch from one element of the environment to the other therefore requires some non-trivial external work.
\end{rem}


\subsubsection{The random situation}

The total work at time $n$ is given by
\begin{eqnarray*}
\Delta E(n) & = & \sum_{k=1}^n \tr_{\fh_\cS\otimes\fh_{\cE_k}}\left[\rho(k-1)\otimes\rho_{\cE_k}\ \left(v_k - \e^{\i\tau_k h_k}v_k\e^{-\i\tau_k h_k}\right)\right]\\
 & & + \tr_{\fh_\cS\otimes\fh_{\cE_{n+1}}}\left[\rho(n)\otimes\rho_{\cE_{n+1}}\ v_{n+1}\right] -\tr_{\fh_\cS\otimes\fh_{\cE_1}}\left[\rho\otimes\rho_{\cE_1}\ v_1\right].
\end{eqnarray*}
The last two terms are typically bounded and will therefore give no contribution to the mean work in the large time limit. Thus we need to understand the large time limit, in the ergodic mean sense, of expectation values of the form
$$
\tr_{\fh_\cS\otimes \fh_\cE(\omega_n)} \left[\rho(n-1,\omega)\otimes\rho_{\cE(\omega_n)}\ A(\omega_n)\right],
$$
where $A(\omega_n)\in\cB(\fh_\cS\otimes \fh_{\cE(\omega_n)}$. The following proposition is an extension of Theorem \ref{thm:randomstatesmall} to the expectation values of such observables.
\begin{prop}\label{prop:randominstantaneuous} Suppose that $\p(\cL(\omega)\in\cM_{(E)})>0$ and $A(\omega_0)\in\cB(\fh_\cS\otimes \fh_{\cE(\omega_0)})$ is a random observable. Then,
\begin{eqnarray*}
\lefteqn{\lim_{N\to\infty} \frac{1}{N}\sum_{n=1}^N \tr_{\fh_\cS\otimes \fh_\cE(\omega_n)} \left[\rho(n-1,\omega)\otimes\rho_{\cE(\omega_n)}\ A(\omega_n)\right]}\\
& = & \E \left(\tr_{\fh_\cS\otimes \fh_\cE(\omega_0)} \left[\rho_{\cS,+}\otimes\rho_{\cE(\omega_0)}\ A(\omega_0)\right]\right) = \tr_{\fh_\cS} \left[\rho_{\cS,+} \times \E\left( A_\cS(\omega_0) \right)\right],
\end{eqnarray*}
where $A_\cS(\omega_0)=\tr_{\fh_{\cE(\omega_0)}} \left[ \one\otimes \rho_{\cE(\omega_0)}\ A(\omega_0) \right]$.
\end{prop}

\noindent {\bf Idea of the proof.} We write
\begin{eqnarray*}
\tr_{\fh_\cS\otimes \fh_\cE(\omega_n)} \left[\rho(n-1,\omega)\otimes\rho_{\cE(\omega_n)}\ A(\omega_n)\right] & = & \tr_{\fh_\cS} \left[ \rho(n-1,\omega) A_\cS(\omega_n) \right]\\
 & & \tr_{\fh_\cS} \left[ \cL(\omega_{n-1})\circ\cdots\circ\cL(\omega_1)(\rho) A_\cS(\omega_n) \right].
\end{eqnarray*}
Introducing the map $\cA_\cS(\omega_0):\cB^1(\fh_\cS)\ni \rho \mapsto \rho A_\cS(\omega_0)$, we have to understand the ergodic average limit of the product $\cA_\cS(\omega_n)\circ  \cL(\omega_{n-1})\circ\cdots\circ\cL(\omega_1)$. In the same way as for Theorem \ref{thm:randomstatesmall} one proves that for any initial state $\rho$
$$
\lim_{N\to\infty} \frac{1}{N}\sum_{n=1}^N \left(\cA_\cS(\omega_n)\circ  \cL(\omega_{n-1})\circ\cdots\circ\cL(\omega_1)\right)(\rho)= \E(\cA_\cS)(\rho_{\cS,+}),
$$
from which the result follows.

\medskip

As a corollary of the previous Proposition we immediately get
\begin{prop} Suppose that $\p(\cL(\omega)\in\cM_{(E)})>0$. Then, the mean work per unit time exists $\P$-almost surely and equals
\begin{eqnarray}\label{randommeanwork}
\Delta W & = & \frac{1}{\E(\tau)} \tr_{\fh_\cS} \Big[ \rho_{\cS,+} \times \E\Big(\tr_{\fh_\cE} \left[ \one\otimes \rho_{\cE} \times \left(v-\e^{\i\tau h}v\,\e^{-\i\tau h}\right) \right] \Big)\Big] \nonumber \\
 & = & \frac{1}{\E(\tau)} \E\left(\tr_{\fh_\cS\otimes\fh_{\cE}} \left[\rho_{\cS,+}\otimes\rho_{\cE} \times \left(v-\e^{\i\tau h}v\,\e^{-\i\tau h}\right)\right]\right).
\end{eqnarray} 
\end{prop}

\begin{exo} Prove that
\begin{equation}\label{randommeanwork2}
\Delta W = \frac{1}{\E(\tau)} \E\left(\tr_{\fh_\cS\otimes\fh_{\cE}} \left[\rho_{\cS,+}\otimes\rho_{\cE} \times \int_0^\tau \e^{ish} \Phi \, \e^{-ish}\d s\right]\right).
\end{equation}
\end{exo}

\begin{exo} In the two cases of Exercise \ref{exo:toyrandom} prove that $\Delta W=0$.
\end{exo}


\subsection{Entropy production}

For a thermodynamic interpretation of the entropy and its relation to the total work, when we will deal with entropy we will always assume that all the initial states $\rho_{\cE_k}$ are Gibbs states at some inverse temperature $\beta_k$.


\subsubsection{General setup}

We fix a reference state $\rho_\cS$ for the small system. Given an initial state $\rho$ of $\cS$ we are interested in the variation of relative entropy of the state $\rho^{\rm tot}(n)$ of the system with respect to the reference state $\rho_0=\rho_\cS\otimes \bigotimes_{k\geq 1}\rho_{\cE_k,\beta_k}$ between time $0$ and time $n$ and where $\ds \rho_{\cE_k,\beta_k}=\frac{\e^{-\beta_k h_{\cE_k}}}{\tr \left(\e^{-\beta_k h_{\cE_k}}\right)}$. 

The relative entropy of two states (density matrices) $\rho$ and $\rho'$ is the (possibly infinite) non-positive quantity
$$
\ent(\rho'|\rho) = \tr (\rho' \log \rho- \rho' \log \rho').
$$
We are interested in the mean entropy production per time unit, i.e.
\begin{equation*}\label{def:entropyprod}
\Delta S:= \lim_{n\to\infty} -\frac{1}{t_n}\left[{\rm Ent}( \rho^{\rm tot}(n)| \rho_0)-{\rm Ent} (\rho^{\rm tot}(0) |\rho_0)\right],
\end{equation*}
and its relation to the external work. (Note that since the relative entropy is negative, the entropy production is indeed a positive quantity.)

As we mentioned in Section \ref{sec:framework} the infinite tensor product $\bigotimes_{k\geq 1} \rho_{\beta_k}$ does not make sense, neither does $\rho^{\rm tot}(n)$. However at time $t_n$ only the $n$ first elements of the chain have interacted with $\cS$ and the other ones are ``at equilibrium''. Therefore the elements with index $k>n$ give no contribution to the entropy production at time $t_n$ and we may write 
\begin{eqnarray}\label{eq:relatentropy}
{\rm Ent}(\rho^{\rm tot}(n)|\rho_0) & := & \tr\left[ \rho^{\rm tot}(n) \log \rho_0 - \rho^{\rm tot}(n) \log (\rho^{\rm tot}(n))\right]\nonumber\\
 & = & \tr_{\fh_\env^{(n)}} \left[\tilde\rho(n) \log (\tilde\rho_0) - \tilde\rho(n) \log (\tilde\rho(n)) \right],
\end{eqnarray}
where
$$
\tilde\rho(n)=\e^{-\i\tau_n \th_n}\cdots\e^{-\i\tau_1 \th_1} \left( \rho\otimes \bigotimes_{k=1}^n \rho_{\cE_k,\beta_k}\right) \e^{\i\tau_1 \th_1}\cdots\e^{\i\tau_n \th_n} \quad \mbox{and} \quad
\tilde\rho_0= \rho_\cS\otimes \bigotimes_{k=1}^n\rho_{\cE_k,\beta_k}.
$$
Using the cyclicity of the trace and the fact that $\tilde\rho_0$ is a product state, we thus have
\begin{eqnarray}\label{eq:epformulahamilt}
\Delta S(n) & = & \ent(\rho^{\rm tot}(n)|\rho_0) - \ent(\rho^{\rm tot}(0)|\rho_0) \nonumber\\
 & = & \tr_{\fh_\env^{(n)}} \left[ \tilde\rho(n) \log (\tilde\rho_0) -\tilde\rho(n) \log (\tilde\rho(n))\right] - \tr_{\fh_\cS} \left[ \rho \log (\rho_\cS) -\rho \log (\rho)\right]\nonumber\\
 & = & \tr_{\fh_\env^{(n)}} \left[ \tilde\rho(n) \log (\tilde\rho_0) -\tilde\rho(0) \log (\tilde\rho(0))\right]- \tr_{\fh_\cS} \left[ \rho \log (\rho_\cS) -\rho \log (\rho)\right]\nonumber\\
 & = & \tr_{\fh_\cS} \left[ \rho(n) \log (\rho_\cS) - \rho\log(\rho)\right]-\sum_{k=1}^n \beta_k \tr_{\fh_\env^{(n)}} \left[\left(\tilde\rho(n)-\tilde\rho(0)\right) h_{\cE_k} \right]\nonumber\\
 & & - \tr_{\fh_\cS} \left[ \rho \log (\rho_\cS) -\rho \log (\rho)\right]\nonumber\\
& = & \tr_{\fh_\cS} \left[ (\rho(n)-\rho) \log (\rho_\cS)\right]-\sum_{k=1}^n \beta_k \tr_{\fh_\env^{(n)}} \left[\left(\tilde\rho(k)-\tilde\rho(k-1)\right) h_{\cE_k} \right]\nonumber\\
& = & \tr_{\fh_\cS} \left[ (\rho(n)-\rho) \log (\rho_\cS)\right]\\
 & & -\sum_{k=1}^n \beta_k \tr_{\fh_\cS\otimes\fh_{\cE_k}}\left[\rho(k-1)\otimes \rho_{\cE_k,\beta_k}  \times \int_0^{\tau_k} \e^{i s h_k} \Phi_k \e^{-is h_k} \, \d s \right].\nonumber
\end{eqnarray}
which can also be rewritten as
\begin{eqnarray}\label{eq:epformulahamilt2}
\Delta S(n) & = & \tr_{\fh_\cS} \left[ (\rho(n)-\rho) \log (\rho_\cS)\right] + \sum_{k=1}^n \beta_k \tr_{\fh_\cS} \left[\left(\rho(k-1)-\rho(k)\right) h_\cS \right]\nonumber\\
 & & + \sum_{k=1}^n \beta_k \tr_{\fh_\cS\otimes\fh_{\cE_k}}\left[\rho(k-1)\otimes \rho_{\cE_k,\beta_k}  \left(\e^{i\tau_k h_k} v_k \e^{-i\tau_k h_k} - v_k \right) \right].
\end{eqnarray}


\subsubsection{Ideal situation}

In the case of identical interactions, equation (\ref{eq:epformulahamilt}) simplifies as
\begin{eqnarray}\label{eq:idealentropy}
\lefteqn{\Delta S(n)}\\
& = &  \tr_{\fh_\cS} \left[ (\rho(n)-\rho) \log(\rho_{\cS}) \right]  - \beta \sum_{k=1}^n \tr_{\fh_\cS\otimes\fh_{\cE}}\left[\rho(k-1)\otimes\rho_{\cE,\beta}\times \int_0^\tau \e^{ish} \Phi \, \e^{-ish}\d s\right].\nonumber
\end{eqnarray}
As a consequence, using also (\ref{idealmeanwork2}), we immediately get
\begin{prop} Suppose Assumption {\bf (E)} is satisfied. Then for any initial state $\rho\in \cB^1(\fh_\cS)$,
\begin{equation}\label{eq:idealmeanentropy}
\Delta S = \lim_{n\to\infty} -\frac{\Delta S(n)}{n\tau} =\beta\Delta W.
\end{equation}
\end{prop}


\subsubsection{Random situation}

A direct application of Proposition \ref{prop:randominstantaneuous} with $\ds A=\beta \int_0^\tau \e^{ish} \Phi \, \e^{-ish}\d s$ gives
\begin{prop} Suppose that $\p(\cL(\omega)\in\cM_{(E)})>0$. Then, for any initial state $\rho$, the mean entropy production per unit time exists $\P$-almost surely and equals
\begin{equation}\label{randommeanentropy}
\Delta S = \frac{1}{\E(\tau)} \E\left(\beta\tr_{\fh_\cS\otimes\fh_{\cE}} \left[\rho_{\cS,+}\otimes\rho_{\cE} \times \int_0^\tau \e^{ish} \Phi \, \e^{-ish}\d s\right]\right).
\end{equation}
\end{prop}

As for the external work, the entropy production is also deterministic. Comparing (\ref{randommeanentropy}) with (\ref{randommeanwork2}), one can recognize a sort of 2nd law of thermodynamics. If in particular the inverse temperature $\beta$ of the various subsystems is not random (i.e. we are at equilibrium), then $\Delta S=\beta \Delta W$ as expected. Of course, if $\beta$ is not constant, we are in a non-equilibrium situation and the fact that there is no clear relation between $\Delta S$ and $\Delta W$ should not come as a surprise.

\begin{exo} Consider the two situations of Exercise \ref{exo:toyrandom}. Prove that in case 1) $\Delta S=0$ and in case 2) $\Delta S= \frac{E_0(1-e_0)}{\tau} {\rm Cov}\left( \beta , \frac{1}{1+\e^{-\beta E_0}} \right)$ where $e_0$ is defined in (\ref{def:e0}). When does then entropy production vanish? 
\end{exo}


\subsection{Energy fluxes}\label{ssec:fluxes}

When several reservoirs are present one is also interested in the energy transfer between the various reservoirs. We shall consider here the following situation : RI systems with $K$ ``beams''. This can be achieved in two ways : either assuming that for each $j=1,\ldots,K$ the $(mK+j)$-th subsystems, $m\in\N$, are identical, or considering a random situation where the underlying probability space is the set $\{1,\ldots,K\}$ with the uniform probablity measure. A third situation will be considered in Section \ref{ssec:leakyri} where, besides the chain, the system $\cS$ is also coupled to an extra reservoir.


\subsubsection{Several beams I : deterministic}\label{sssec:noneq-deterministic}

The first non-equilibrium situation we consider is that of $K$ ideal ``beams'', or subchains. More precisely, $\fh_{\cE_{mK+j}}\equiv \fh_{\cE_j}$, $h_{\cE_{mK+j}}\equiv h_{\cE_j}$, \ldots We denote by $\cL_1$, $\cL_2$,... the corresponding RDM's. The state of $\cS$ at time $n$ is therefore
\begin{equation}\label{eq:rho-noneq}
\rho(n)= (\cL_j\circ\cL_{j-1}\circ\cdots\cL_1\circ\cL_K\circ\cdots \circ\cL_{j+1})^m \left(\cL_j\circ\cdots\circ\cL_1(\rho)\right), \ n=mK+j.
\end{equation}

Obviously, the $j$-th beam can exchange energy only when it interacts with $\cS$, that is in the time intervals $[t_{mK+j-1}, t_{mK+j})$. During such an interval, and if the system $\cS$ is in a state $\rho$ at the beginning of the interaction, the amount of energy lost by that beam is
\begin{equation}\label{def:jthenergyvar}
\delta E_j(n) =  -\tr_{\fh_\cS\otimes\fh_{\cE_j}} \left[ \rho\otimes \rho_{\cE_j} \times \int_0^{\tau_j} \e^{i s h_j} \Phi_j\, \e^{-is h_j} \, \d s \right].
\end{equation}
The total amount of energy lost by the $j$-th beam between time $0$ and time $t_n$ is therefore
\begin{equation}\label{eq:jthenergytotalvar}
\Delta E_j(n)= - \sum_{m=0}^{[n/K]} \tr_{\fh_\cS\otimes\fh_{\cE_j}} \left[ \rho(mK+j-1)\otimes \rho_{\cE_j} \times \int_0^{\tau_j} \e^{i s h_j} \Phi_j\, \e^{-is h_j} \, \d s \right],
\end{equation}
and the energy flux in the $j$-th beam is therefore (if it exists)
$$
\phi_j = \lim_{n\to \infty} \frac{\Delta E_j(n)}{t_n}.
$$

\medskip

Generically, in non-equilibrium situation, the usual limit $n\to\infty$ of $\rho(n)$ does not exist and one has to resort to limits in the ergodic mean. However here, in view of (\ref{eq:jthenergytotalvar}), we are not only interested in the limit of $\rho(n)$ but also in the large $m$ limit of the $K$ subsequences $\rho(mK+j)$. The latter will depend on the spectral properties of the maps
$$
\tilde\cL_j=\cL_j\circ\cL_{j-1}\circ\cdots\cL_1\circ\cL_K\circ\cdots \circ\cL_{j+1}.
$$
\begin{prop} Assume the map $\tilde\cL_j$ satisfies Assumption {\bf (E)}, and let $\rho_{\cS,+}^j$ denote its unique invariant state. Then
$$
\phi_j= \frac{-1}{\tau_1+\cdots +\tau_K}\tr_{\fh_\cS\otimes\fh_{\cE_j}} \left[ \rho_{\cS,+}^j\otimes \rho_{\cE_j} \times \int_0^{\tau_j} \e^{i s h_j} \Phi_j\, \e^{-is h_j} \, \d s \right].
$$
\end{prop}

In this situation, using (\ref{eq:totalworkham}) and (\ref{eq:jthenergytotalvar}), the total work between time $0$ and time $t_n$ becomes
\begin{eqnarray*}
\Delta E(n) & = & -\sum_{j=1}^K \Delta E_j(n) + \tr_{\fh_\cS} \left[ \left(\rho(n)-\rho(0)\right) h_\cS\right]\nonumber \\
 & &  + \tr_{\fh_\cS\otimes\fh_{\cE_{n+1}}} \left[ \rho(n)\otimes \rho_{\cE_{n+1}}\, v_{n+1}\right] - \tr_{\fh_\cS\otimes\fh_{\cE_1}} \left[ \rho(0)\otimes \rho_{\cE_1}\, v_1\right].\nonumber
\end{eqnarray*}
Similarly, if furthermore the initial states $\rho_{\cE_j}$ are Gibbs states, the variation of entropy production between time $0$ and time $t_n$ is
$$
\Delta S(n)= \tr_{\fh_\cS} [(\rho(n)-\rho)\log(\rho_\cS)] -\sum_{j=1}^K \beta_j\Delta E_j(n).
$$

As a consequence we get
\begin{prop} Assume the maps $\tilde\cL_j$ satisfy Assumption {\bf (E)}. Then the mean work $\Delta W$ and mean entropy production $\Delta S$ exist. Moreover one has
$$
\Delta W = -\sum_{j=1}^K \phi_j \quad \mbox{ and } \quad \Delta S = -\sum_{j=1}^K \beta_j\phi_j.
$$ 
\end{prop}

\begin{rem} The non-equilibrium steady state of the system is $\rho_{\cS,+}=\frac{1}{K}\sum_{j=1}^K \rho_{\cS,+}^j$. 
\end{rem}

\begin{exo}\label{exo:toynoneq} Consider again Example \ref{ex:toy-description-rdm}. Prove that $\tilde\cL_j$ satisfies {\bf (E)} iff $\nu\tau\notin 2\pi\N$. In that case, prove that
\begin{equation*}\label{def:partialness}
\rho_{\cS,+}^j=\frac{1-e_0}{1-e_0^K}\left( \rho_{\cS,\beta^*_j}+e_0\rho_{\cS,\beta^*_{j-1}}+\cdots +e_0^{K-1}\rho_{\cS,\beta^*_{j+1}}\right).
\end{equation*}
Prove that $\phi_j=\frac{E_0(1-e_0)^2}{K\tau(1-e_0^K)}\sum_{k=1}^K (Z_{\beta_k}^{-1}-Z_{\beta_j}^{-1}) e_0^{[j-k-1]}$, where $[n]$ denotes the residue class mod $K$. Calculate $\Delta W$ and $\Delta S$. When does $\Delta S$ vanishes? 
\end{exo}


\subsubsection{Several beams II : random}\label{sssec:noneq-random}

As we mentioned, another way to have a non-equilibrium situation with $K$ beams is to consider a random situation where the underlying probability space is the set $\{1,\ldots,K\}$ with the uniform probablity measure. The calculation of the mean external work $\Delta W$ and of the mean entropy production $\Delta S$ are thus particular cases of (\ref{randommeanwork2}) and (\ref{randommeanentropy}). It remains to define the various energy fluxes and relate them to $\Delta W$ and $\Delta S$.

According to (\ref{def:jthenergyvar}), the energy lost by the $j$-th beam during the time interval $[t_{-1}n,t_n)$, i.e. during the $n$-th interaction, is 
\begin{equation}
\delta E_j(n,\omega) =  \left\{ \begin{array}{ll} -\tr_{\fh_\cS\otimes\fh_{\cE_j}} \left[ \rho(n-1,\omega)\otimes \rho_{\cE_j} \times \int_0^{\tau_j} \e^{i s h_j} \Phi_j\, \e^{-is h_j} \, \d s \right], & {\rm if} \ \cE(\omega_n)=\cE_j,\\
0, & {\rm otherwise}. \end{array}\right.
\end{equation}
The total amount of energy lost by the $j$-th beam between time $0$ and time $t_n$ is therefore
\begin{equation}
\Delta E_j(n,\omega) =  \sum_{k=1}^n \delta E_j(k,\omega).
\end{equation}
Introducing the random observable
$$
\tilde\Phi_j(\omega_0)= \left\{ \begin{array}{ll} \int_0^{\tau_j} \e^{i s h_j} \Phi_j\, \e^{-is h_j} \, \d s & {\rm if} \ \cE(\omega_0)=\cE_j,\\
0, & {\rm otherwise}, \end{array}\right.
$$
we have
$$
\Delta E_j(n,\omega) = -\sum_{k=1}^n \tr_{\fh_\cS\otimes \fh_{\cE(\omega_k)}} \left[ \rho(k-1,\omega)\otimes \rho_{\cE(\omega_k)} \tilde\Phi_j(\omega_k)\right].
$$
Using Proposition \ref{prop:randominstantaneuous} we therefore have
\begin{prop} Suppose that $\p(\cL(\omega)\in\cM_{(E)})>0$. Then, for any initial state $\rho$, the energy flux in the $j$-th beam exists $\P$-a.s. and is given by
\begin{eqnarray}
\phi_j & = & - \frac{1}{\E(\tau)} \E\left( tr_{\fh_\cS\otimes \fh_\cE} \left[ \rho_{\cS,+}\otimes \rho_\cE \times \tilde\Phi_j\right] \right)\\
 & = & \frac{-1}{\tau_1+\cdots+\tau_K} \tr_{\fh_\cS\otimes\fh_{\cE_j}} \left[ \rho_{\cS,+}\otimes \rho_{\cE_j} \times \int_0^{\tau_j} \e^{i s h_j} \Phi_j\, \e^{-is h_j} \, \d s \right].
\end{eqnarray}
Moreover one has
$$
\Delta W = -\sum_{j=1}^K \phi_j \quad \mbox{ and } \quad \Delta S = -\sum_{j=1}^K \beta_j\phi_j.
$$ 
\end{prop}

\begin{rem} Note the difference between the energy flux in this random situation compared to the previous deterministic situation. In both case one calculates the expectation value of the observable $\ds \int_0^{\tau_j} \e^{i s h_j} \Phi_j\, \e^{-is h_j} \, \d s$ but not in the same state. 
\end{rem}

\begin{exo} Consider the random version of Exercise \ref{exo:toynoneq}. Prove that
\begin{equation}\label{eq:spinrandomfluxes}
\phi_j=\frac{E_0(1-e_0)}{K^2\tau}\sum_{k=1}^K (Z_{\beta_k}^{-1}-Z_{\beta_j}^{-1}).
\end{equation}
Calculate $\Delta W$ and $\Delta S$. When does $\Delta S$ vanishes? 
\end{exo}

One interest in considering the random situation rather than the deterministic non-equilibirum situation is when one turns to linear response theory (and beyond). For example, do the Green-Kubo formula and the Onsager reciprocity relations hold for repeated interaction systems? In the deterministic setting, one should not expect neither the Green-Kubo formula nor the Onsager reciprocity relations to hold because the system is not at all \emph{time reversal invariant}. Indeed, the various beams interact with the system $\cS$ in a precise order : $1,2,\ldots,K,1,2,\ldots$ It is therefore reasonnable to expect that a change (of temperature say) in beam $1$ will have a greater influence on the energy flux in beam $2$ than on the one in beam $K$. In the same spirit, one expects that a change in beam $1$ will have a greater influence on the flux in beam $2$ than a change in beam $2$ on the flux in beam $1$. More precisely if, for $j=1,\ldots,K$, $X_j=\beta-\beta_j$ denote the thermodynamical forces and $L_{jk}=\frac{\partial \phi_j}{\partial X_k}\lceil_{X=0}$ are the kinetic coefficients one would expect e.g. $|L_{21}|>|L_{12}|$, and the larger $K$ the greater should be the difference (beam $2$ arrives right after beam $1$ while we need to wait an amount of time $(K-2)\tau$ before beam $1$ comes back after beam $2$). More generically, the further beam $j$ arrives after beam $k$ the smaller $|L_{jk}|$ should be.

\begin{exo} Calculate the kinetic coefficients in the deterministic non-equlibrium situation of the toy model of Example \ref{ex:toy-description-rdm} (the fluxes $\phi_j$ are given in Exercise \ref{exo:toynoneq}). What do we observe? 
\end{exo}

This is why random situation is relevant : to restore symmetry and have a hope to prove Green-Kubo formula and Onsager relations.  

\begin{exo} Calculate the kinetic coefficients in the random non-equlibrium situation of the toy model of Example \ref{ex:toy-description-rdm} (the fluxes $\phi_j$ are given in (\ref{eq:spinrandomfluxes})). What do we observe? 
\end{exo}

\begin{open} Study the linear response theory of repeated interaction systems. Do the Green-Kubo formula and the Onsager reciprocity relations? In which form? What is the fluctuation theory for such systems (large deviations for entropy production, Evans-Searles and/or Gallavotti-Cohen symmetry)?
\end{open}

\begin{open} Analyse the non-equilibrium situation (several beams) of concrete models like the One-atom maser model of Section \ref{ssec:qedcavity}:  existence of a NESS, linear response theory, fluctuation relations, etc. 
\end{open}


\section{The Liouvillian description and application to leaky RI systems}\label{sec:leaky}


\subsection{$C^*$-dynamical systems and the Liouvillian approach}\label{ssec:riliouv}

In this section we give an alternative description of RI systems using the language of algebraic quantum statistical mechanics, and starting from the $C^*$- dynamical system formalism. The main reason is to include an extra-reservoir $\cR$ with which $\cS$ will interact (leaky RIS) and get a unified description of the full model. Besides adding an extra reservoir, this also allows for more general systems, e.g. take the subsystems $\cE_n$ to be thermal reservoirs described by infinitely extended Fermi gas, and even if it is not our main concern, show how to construct the full system, including the infinite tensor product.

We first briefly recall some basic concepts of algebraic quantum statistical mechanics that we need here. We refer to e.g. \cite{BR,P} for a more complete introduction to the subject. A $C^*$- dynamical system is a pair $(\fa,\alpha^t)$ where $\fa$ is a $C^*$- algebra (describing the observables of the physical system under consideration) and $t\mapsto \alpha^t$ is a strongly continuous group of $*$-automorphisms of $\fa$ (describing the evolution of the observables). A state of the system is described by a positive linear functional $\varrho$ on $\fa$ satisfying $\varrho(\one)=1$. Following \cite{JP2}, a triple $(\fa, \alpha^t,\varrho)$, where $\varrho$ is an invariant state (i.e. $\varrho\circ \alpha^t\equiv \varrho$), is called a quantum dynamical system.

\begin{ex}\label{ex:cstar-finite} (Finite systems) Consider a quantum system described by the Hilbert space $\fh=\C^n$, the Hamiltonian $h$ and the invariant state $\rho=\sum \rho_j |\psi_j\ket\bra\psi_j|$ where $\{\psi_j\}$ is an orthonormal basis of eigenvectors of $h$. The corresponding quantum dynamical system is described by the algebra of observables $\fa=M_n(\C)$, the dynamics $\alpha^t(A)=\e^{\i t h}A\, \e^{-\i th}$ and state $\varrho(A)=\tr(\rho A)$.
\end{ex}

\begin{ex}\label{ex:cstar-fermi} (Free Fermi gas) Let $\fh$ be a Hilbert space, later refered to as the one-particle space, and $h$ a self-adjoint operator on $\fh$. The pair $(\fh,h)$ describes one fermion. The Hilbert space describing a gas of non-interacting fermions is the Fermionic Fock space $\Gamma_-(\fh)=\bigoplus_{n\geq 0} \wedge^n\,\fh$. The algebra of observables is the $C^*$-algebra of operators $\mathfrak A$ generated by $\{a^\#(f)\, |\, f\in\fh\}$ where $a/a^*$ denote the usual annihilation/creation operators on $\Gamma_-(\fh)$. The dynamics is the Bogoliubov dynamics generated by the one-particle Hamiltonian $h$, i.e. given by $\alpha^t(a^\#(f))=a^\#(\e^{\i th}f)$. It is well known (see e.g. \cite{BR}) that for any
$\beta>0$ there is a unique $(\alpha^t,\beta)-$KMS state $\rho_\beta$ on $\mathfrak A$, determined by the two point function $\rho_{\beta}(a^*(f)a(f))=\langle f, (1+\e^{\beta_\cR h_\cR})^{-1}f\rangle$. The triple $(\fa, \alpha^t,\rho_\beta)$ is a quantum dynamical system describing a free Fermi gas in thermal equilibrium at inverse temperature $\beta$. 
\end{ex}

Each component $\#=\cS, \cE_n$ of the RI system will be described by a quantum dynamical system $(\fa_\#, \alpha_\#^t,\varrho_\#)$. The ``reference'' states $\varrho_\#$ determine the macroscopic properties of the systems, e.g. they are KMS states at some inverse temperature $\beta_\#$.  We also assume that they are faithful states, i.e. for any $A\in \fa_\#$, $\varrho_\#(A^*A)=0 \Rightarrow A=0$ (this would correspond to $\rho>0$).

To analyze the (time) asymptotic behaviour of the system, we will use a spectral approach. For that purpose, it is convenient to have a ``Hilbert space description'' of the system. Such a description is easy to obtain via the GNS-representation\footnote{We recall that the GNS (Gelfand-Naimark-Segal) representation of a $C^*$- algebra $\fa$ associated to a state $\varrho$ is a triple $(\cH,\pi,\Psi)$ where $\cH$ is a Hilbert space, $\pi$ a $*$- algebra morphism from $\fa$ to $\cB(\cH)$, and $\Psi$ a unit vector in $\cH$ such that $\{ \pi(A)\Psi, \ A\in \fa\}$ is dense in $\cH$ and $\varrho(A)=\bra \Psi, \pi(A)\Psi\ket$ for any $A\in\fa$.} $(\cH_\#,\pi_\#,\Psi_\#)$ of the algebras $\fa_\#$ associated to the states $\varrho_\#$. Since the $\varrho_\#$ are faithful, the $\pi_\#$ are injections and we can identify $\fa_\#$ and $\pi_\#(\fa_\#)$ (we will therefore simply write $A$ for $\pi(A)$). We set $\fm_\#=\pi_\#(\fa_\#)''\subset {\cal B}(\cH_{\#})$, where $''$ denotes the double commutant. The $\fm_{\# }$ form the von Neumann algebras of observables. Finally, by construction the representative vectors $\Psi_\#$ are cyclic for $\fm_\#$, and we assume that they are also separating vectors for $\fm_\#$, i.e. $A\Psi_\#=0\ \Rightarrow A=0$ for any $A\in\fm_\#$ (note that since $\varrho_\#$ is faithful, this is automatic when $A\in \pi_\#(\fa_\#)$). Typically, the $\Psi_\#$ describe the equilibrium states at some fixed temperature $T_\#>0$.

The free dynamics $\alpha_\#^t$ of each constituent is implemented in the GNS-representation by self-adjoint operators $L_{\#}$ called Liouvillians, i.e. the Heisenberg evolution of an observable $A\in\fm_{\#}$ at time $t$ is given by $\e^{\i t L_{\#}} A \e^{-\i t L_{\#}}$. In other words we have $\pi_\#(\alpha_\#^t(A))= \e^{\i t L_{\#}} \pi_\#(A) \e^{-\i t L_{\#}}$. Since the $\varrho_\#$ were invariant states, one can also chose the Liouville operators $L_\#$ so that $L_\#\Psi_\#=0$ (actually such an $L_\#$ is unique).

\begin{ex}\label{ex:gns-finite} (Finite system, continuation of Example \ref{ex:cstar-finite}) In the GNS representation, the Hilbert space, the observable algebra and the Liouville operator are given by
\begin{equation}
\cH= \fh\otimes\fh,\qquad \fm =\cB(\fh)\otimes\one, \qquad L=h\otimes\one - \one\otimes h,
\label{m4}
\end{equation}
and the representative vector $\Psi$ by $\Psi = \sum \sqrt{\rho_j}\,  \psi_j\otimes\psi_j$. (The morphism $\pi$ is defined as $\pi(A)=A\otimes\one$.)
\end{ex}

\begin{ex}\label{ex:gns-fermi} (Free Fermi gas, continuation of Example \ref{ex:cstar-fermi}) The GNS representation of a Free Fermi gas is given by the so-called Araki-Wyss representation \cite{AW}. Namely, if $\Omega$ denotes the Fock vacuum and $N$ the number operator on $\Gamma_-(\fh)$, the Hilbert space, the observable algebra and the Liouville operator are given by
\begin{equation*}
\cH= \Gamma_-(\fh) \otimes \Gamma_-(\fh),\qquad \fm= \pi_\beta\left( \mathfrak A\right)'', \qquad L=\d\Gamma(h)\otimes\one -\one\otimes \d\Gamma(h),
\end{equation*}
where
\begin{equation}\label{def:pibeta}
\begin{array}{l}
\pi_{\beta}(a(f)) = a\left(\frac{\e^{\beta h/2}}{\sqrt{1+\e^{\beta h}}}f \right)\otimes \one+(-1)^N\otimes a^*\left(\frac{1}{\sqrt{1+ \e^{\beta  h}}} \bar{f} \right)=:a_{\beta}(f),\\
\pi_{\beta}(a^*(f)) = a^*\left(\frac{\e^{\beta h/2}}{\sqrt{1+\e^{\beta h}}}f \right)\otimes \one+(-1)^N\otimes a\left(\frac{1}{\sqrt{1+\e^{\beta h}}} \bar{f} \right)=:a^*_{\beta}(f),
\end{array}
\end{equation}
and the representative vector is $\Psi=\Omega \otimes \Omega$. 
\end{ex}

Each component of the RI system is thus now described by a von Neumann algebra (of observables) $\fm_\#$ acting on the Hilbert space $\cH_\#$, a self-adjoint operator $L_\#$ on $\cH_\#$ which implements the dynamics and a unit vector $\Psi_\#\in\cH_\#$ which represents some reference invariant state. The Hilbert space $\cH_\env$ for the environment is then the infinite tensor product of factors $\cH_{\cE_n}$, taken with respect to the stabilizing sequence $(\Psi_{\cE_n})_n$. The vector $\Psi_\env = \otimes_{n\geq 1} \Psi_{\cE_n}$ is the reference vector for the environment, and the algebra of observables $\fm_\env$ of the environment is the von Neumann algebra $\ds \fm_\env =\otimes_{n\geq 1}\fm_{\cE_n}$ acting on $\cH_\env$, which is obtained by taking the weak closure of finite linear combinations of operators $\otimes_{n\geq 1} A_n$, where $A_n\in\fm_{\cE_n}$ and $A_n=\one_{\cH_{\cE_n}}$ except for finitely many indices.

To summarize, the non-interacting system is described by a von Neumann algebra $\fm=\fm_\cS \otimes \fm_\env,$ acting on the Hilbert space $\cH = \cH_\cS\otimes\cH_\env$, and its dynamics is generated by the (free) Liouvillian
\begin{equation*}
L_0 = L_\cS+ \sum_{n\geq 1} L_{\cE_n}.
\label{-3}
\end{equation*}
The operators governing the couplings between $\cS$ and $\cE_n$ are given by operators
$$
V_n \in \fm_\cS \otimes \fm_{\cE_n}.
$$
(If the system is initially given in the Hamiltonian formalism, then $V_n=\pi_\cS\otimes\pi_{\cE_n}(v_n)$.) The evolution of the interacting system is thus generated by the Liouvillian
\begin{equation*}\label{def:totalliouv}
L(t)=L_0+\sum_{n\geq 1} \chi_n(t) V_n.
\end{equation*}
In the same way as for the Hamiltonian description, we will denote
\begin{equation}
L_n:=L_\cS+L_{\cE_n}+ V_n, \quad \mbox{and}\quad \tL_n=L_n+\sum_{k\neq n} L_{\cE_k},
\label{m3}
\end{equation}
so that $L(t)\equiv \tL_n$ when $t\in[t_{n-1},t_n)$. We will also denote by $U(t,0)$ the associated propagator, i.e. for $t\in[t_n,t_{n+1})$ one has
\begin{equation}
\label{eq:liouvpropagator}
U(t,0)=\e^{-\i (t-t_n) \tL_{n+1}} \e^{-\i\tau_n \tL_n}\cdots \e^{-\i\tau_1 \tL_1}.
\end{equation}
Finally we denote by
\begin{equation*}\label{eq:heisenberg}
\alpha^t_\RI(A):=U(t,0)^*AU(t,0)
\end{equation*}
the evolution of an observable $A\in\fm$ at time $t$.

As in the Hamiltonian description, we now explain how to reduce the analysis of expectation values of observables on $\cS$ to the product of ``Reduced Dynamics Operators'' acting on $\cH_\cS$ only. In order not to muddle the essence of the argument, let us assume that the initial state of the entire system is given by the vector $\Psi_0=\Psi_\cS\otimes\Psi_\env$ (see \cite{BJM1,BJM3} for more details). If $A_\cS\in\fm_\cS$ we thus want to calculate
\begin{eqnarray}
\label{evol}
\bra A_\cS\ket(n) & := & \bra \Psi_0, \alpha^{t_n}_\RI(A_\cS\otimes \one_\env) \Psi_0\ket \\
 & = & \bra \Psi_0, \e^{\i\tau_n \tL_n}\cdots\e^{\i\tau_1 \tL_1} \ A_\cS\otimes \one_\env \ \e^{-\i\tau_1 \tL_1}\cdots \e^{-\i\tau_n \tL_n}\Psi_0\ket.\nonumber
\end{eqnarray}

The first step consists in the following decomposition which serves to isolate the dynamics of the elements $\cE$ which do not interact at a given time, and which is the equivalent of (\ref{eq:dynsplit}):
\begin{equation*}
\e^{-\i\tau_n\tL_n}\cdots\e^{-\i\tau_1 \tL_1}=  U^-_m \e^{-\i\tau L_n}\cdots\e^{-\i\tau L_1}U^+_n,
\end{equation*}
where
$$
U^-_n = \exp\left(-\i \sum_{k=1}^{n-1} (t_n-t_k)L_{\cE_k}\right),\quad
U^+_n = \exp\left(-\i \sum_{k=2}^n t_{k-1} L_{\cE_k}-\i t_n \sum_{k>n} L_{\cE_k}\right).
$$
One easily sees that $U_n^+\Psi_0=\psi_0$ and that $U_n^-$ commutes with $A_\cS\otimes \one_\env$, so that (\ref{evol}) can be written as
\begin{eqnarray}\label{evol2}
\bra A_\cS \ket(n) & = & \bra \Psi_0, \e^{\i\tau_1 L_1}\cdots \e^{\i\tau_n L_n} A_\cS\otimes\one_\env\ \e^{-\i\tau_n L_n}\cdots \e^{-\i\tau_1 L_1} \Psi_0\ket.
\end{eqnarray}

The second step is to replace, for all $n$, the Liouvillean $L_n$ by another (non selfadjoint) generator $K_n$ of the interacting dynamics, called a C-Liouville operator, which satisfies the following additional property:
\begin{equation}\label{eq:cliouv}
K_n \ \Psi_\cS\otimes\Psi_{\cE_n}=0,
\end{equation}
i.e. it ``kills'' the reference vector. The C-Liouville operator has been introduced in \cite{JP2} to study non-equilibrium steady states (NESS).

\noindent {\it Remark.\ } For the existence of such a generator, we refer to e.g. \cite{JP2}. One can also get an explicit expression for it in terms of the Liouvillean and the modular data of the pair $(\fm_\cS\otimes\fm_{\cE_n}, \Psi_\cS\otimes\Psi_{\cE_n})$ \cite{AJP,BR,JP2}.

Since the operators $K_n$ are also generators of the dynamics, and using (\ref{eq:cliouv}), (\ref{evol2}) becomes 
\begin{equation}\label{evol3}
\bra A_\cS \ket(n)=\bra \Psi_0, \e^{\i\tau_1 K_1}\cdots \e^{\i\tau_n K_n} (A_\cS\otimes \one_\env)\Psi_0\ket.
\end{equation}

The last step is to use the independance of the various elements of the environment and to rewrite (\ref{evol3}) in terms of a product of ``reduced dynamics operators'' $M_n$. Let
\begin{equation}
P:=\one_{\cH_\cS}\otimes|\Psi_\env\rangle\langle\Psi_\env|
\label{mP}
\end{equation}
denote the orthogonal projection onto $\cH_\cS \otimes \C\Psi_\env \cong \cH_\cS$. If $B$ is an operator acting on $\cH$ then we identify $PBP$ as an operator acting on $\cH_\cS$. Note that $P\Psi_0=\Psi_0$, hence
\begin{equation*}\label{evol3bis}
\bra A_\cS \ket(n)=\bra \Psi_0, P\e^{\i \tau_1 K_1}\cdots \e^{\i\tau_n K_n}P (A_\cS\otimes \one_\env) \Psi_0\ket.
\end{equation*}
The structure of RI systems gives
$$
P\e^{\i \tau_1 K_1}\cdots \e^{\i\tau_n K_n} P = (P\e^{\i \tau_1 K_1}P) \times (P\e^{\i \tau_2 K_2}P)\times\cdots\times (P\e^{\i\tau_n K_n}P),
$$
(this is nothing but the Markov property). Hence, introducing $M_j:=P\e^{\i\tau_j K_j}P$ (considered as an operator acting on $\cH_\cS$), we finally get
\begin{equation}\label{eq:rdoreduction}
\bra A_\cS\ket(n)=\bra \Psi_\cS, M_1\cdots M_n A_\cS \Psi_\cS\ket.
\end{equation}
We have thus reduced the analysis to the one of the product of the operators $M_1\cdots M_n$.

At first sight, and despite the fact that the operators $M_n$ give the desired reduction procedure, their definition may look quite obscure. Actually they are nothing but the GNS version of the dual $\cL_n^*$ of the RDM's $\cL_n$ as we shall now explain, see (\ref{eq:gnsrdo}). We suppose that the RI system is given in the Hamiltonian formalism. Let $A,B\in \fa_\cS=\cB(\fh_\cS)$. We consider the quantity $\bra \pi_\cS(B)\Psi_\cS, M\pi_\cS(A)\Psi_\cS\ket$ (we drop the index $n$ to simplify notation). One can then write
\begin{eqnarray*}
\bra \pi_\cS(B)\Psi_\cS, M\pi_\cS(A)\Psi_\cS\ket & = & \bra \pi_\cS(B)\otimes \one \ \Psi_\cS\otimes\Psi_\env, \e^{\i\tau K} \pi_\cS(A)\otimes \one \ \Psi_\cS\otimes\Psi_\env\ket\\
 & = & \bra \pi(B\otimes \one) \Psi_\cS\otimes\Psi_\env, \e^{\i\tau L}\pi(A\otimes\one)\e^{-\i\tau L} \Psi_\cS\otimes\Psi_\env\ket\\
 & = & \bra \Psi_\cS\otimes\Psi_\env, \pi(B^*\otimes\one)\pi(\e^{\i \tau h}\, A\otimes\one \, \e^{-\i \tau h}) \Psi_\cS\otimes\Psi_\env\ket\\
 & = & \tr \left( \rho_\cS\otimes \rho_\env\times B^*\otimes\one \times \e^{\i \tau h}\, A\otimes\one \, \e^{-\i \tau h}\right)\\
 & = & \tr\left( \rho_\cS B^* \otimes \rho_\env \times \e^{\i \tau h}\, A\otimes\one \, \e^{-\i \tau h}\right)\\
 & = & \tr\left( \rho_\cS B^* \cL^*(A)\right)\\
 & = & \bra \Psi_\cS, \pi_\cS(B^* \cL^*(A))\Psi_\cS\ket\\
 & = & \bra \pi_\cS(B)\Psi_\cS, \pi_\cS(\cL^*(A))\Psi_\cS\ket.
\end{eqnarray*}
Since $\Psi_\cS$ is a cyclic vector this proves that, for any $A\in\fa_\cS$,
\begin{equation}\label{eq:gnsrdo}
M: \pi_\cS(A)\Psi_\cS \mapsto \pi_\cS(\cL^*(A))\Psi_\cS.
\end{equation}
Of course, the properties of a RDM $\cL$ immediately translate into properties of $M$
\begin{prop}\label{prop:Mproperties} The operator $M$ is a contraction on the Banach space $\cC=\{ A\Psi_\cS\ |\ A\in \fa_\cS\}$ endowed with the norm $||| \phi|||=|||A\Psi_\cS|||:= \|A\|.$ Moreover $1$ is an eigenvalue for $M$ with corresponding eigenvector $\Psi_\cS$. 
\end{prop}
These two properties correspond respectively to the contracting and trace preserving properties of $\cL$. Note also that when the small system has finite dimension the Banach space $\cC$ is simply $\cH_\cS$.


\subsection{Leaky RI systems}\label{ssec:leakyri}

In this section, we consider the situation where, besides the repeated interactions with the subsystems $\cE_k$, the system $\cS$ also interacts with another reservoir $\cR$ in a continuous way. Since the reservoir will consist in a infinitely extended free Fermi gas, it is more appropriate in this section to use the Liouvillian description of RI systems. 

The Liouvillian of the full system is thus now of the form
\begin{equation}\label{def:leakyliouv}
L=L_\cS+\sum_{n\geq 1} L_{\cE_n}+ \sum_{n\geq 1} \chi_n(t) V_{\cS\cE_n} + L_\cR +V_{\cS\cR},
\end{equation}
where $L_\cR$ is the generator of the free dynamics of the reservoir and $V_{\cS\cR}$ describes the interaction between the system $\cS$ and the reservoir $\cR$. Note that $\cR$ is not directly coupled to the subsystems $\cE_n$. We will also stick to the situation where the repeated interactions are identical.

The motivation to study such systems is twofold. First, it describes for example a ``One-Atom Maser'' in which one also takes into account some losses in the cavity, the latter being not completely isolated from the exterior world, e.g. from the laboratory \cite{FJM}. The assumption ``$\cR$ is not directly coupled to the subsystems $\cE_n$'' is physically reasonable. Indeed, again for the ``one-atom maser'' experiment, the idea is that the atoms are ejected from an oven one by one just before they interact with the cavity and moreover the atom-field interaction time $\tau$ is typically much smaller than the damping time due to the presence of the heat reservoir. Therefore, the atoms do not have enough time to feel the effects of the reservoir before and during their interaction with the field.

A second motivation is the study of non-equilibrium quantum systems. Suppose $\cS$ is brought into
contact with several reservoirs $\cR_i$, each of them being in a thermal equilibrium state but with different intensive thermodynamic parameters. The interaction between $\cS$ and the various reservoirs is most often ``continuous'', i.e. $\cS$ and the $\cR_i$ interact for all time (said differently, the generator of the interacting dynamics is time-independent). We have also considered in the previous section the case where the various reservoirs are all of the repeated interaction type (chosing e.g. reference states which are randomly distibuted with uniform distribution over a fixed set $\rho_1,\ldots,\rho_K$). In the system considered in this section we have a situation with two reservoirs of different nature: one is described by a RI system and the other one interacts with $\cS$ in a continuous way, and we want to understand the relative effects of these two reservoirs.

In a sense, one can consider this entire system as a RI system but where $\cS$ has been replaced by $\cS+\cR$, i.e the ``small'' system becomes large as well. The general approach to RI systems, as described in Section \ref{ssec:riliouv}, can therefore be used. However, the reduced dynamics operator $M$ now acts on the space $\cH_\cS\otimes\cH_\cR$ and, as we shall see, its spectral properties are of course much more complicated. The results presented in this section come from \cite{BJM4}.

\subsubsection{The additional reservoir}

The reservoir $\cR$ is a thermal reservoir of free Fermi particles at a temperature $T_\cR>0$, in the thermodynamic limit. Its description was originally given in the work by Araki and Wyss \cite{AW} (see also \cite{JP1,BJM4}).

The Hilbert space is the anti-symmetric Fock space $\cB(\cH_\cR)=\Gamma_-({\fh})$ 
where $\mathfrak G$ is an `auxiliary space' (typically an angular part like $L^2(S^2)$). In this representation, the one-particle Hamiltonian $h$ is the operator of multiplication by the radial variable (extended to negative values ) $s\in\mathbb R$ of $\fh$. The Liouville operator is the second quantization of $h$,
\begin{equation*}\label{def:reservoirliouv}
L_\cR = \d\Gamma(h):= \bigoplus_{n\geq 0} \sum_{j=1}^n h^{(j)},
\end{equation*}
where $h^{(j)}$ is understood to act as $h$ on the $j$-th factor of $\wedge^n\,\fh$ and trivially on the other ones. The von Neumann algebra $\fm_\cR$ is the subalgebra of ${\cal B}(\cH_\cR)$ generated by the thermal fermionic field operators (at inverse temperature $\beta_\cR$), represented on $\cH_\cR$ by
\begin{equation*}\label{def:thermalfield}
\varphi(g_{\beta_\cR}) = \frac{1}{\sqrt 2}\big[ a^*(g_{\beta_\cR}) +a(g_{\beta_\cR})\big].
\end{equation*}
Here, for $g\in L^2({\mathbb R}_+,{\mathfrak G}),$ we define $g_\beta\in \fh$ by
\begin{equation*}
g_{\beta_\cR}(s) = \sqrt{\frac{1}{\e^{-\beta_\cR s}+1}} \left\{
\begin{array}{ll} g(s) & \mbox{if $s\geq 0$} \\ \overline{g}(-s) & \mbox{if $s<0$.} \end{array}
\right.
\end{equation*}
Finally, we choose the reference state to be the thermal equilibrium state, represented by the vacuum vector of $\cH_\cR$,
\begin{equation*}
\Psi_\cR = \Omega.
\end{equation*}

\begin{ex}\label{ex:fermireservoir} Consider a bath of non-interacting and non-relativistic fermions at inverse temperature $\beta_\cR$. The one particle space is $\fh_\cR=L^2(\R^3, \d^3 k)$ and the one-particle energy operator $h_\cR$ is the multiplication operator by $|k|^2$. The Araki-Wyss representation of this free Fermi gas is then, see Example \ref{ex:gns-fermi}, $\tilde{\cH}_\cR= \Gamma_-(L^2(\R^3, \d^3 k)) \otimes \Gamma_-(L^2(\R^3, \d^3 k))$, $\tilde{\fm}_\cR = \pi_{\beta_\cR}\left( \mathfrak A\right)''$ where $\pi_\beta$ is defined in (\ref{def:pibeta}), $\tilde{\Psi}_\cR=\Omega \otimes \Omega$ and the Liouvillean is $\tilde{L}_\cR=\d\Gamma(h_\cR)\otimes\one -\one\otimes \d\Gamma(h_\cR)$.

We now show how to get a description of the Fermi gas of the above form using the Jak\v si\'c-Pillet gluing method \cite{JP0}. We consider the isomorphism between $L^2(\R^3,\d^3k)$ and $L^2(\R^+\times S^2, \frac{\sqrt{r}}{2} \d r \d\sigma)\simeq L^2(\R^+,\frac{\sqrt{r}}{2} \d r; \frak G)$, where ${\frak G}=L^2(S^2,\d\sigma)$, so that the operator $h_\cR$ becomes multiplication by $r\in \R^+$. The Hilbert space $\tilde{\cH}_\cR$ is thus isomorphic to
\begin{equation}\label{eq:awspace2}
\Gamma_-\left( L^2(\R^+,\frac{\sqrt{r}}{2} \d r; \frak G)\right)\otimes \Gamma_-\left( L^2(\R^+,\frac{\sqrt{r}}{2} \d r; \frak G)\right).
\end{equation}
Next, we make use of the maps $a^\#(f)\otimes \one \mapsto a^\#(f\oplus 0)$ and $(-1)^N \otimes a^\#(f)\mapsto a^\#(0\oplus f)$ to define an isometric isomorphism between (\ref{eq:awspace2}) and
\begin{equation*}\label{eq:awspace3}
\Gamma_-\left( L^2(\R^+,\frac{\sqrt{r}}{2} \d r; {\frak G}) \oplus L^2(\R^+,\frac{\sqrt{r}}{2} \d r; \frak G) \right).
\end{equation*}
A last isometric isomorphism between the above Hilbert space and $\cH_\cR:= \Gamma_-\left( L^2(\R,\d s; {\frak G})\right)$
is induced by the following isomorphism between the one-particle spaces $L^2(\R^+,\frac{\sqrt{r}}{2} \d r; {\frak G}) \oplus L^2(\R^+,\frac{\sqrt{r}}{2} \d r; \frak G)$ and $L^2(\R,\d s; {\frak G})=:\fh$
\begin{equation*}
f\oplus g \mapsto h, \ {\rm where} \ h(s)=\frac{|s|^{1/4}}{\sqrt{2}}\left\{ \begin{array}{ll} f(s) & {\rm if} \ s\geq 0,\\ g(-s) & {\rm if} s<0.  \end{array}\right.
\end{equation*}
Using these isomorphisms, one indeed gets the desired description of the Fermi gas.
\end{ex}


\subsubsection{Translation analyticity}

As already mentioned, the reduced dynamics operator $M$ is now defined as an operator on the larger space $\cH_\cS\otimes\cH_\cR$ and will have more complicated spectral properties. To understand why, let's switch off the interactions. Then, clearly $M=\e^{\i\tau L_\cS}\otimes\e^{\i\tau L_\cR}$ which, besides some eigenvalues, has continuous spectrum equal to the whole circle $S^1$. When turning on the interaction, this continuous spectrum survives. In order to seperate it from the eigenvalues, we use analytic spectral deformation methods, see e.g. \cite{BFS,JP2,MMS,RS4}, and will resort to perturbation theory. For that purpose, we therefore add coupling constants in the interaction, i.e. (\ref{def:leakyliouv}) becomes
\begin{equation*}\label{def:leakyliouv2}
L=L_\cS+ L_\cR +\sum_{n\geq 1} L_{\cE_n}+\lambda_{\cS\cR} V_{\cS\cR} +\lambda_{\cS\cE}\sum_{n\geq 1} \chi_n(t) V_{\cS\cE_n},
\end{equation*}
and the perturbation will be in term of the coupling constant $\lambda:=(\lambda_{\cS\cR},\lambda_{\cS\cE})\in\R^2$. We will also write $V(\lambda)=\lambda_{\cS\cR} V_{\cS\cR}+\lambda_{\cS\cE} V_{\cS\cE}$ and denote by $M(\lambda)$ the reduced dynamics operator.

The reduction process described in Section \ref{ssec:riliouv} makes use of another generator of the interacting dynamics than the Liouvillian, the so-called C-Liouville operator. Its explicit form involves the modular data $(J,\Delta)$ of the pair $(\fm_\cS\otimes\fm_\cR\otimes\fm_\cE,\Psi_\cS\otimes\Psi_\cR\otimes\Psi_\cE)$, see e.g. \cite{AJP,BR}. More precisely, it can be written as
\begin{equation}\label{def:cliouv}
K=L_\cS+L_\cR +L_\cE+V(\lambda)-J\Delta^{1/2}V(\lambda)\Delta^{-1/2}J.
\end{equation}
In order to make it simple, we shall assume that
\bigskip
\begin{itemize}
 \item[\!\!\!\!\! {\bf (H')}] $\quad \Delta^{1/2}V(\lambda)\Delta^{-1/2}\in \fm_\cS\otimes\fm_\cR\otimes\fm_\cE$.
\end{itemize}
\bigskip
This ensures that $K$ generates a strongly continuous group $\e^{\i tK}$ of bounded operators on $\cH_\cS\otimes\cH_\cR\otimes\cH_\cE$ (this assumption can certainly be relaxed, see \cite{JP2}).

Moreover, since we will be using analytic spectral deformation methods on the factor $\cH_\cR$ of $\cH$, we need to make a regularity assumption on the interaction. Let $\R\ni \theta\mapsto T(\theta)\in \cB(\cH_\cR)$ be the unitary group defined by
\begin{equation*}\label{def:translation}
T(\theta)=\Gamma(e^{-\theta \partial_s}) \ \ \mbox{on} \ \ \Gamma_-(L^2(\R,{\mathfrak G})),
\end{equation*}
where for any $f\in L^2(\R,{\mathfrak G})$,
$$
(e^{-\theta \partial_s}f)(s)=f(s-\theta),
$$
i.e. we use the generator of translation. In the following, we will abuse notation and (for simplicity) also write $T(\theta)$ for $\one_\cS\otimes T(\theta)\otimes \one_\cE$ and $\one_\cS\otimes T(\theta)\otimes \one_\env$. Note that $T(\theta)$ commutes with all observables acting trivially on $\cH_\cR$, in particular with $P_{\cal SR}=\one_{\cS}\otimes\one_\cR\otimes |\Psi_\env\ket\bra\Psi_\env|$. Also, we have $T(\theta)\Psi_\cR=\Psi_\cR$ for all $\theta$. The spectral deformation technique relies on making the parameter $\theta$ complex.
\bigskip
\begin{itemize}
\item[\!\!\!\!\!{\bf (A)}] The coupling operator $W_{\cS\cR}:= V_{\cS\cR}-J\Delta^{1/2} V_{\cS\cR}\Delta^{-1/2}J$ is translation analytic in a strip $\kappa_{\theta_0}=\{ z\ :\ 0<\Im z<\theta_0\}$ and strongly continuous on the real axis. More precisely, there is a $\theta_0>0$ such that the map
$$
\R\ni\theta \mapsto T^{-1}(\theta) W_{\cS\cR}T(\theta)= W_{\cS\cR}(\theta)\in \fm_\cS\otimes\fm_\cR,
$$
admits an analytic continuation into $\theta\in\kappa_{\theta_0}$ which is strongly continuous as $\Im \theta\downarrow 0$, and which satisfies
$$
\sup_{0\leq \Im\theta<\theta_0} \| W_{\cS\cR}(\theta)\| < \infty.
$$
\end{itemize}
\bigskip
The reduced dynamics operator will also be deformed as
$$
M_\theta(\lambda):=T(\theta)^{-1}M(\lambda)T(\theta).
$$
The ergodicity assumption {\bf (E)} will now be written for this deformed operator $M_\theta(\lambda)$. More precisely, we will assume that the following Fermi Golden Rule condition holds
\bigskip
\begin{itemize}
\item[\!\!\!\!\!{\bf (FGR)}] There is a $\theta_1\in\kappa_{\theta_0}$ and a $\lambda_0>0$ (depending on $\theta_1$ in general) such that, for all $\lambda$ with $0<|\lambda|< \lambda_0$, $M_{\theta_1}(\lambda))$ satisfies {\bf (E)}.
\end{itemize}
\bigskip

Of course, an important issue in the analysis of concrete models is the verification of this {\it Fermi Golden Rule} assumption {\bf (FGR)}. Let us denote the eigenvalues of $h_\cS$ by $E_1, \cdots, E_d.$ When $\theta\in\kappa_{\theta_0}$ and $\lambda_{\cS\cR}=\lambda_{\cS\cE}=0$. Then
$$
M_{\theta}(0)=\e^{\i\tau(L_\cS+L_\cR+\theta N)}=\e^{\i\tau L_\cS}\otimes\e^{\i\tau L_\cR}\e^{\i\tau\theta N},
$$
where $N$ denotes the number operator on $\Gamma_-(\fh_\cR)$, and hence
\begin{equation*}\label{smo}
\sp({M_{\theta}(0)})=\{ \e^{\i\tau(E_j-E_k)}\}_{j,k\in \{1,\cdots, d\}}\cup\{\e^{\i l}\e^{-\tau j\Im\theta}, \ l\in\R \}_{j\in \N^*}.
\end{equation*}
The effect of the analytic translation is to push the continuous spectrum  of $M_{\theta}(0)$ onto circles with radii $\e^{-\tau j\Im\theta}$, $j=1,2,\ldots$, centered at the origin. Hence the discrete spectrum of $M_\theta(0)$, lying on the unit circle, is separated from the continuous spectrum by a distance
$1-e^{-\tau\Im \theta}$. Analytic perturbation theory in the parameters $\lambda_{\cS\cR}$, $\lambda_{\cS\cE}$ guarantees that the discrete and continuous spectra stay separated for small coupling. 
As a consequence a verification of {\bf (FGR)} for concrete models, like the one of Example \ref{ex:toyleaky}, is done via (perturbative) analysis \emph{only of the discrete eigenvalues} of $M_\theta(\lambda)$.


\subsubsection{Asymptotic state}

\begin{defn}\label{def:analyticobs} An observable $O$ is called analytic if the map $\theta\to T(\theta)^{-1} O \Psi_0$, where $\Psi_0=\Psi_\cS\otimes\Psi_\cR\otimes\Psi_\env$, has an analytic extension to $\theta\in\kappa_{\theta_0}$ which is continuous on the real axis. 
\end{defn}
\noindent Note that for an obervable $O=O_\cS\otimes O_\cR\otimes O_\env$, since $T$ acts on $\cH_\cR$ only, this is equivalent to $T(\theta)^{-1}O_{\cR}\Psi_0$ having such an extension. In particular, any observable on the small system is analytic.

\begin{thm}\label{thm:leakystate} Assume that assumptions {\bf (H')}, {\bf (A)} and {\bf (FGR)} are satisfied. Then there is a $\lambda_0>0$ s.t. if $0<|\lambda|<\lambda_0$, the following holds. There exists a state $\rho_{+,\lambda}$ on $\fm_\cS\otimes\fm_\cR$ such that for any normal initial state $\varrho$ on $\fm$, and any analytic observable $O_{\cS\cR}\in\fm_\cS\otimes\fm_\cR$,
\begin{equation*}\label{eq:leakystate}
\lim_{n\to\infty} \varrho \Big( \alpha^{n\tau}_\RI\big(O_{\cS\cR}\big)\Big) =
\rho_{+,\lambda}\big( O_{\cS\cR}\big):= \bra\psi^*_{\theta_1}(\lambda)| T(\theta_1)^{-1} A_{\cS\cR} \Psi_\cS\otimes\Psi_\cR\ket,
\end{equation*}
where $\psi_{\theta_1}^*(\lambda)$ is the unique invariant vector of the adjoint operator $[M_{\theta_1}(\lambda)]^*$, normalized as $\langle \psi_{\theta_1}^*(\lambda) | \Psi_\cS\otimes\Psi_\cR\rangle=1$.
\end{thm}

\begin{ex}\label{ex:toyleaky} We consider the leaky version of the system described in Example \ref{ex:toy-description-rdm}, i.e. the subsystems $\cS$ and $\cE_k$'s are 2-level systems and the reservoir is chosen as in Example \ref{ex:fermireservoir}. The interaction $V_{\cS\cR}$ between the system $\cS$ and the reservoir is given by
\begin{equation*}
V_{\cS\cR}=\lambda_1(\sigma_x\otimes \one_{\C^2}) \otimes \varphi(f_{\beta_\cR}) \in \fm_\cS\otimes \fm_\cR,
\end{equation*}
where $\sigma_x$ is the first Pauli matrix, $f\in L^2(\R^3,\d^3k)$ is a form factor, and $f_{\beta_\cR}\in \fh=L^2(\R,\d s; L^2(S^2,\d \sigma))$ is related to $f\in L^2(\R^3,\d^3k)$ as follows
\begin{equation}\label{eq:awformfactor}
\left(f_{\beta_\cR}(s)\right)(\sigma)=\frac{1}{\sqrt{2}}\frac{|s|^{1/4}}{\sqrt{1+\e^{-\beta_\cR s}}}\left\{  \begin{array}{ll} f(\sqrt{s}\;\sigma) & {\rm if} \ s\geq 0, \\ \bar{f}(\sqrt{-s}\;\sigma)& {\rm if} \ s< 0. \end{array}\right.
\end{equation}
We will denote by $\lambda_2$ instead of $\lambda$ the coupling constant in the interaction term between $\cS$ and $\cE$, see Example \ref{ex:toy-description-rdm}.

In order to satisfy the analyticity assumption {\bf (A)} we need some assumption on the form factor $f$.
\begin{itemize}
\item[{\bf (A')}] Let $f_0$ be defined by \fer{eq:awformfactor}, with $\beta_\cR=0$. There is a $\delta>0$ s.t. $\e^{-\beta_\cR s/2} f_0(s)\in H^2(\delta)$, the Hardy class of analytic functions $h: \{z\in\C,\, |\Im(z)|<\delta\}\to {\mathfrak G}$ which satisfy
$$
\|h\|_{H^2(\delta)}:= \sup_{|\theta|<\delta} \ \int_\R \|h(s+i\theta)\|_{\mathfrak G}^2 \d s < \infty.
$$
\end{itemize}
If $f$ satisfies {\bf (A')},  $\|f(\sqrt{E})\|_{\mathfrak G}^2:=\int_{S^2} |f(\sqrt{E}\; \sigma)|^2\d\sigma\neq 0$ and $\tau(E_0-E)\notin 2\pi \Z^*$, then Theorem \ref{thm:leakystate} holds and the asymptotic state $\rho_{+,\lambda}$ is given by
$$
\rho_{+,\lambda}= \left( \gamma \rho_{\beta_\cR,\cS}+ (1-\gamma) \rho_{\beta_\cE^*,\cS} \right)\otimes \rho_{\beta_\cR,\cR} +O(\lambda),
$$
where $\rho_{\beta,\#}$ is the Gibbs state of $\#$, $\#=\cS,\cR$, at inverse temperature $\beta_\cE^*=\frac{E_0}{E}\beta_\cE$, and where $\gamma$ is given by
\begin{equation}\label{eq:gammas}
\gamma =  \frac{\lambda_1^2\gamma_{\rm th}}{\lambda_1^2\gamma_{\rm th} + \lambda_2^2\gamma_{\rm ri}}, \quad 
\gamma_{\rm th}= \frac{\pi}{2} \sqrt{E}\|f(\sqrt{E})\|_{\mathfrak G}^2, \quad \gamma_{\rm ri}=\frac{\tau}{8}\sinc^2\left( \frac{\tau(E_0-E)}{2}\right).
\end{equation}
\end{ex}

\begin{open} Consider the leaky version of the One-atom maser model of Section \ref{ssec:qedcavity}. Besides the atomic beam, the cavity is coupled to an extra reservoir which traduces the fact that the cavity is not perfectly isolated. This is particularly important if the atoms have a higher probability to be in their excited state that in their ground state. Can one prove convergence to some stationnary state? One particularly relevant question is what is the statistics of the photon number in that stationnary state. 
\end{open}

\begin{open} A similar model of a cavity interacting with an atomic beam with or without leaks has recently been investigated in \cite{NVZ} but with an atom-field interaction of the form $\lambda (a+a^*)\otimes b^*b$. This interaction has the advantage that it leaves the atom state invariant and makes the mathematical analysis more tractable. Moreover, the leak is described in the Kossakowski-Lindblad extension of the Hamiltonian dynamics, i.e. by adding a dissipative part to the field hamiltonian, see e.g. \cite{AJP}. Can one analyze a purely hamiltonian version of this model adding an extra reservoir to describe the leak.
\end{open}

\begin{open} In the physics litterature on the one-atom maser, it is argued that once the leaks are taken into account the qualitative aspects of the photon number statistics is the same if one considers that the interaction times of the various atoms with the cavity is constant or random \cite{FJM}. Study the leaky version of the One-atom maser + random interaction time and compare the photon statistics to the one obtained with fixed interaction time. 
\end{open}

\begin{open} As a first step towards the understanding of the leaky cavity with random interaction times, analyze RI systems with leaks and randomness for finite dimensional small systems (under appropriate general assumptions).\\
\end{open}


\subsubsection{Thermodynamics}

\noindent \emph{External work}

When we consider leaky RI systems, we need to turn to the the Liouvillian description. It is then natural to define the work observable as
\begin{equation}\label{def:worktimen}
W(n):= \pi(w(n))= U(t_n,0)^*\left(V_{n+1}-V_n\right)U(t_n,0)= \alpha^{t_n}_\RI(V_{n+1}-V_n)
\end{equation}
If $\varrho$ is the initial state of the (entire) system, the power delivered to the system is therefore (if it exists)
\begin{equation}\label{eq:totalwork}
\Delta W=\lim_{n\to\infty} \frac{1}{t_n}\sum_{k=1}^n \varrho(W(n)).
\end{equation}

\bigskip

\noindent \emph{Entropy production}

We first need to generalize (\ref{eq:epformulahamilt}) to the case where the system is not described via the hamiltonian formalism.

If $\varrho$ and $\varrho_0$ are two normal states\footnote{A state $\varrho$ on a von Neumann algebra $\fm$ is normal if it is $\sigma$-weakly continuous} on $\fm$, the relative entropy of Araki of the state $\varrho$ with respect to $\varrho_0$ is denoted by ${\rm Ent}(\varrho|\varrho_0)$\footnote{For finite systems, and if $\varrho$ and $\varrho_0$ are given by density matrices $\rho$ and $\rho_0$ respectively, then ${\rm Ent}(\varrho|\varrho_0)=-\tr (\rho \log \rho -\rho\log \rho_0)$.}. We here adopt the same convention as in \cite{BR,JP2}, so that ${\rm Ent}(\varrho|\varrho_0)\leq 0$. The reference state $\varrho_0$ will naturally be the vector state on $\fm$ determined by the vector $\Psi_0=\Psi_\cS\otimes\Psi_\env$.

The analysis of the entropy production relies on the so-called {\em entropy production formula} \cite{JP3} (see \ref{eq:epformula}) which we recall here for the sake of completeness. Consider a quantum dynamical system $(\fa,\alpha^t,\omega)$. We moreover assume that $\omega$ is a $(-1,\sigma_\omega^t)$--KMS state for some $C^*$- dynamics $\sigma_\omega^t$ with generator $\delta_\omega$. Let $V\in {\rm Dom}(\delta_\omega)$ and consider the perturbed dynamics $\alpha_V^t$ defined in the natural way:
$$
\alpha_V^t(A):=\alpha^t(A)+\sum_{n\geq 1}\i^n \int_0^t\d t_1\int_0^{t_1}\d t_2\cdots\int_0^{t_{n-1}}\d t_n [\alpha^{t_n}(V),[\cdots[\alpha^{t_1}(V),A]\cdots],
$$
(if $\delta_\alpha$ is the generator of $\alpha^t$, the one of $\alpha_V^t$ is $\delta_\alpha+\i[V,\cdot]$). Then for any state $\eta$:
\begin{equation}\label{eq:epformula}
{\rm Ent}(\eta\circ\alpha^t_V| \omega)-{\rm Ent} (\eta|\omega)=-\int_0^t \eta\circ \alpha_V^s(\delta_\omega(V))\d s.
\end{equation}
In the particular case of a composite system ($\fa=\otimes_k \, \fa_k$ and $\alpha^t=\otimes_k\, \alpha^t_k$) where the reference state $\omega$ is of the form $\omega=\otimes_k\, \omega_k$ and where the $\omega_k$ are $(\beta_k,\alpha_k^t)$-KMS states, one can take $\sigma^t=\otimes_k \, \alpha_k^{-\beta_k t}$. In the GNS-representation, if the $\alpha_k^t$ are implemented by Liouvillians $L_k$ then $\sigma^t$ is generated by $\ds L=-\sum_k \beta_k L_k$, so that $\delta_\omega(V)$ becomes $\ds -\i\sum_k \beta_k[L_k,V]$.

\medskip

In the RI setting (\ref{eq:epformula}) translates into the following formula which is the exact generalization of (\ref{eq:epformulahamilt2}) with $\ds \rho_\cS=\frac{\e^{-\beta_\cS h_\cS}}{\tr (\e^{-\beta_\cS h_\cS})}$ :
\begin{eqnarray}
\lefteqn{{\rm Ent}(\varrho\circ\alpha^{t_n}_\RI| \varrho_0)-{\rm Ent} (\varrho|\varrho_0)}\nonumber\\
 & = & \sum_{k=1}^n \left[\beta_{\cE_k} \varrho \left( \alpha^{t_k}_\RI(V_k)-\alpha^{t_{k-1}}_\RI(V_k)\right)+(\beta_{\cE_k}-\beta_\cS)\varrho \left( \alpha^{t_k}_\RI(L_\cS)-\alpha^{t_{k-1}}_\RI(L_\cS)\right)\right].\nonumber
\end{eqnarray}
The latter can also be written as
\begin{eqnarray}\label{eq:epformulari}
\lefteqn{{\rm Ent}(\varrho\circ\alpha^{t_n}_\RI| \varrho_0)-{\rm Ent} (\varrho|\varrho_0)} \\
 & = & - \sum_{k=1}^n \beta_{\cE_k} \varrho \left(W(k)\right) + \sum_{k=1}^n \beta_{\cE_k} \varrho\left(\alpha^{t_k}_\RI(V_{k+1})- \alpha^{t_{k-1}}_\RI(V_k) \right)\nonumber \\
 & & +\sum_{k=1}^n (\beta_{\cE_k}-\beta_\cS) \varrho\left( \alpha^{t_k}_\RI(L_\cS)-\alpha^{t_{k-1}}_\RI(L_\cS)\right),\nonumber
\end{eqnarray}

Note that $L_\cS$ is a priori not an observable ($L_\cS\notin \fm$) and neither is $\alpha_\RI^t(L_\cS)$. However the differences $\alpha^{t_k}_\RI(L_\cS)-\alpha^{t_{k-1}}_\RI(L_\cS)$ are observables. This follows from the fact that $\e^{\i \tau_k L_k}L_\cS \,\e^{-\i \tau_k L_k}-L_\cS\in \fm_\cS\otimes \fm_{\cE_k}$, which in turn is proven by noting that 
$$
\e^{\i \tau_kL_k}L_\cS \,\e^{-\i \tau_kL_k}-L_\cS = \int_0^{\tau_k} \e^{\i t L_k}[\i L_k,L_\cS] \e^{-\i t L_k} \d t = \int_0^{\tau_k} \e^{\i t L_k}[\i V_k,L_\cS] \e^{-\i t L_k} \d t,
$$
where $[\i V_k,L_\cS]=-\frac{\d}{\d t} \e^{\i t L_\cS}V_k\e^{-\i t L_\cS}|_{t=0}\in \fm_\cS\otimes \fm_{\cE_k}$.

\bigskip

\noindent \emph{Fluxes}

Besides the external work and the entropy production, the presence of two environments/reservoirs induces other quantities of interest, namely the heat fluxes. If the system were described in the Hamiltonian formalism, one would define the variation of energy in the environment $\env=\cE_1+\cE_2+\cdots$ and the reservoir $\cR$ between time $n\tau$ and $(n+1)\tau$ as
\begin{eqnarray*}
\delta e^{\cR}(n) & := & u((n+1)\tau,0)^* h_\cR u((n+1)\tau,0)-u(n\tau,0)^* h_\cR u(n\tau),\label{def:energyr}\\
\delta e^\env(n) & := & u((n+1)\tau,0)^* h_{\cE_{n+1}} u((n+1)\tau,0)-u(n\tau,0)^* h_{\cE_n} u(n\tau).\label{def:energychain}
\end{eqnarray*}
For the energy variation in the environment, recall that between time $n\tau$ and $(n+1)\tau$ only the $(n+1)$-th subsystem interacts with $\cS$ and can thus exchange energy. In the Liouvillian description one thus defines $\delta E^\#(n)=\pi(\delta e^\#(n))$ for $\#=\cR,\env$. As a consequence of Theorem \ref{thm:leakystate}, we have the following
\begin{prop}\label{prop:leakythermo} If {\bf (H')}, {\bf (A)} and {\bf (FGR)} are satisfied and if the commutators $[V_{\cS\cR},L_\cS]$ and $[V_{\cS\cR},L_\cR]$ define analytic observables, then for any normal initial state $\varrho$
$$
\Delta E^\#:=\lim_{n\to\infty}\frac{1}{n\tau}\sum_{k=1}^n \varrho (\Delta E^\#(n))= \frac{1}{\tau}\rho_{+,\lambda}(P_{\cS\cR}j^\#P_{\cS\cR}), \qquad \#=\cR,\env,
$$
where
\begin{equation*}
j^\env = \i\int_0^\tau \alpha^t_\RI ([\lambda_{\cS\cE}V_{\cS\cE}, L_\cE])\,\d t, \qquad 
j^{\cR} = \i\int_0^\tau \alpha^t_\RI ([\lambda_{\cS\cR}V_{\cS\cR}, L_\cR])\,\d t. \label{def:energyobs}
\end{equation*}
The external work is
$$
\Delta W=\frac{1}{\tau} \rho_{+,\lambda} \Big(P_{\cS\cR}V(\lambda)P_{\cS\cR}- P_{\cS\cR}\alpha^\tau_\RI(V(\lambda))P_{\cS\cR}\Big),
$$
and we have
$$
\Delta W=\Delta E^\cR+\Delta E^\env.
$$
If moreover, the reference states are $KMS$-states at inverse temperatures $\beta_\cS$, $\beta_\cR$ and $\beta_\cE$, the entropy production $\Delta S$ exists and satisfies
$$
\Delta S=\beta_\cE \Delta E^\env+\beta_\cR \Delta E^\cR.
$$
\end{prop}
As expected, the energy gain in the system (due to the external work) is shared between the reservoir $\cR$ and the environment. However, contrary to the ideal case, the external work may be positive or negative (one can pump energy from the reservoir $\cR$).

\begin{ex}\label{ex:toyleakytthermo} Consider the system described in Example \ref{ex:toyleaky}. Under the same assumptions, i.e. if $f$ satisfies {\bf (A')}, $\|f(\sqrt{E})\|_{\mathfrak G}^2:=\int_{S^2} |f(\sqrt{E}\; \sigma)|^2\d\sigma\neq 0$ and $\tau(E_0-E)\notin 2\pi \Z^*$, then Proposition \ref{prop:leakythermo} holds. Moreover
\begin{eqnarray*}
\Delta E^\env & = & \kappa E_0 \left( \e^{-\beta_\cR }- \e^{-\beta_\cE^* E} \right)+O(\lambda^3),\\
\Delta E^\cR & = & \kappa E \left( \e^{-\beta_\cE^* E}- \e^{-\beta_\cR E} \right)+O(\lambda^3),\\
\Delta W & = & \kappa (E_0-E) \left( \e^{-\beta_\cR E}- \e^{-\beta_\cE^* E} \right)+O(\lambda^3),\\
\Delta S & = & \kappa (\beta_\cE^* E-\beta_\cR E) \left( \e^{-\beta_\cR E}- \e^{-\beta_\cE^* E} \right)+O(\lambda^3),
\end{eqnarray*}
where $\ds \kappa=\frac{1}{1+\e^{-\beta_\cR E}}\times\frac{1}{1+\e^{-\beta_\cE^* E}}\times \frac{\lambda_1^2 \gamma_{\rm th} \times \lambda_2^2 \gamma_{\rm ri}}{\lambda_1^2 \gamma_{\rm th}+\lambda_2^2 \gamma_{\rm ri}}.$
\end{ex}

\begin{rem} 1. The constant $\kappa$ is positive and of order $\lambda^2$. Moreover it is zero if at least one of the two coupling constants vanishes (we are then in an equilibrium situation and there is no energy flux neither entropy production).

2. The energy flux $\Delta E^\env$ is positive (energy flows {\it into} the environment) if and only if the reservoir temperature $T_\cR=\beta_\cR^{-1}$ is greater than the renormalized temperature $T_\cE^*=(\beta_\cE^*)^{-1}$ of the environment, i.e. iff the reservoir is ``hotter''. A similar statement holds for the energy flux $\Delta E^\cR$. Note that, as for the ideal case, it is not the temperature of the environment which plays a role but its renormalized value.

3. When both the reservoir and the environment are coupled to the system $\cS$, i.e. $\lambda_1 \lambda_2\neq 0$, the entropy production vanishes (at the leading order) if and only if the two temperatures $T_\cR$ and $T_\cE^*$ are equal, i.e. if and only if we are in an equilibrium situation. Once again, it is not the initial temperature of the chain which plays a role but the renormalized one.

4. As mentioned above, the external work can be either positive or negative depending on the parameters of the model.
\end{rem}

\newcommand{\bbbone}{\one}
\renewcommand{\cP}{\cE}

\section{RI systems and the quantum measurement process}

\subsection{Introduction and main results}

\newcommand{\M}{{\cal M}}

Many experiments in physics are based on scattering mechanisms. A system of interest (the scatterer) is subject to a beam of scattering probes, interacting one by one  with the scatterer. Before the interaction, the probes are prepared in a desired state and after the interaction they carry some information of the scatterer. A concrete physical setup is given by atoms (probes) shot through a cavity containing an electromagnetic field, the modes which interact with the atoms forming the scatterer, see also Section \ref{ssec:qedcavity}. We can describe this situation with a repeated interaction model: the interaction of each probe with the system is governed by a fixed interaction time $\tau>0$ and a fixed interaction operator $V$. To obtain a `readout' of the probes, we perform a quantum measurement on the outcoming probes. The result of the measurement of the $n$-th probe is a random variable, denoted $X_n$, and the stochastic process $\{X_n\}_{n\geq 1}$ is the measurement history. We consider the incident probes to be independent (unentangled) and in a stationary state with respect to their own dynamics. However, due to the entanglement of the probes with the scatterer during their interaction, the $X_n$ are not independent random variables. We study systems with only finitely many degrees of freedom involved in the scattering process. This means that the Hilbert spaces of pure states both of the system and each probe is finite-dimensional. The measurement of a probe is a von Neumann, or projective, measurement associated to a self-adjoint probe observable $\M$. The eigenvalues $m$ of $\M$ are the possible measurement outcomes. The random variables $X_n$ have finite range (${\rm spec}(\M)$).

In this section, we analyze the asymptotic properties of the measurement process. We show that it is generically not convergent and we analyze the fluctuations of the measurement history, provoked by the scattering process, by analyzing the measurement frequencies. The results presented in this section are found in \cite{Merkli-Penney}.

It is assumed that the interaction allows for {\em energy exchanges} between the probes and the scatterer. More precisely, we suppose that condition {\bf (E)} of Section \ref{ssec:idealri} is satisfied.

Note that according to Theorem \ref{thm:idealsmall}, the approach to the final state is exponentially quick. An important consequence of assumption {\bf (E)} is that the scatterer loses its memory. Suppose that we make a measurement at time $l$ and a second one at time $m>l$. During the time span between the two measurements, the scatterer follows the process of relaxation to its asymptotic state. It therefore erases correlations between the two measurements, and this more and more as $m-l$ increases. It is thus plausible that the outcomes $X_l$ and $X_m$ are becoming `more and more independent' with growing time-distance $m-l$. Let $P$ be the probability measure associated with the process $\{X_n\}_{n\geq 1}$. A measure for the independence of $X_l$ and $X_m$ is 
$$
|P(X_l\in A,X_m\in B) - P(X_l\in A)P(X_m\in B)|,
$$
for subsets $A,B\subset {\rm spec}(\M)$. The smaller this number is, the `more independent' the random variables $X_l, X_m$ are.

We make these ideas precise in the next result. Let $\sigma(X_r,\ldots,X_s)$ be the {\it sigma-algebra generated by the random variables $X_r,\ldots,X_s$}, $1\leq r\leq s\leq \infty$.
\begin{thm}[Decay of correlations]
\label{corrdeclemma'}
Suppose that Condition {\rm \bf (E)} holds. Then there are constants $c$, $\gamma>0$, such that for $1\leq k\leq l < m\leq n<\infty$, $A\in\sigma(X_k,\ldots X_l)$ and $B\in\sigma(X_m,\ldots,X_n)$, we have 
\begin{equation}
\left| P(A\cap B)-P(A)P(B)\right| \leq c P(A) \ \e^{-\gamma(m-l)}.
\label{010'}
\end{equation}
\end{thm}
As expected from the above discussion, the rate $\gamma$ in \fer{010'} is linked to the convergence rate of the dynamics without measurement (Theorem \ref{thm:idealsmall}).

The {\it tail sigma-algebra} is defined by ${\cal T}= \cap_{n\geq 1} \sigma(X_n, X_{n+1},\ldots)$. Decaying correlations imply the Kolmogorov zero-one law: {\em 
Assume the decay of correlations \fer{010'}. Then any tail event $A\in{\cal T}$ satisfies $P(A)=0$ or $P(A)=1$.}  In textbooks, the Kolmogorov zero-one law is usually presented for {\it independent} random variables \cite{Bmm}. However, an adaptation of the proof yields the result for random variables with decaying correlations. It is not necessary that the correlations decay exponentially quickly for this result to hold, all that is needed is that the left side of \fer{010'} tends to zero as $m-l$ tends to infinity, see \cite{Amm}. The tail sigma-algebra captures convergence properties. For instance, given any possible outcome $m\in{\rm spec}(\M)$, the set $\{\lim_nX_n=m\}$ is a tail event. According to the zero-one law, it has probability either zero or one. Generically, the probability of convergence is zero. This is due to the transmission of statistical uncertainty of the incoming probes to the outgoing ones. One can explain this mechanism as follows. 

Let $\omega_{\rm in}$ be the state of the incoming probes. Denote by $E_S$ the spectral projection of the measurement operator $\M$ associated to $S\subset{\rm spec}(\M)$ and write $E_m=E_{\{m\}}$ for $m\in{\rm spec}(\M)$. In absence of interaction (when $V=0$ or $\tau=0$), the $X_j$ are independent random variables. We show in Proposition \ref{propx2} that the dependence generated by the interaction with the scatterer is small for small interactions, {\it uniformly in time} $n\geq 0$, which means that
$$
P(X_n=m)=\omega_{\rm in}(E_m)+ O(\|V\|)
$$
and that 
$$
P(X_{n+1}=m, X_n=m)=P(X_{n+1}=m)P(X_n=m) +O(\|V\|).
$$ 
It follows that we have
\begin{eqnarray*}
P(X_{n+1}=X_n) &=& \sum_{m\in{\rm spec}(\M)} P(X_{n+1}=m, X_n=m)\\
&=& \sum_{m\in{\rm spec}(\M)}\omega_{\rm in}^2(E_m) +O(\|V\|).
\end{eqnarray*}
The numbers $\omega_{\rm in}(E_m)$ are probabilities. Therefore, $\sum_m\omega_{\rm in}^2(E_m)=1$ if and only if  $\omega_{\rm in}(E_{m_0})=1$ for a single $m_0$, while for all other $m$, $\omega_{\rm in}(E_m)=0$. This means that $P(X_{n+1}=X_n)<1$ for small $V$, whenever there are several $m$ with $\omega_{\rm in}(E_m)>0$. Together with the zero-one law, this implies that {\em $P(X_n {\rm \ converges})=0$ whenever the incoming state is not localized in a single subspace of $\M$ (and $V$ is small enough).} Note that if $m$ is a simple eigenvalue of $\M$ with associated eigenvector $\psi_m$, then $\omega_{\rm in}(E_m)=1$ is equivalent to $\omega_{\rm in}(\cdot)=\scalprod{\psi_m}{\cdot \,\psi_m}$. Statistical fluctuations in the incoming probes (mixture of states localized w.r.t. measurement values) thus get transferred to outcoming probes, even in the limit of large times. The following is a more general statement of this fact.
\begin{thm}
\label{propx1} Assume Condition {\em{\bf(E)}} holds. There is a constant $C$ s.t., for any $S\subset {\rm spec}(\M)$ with $\omega_{\rm in}(E_S)\neq 1$, if $\|V\|\leq C(1-\omega_{\rm in}(E_S))$, then 
$$
P(X_n\in S \rm{\ eventually})=0.
$$
\end{thm}
The result on non-convergence of $X_n$ explained before Theorem \ref{propx1} is a special case of Theorem \ref{propx1}, when $S=\{m\}$, $m\in{\rm spec}(\M)$. We mention that our analysis also gives a condition under which $P(X_n\in S \rm{\ eventually})=1$, see \cite{Merkli-Penney}.

\medskip
The process $X_n$ carries information about the scattering process, encoded in the relative occurrence of a particular measurement outcome. We define the {\it frequency} of $m\in{\rm spec}(\M)$ by 
\begin{equation*}
f_m = \lim_{n\rightarrow\infty} \frac 1n\big\{\mbox{number of $k\in\{1,\ldots,n\}$ s.t. $X_k=m$}\big\}.
\end{equation*}
$f_m$ is a random variable and the limit is in the almost everywhere sense. The following result analyzes the influence of the scattering process on the frequencies.

\begin{thm}[Frequencies]
\label{theorem2}
Assume that condition {\em{\bf (E)}} holds and denote the resulting asymptotic state of the scatterer without measurements by $\omega_+$. Then $f_m$ exists as an almost everywhere limit and is deterministic (not random), given by
$$
f_m = \omega_+\otimes\omega_{\rm in}(\e^{\i\tau H} E_m\e^{-\i\tau H}).
$$
Here, $H$ is the Hamiltonian of the interacting scatterer-probe system.  
\end{thm}

Suppose the interaction operator is of the form $\lambda V$, where $\lambda\in\mathbb R$ is a coupling constant, i.e., that $H=H_0+\lambda V$, where $H_0=H_\cS+H_\cP$ is the non-interacting Hamiltonian. Then $f_m=f_m(\lambda)$ is a holomorphic function in $\lambda$ at $\lambda=0$ (note that $\omega_+$ also depends on $\lambda$) and 
$$
f_m(\lambda) = \omega_{\rm in}(E_m) +\lambda\, \omega_\cS\otimes\omega_{\rm in}\left( \i \tau [V,\overline{\! E}_m(\tau)]\right)+O(\lambda^2),
$$ 
where the linear term in $\lambda$ is the {\it flux of the observable}
$$
\overline{\! E}_m(\tau) =\frac 1\tau\int_0^\tau \e^{\i sH_{\cP}}E_m \e^{-\i sH_{\cP}}\d s
$$
{\it in the initial scatterer-probe state} $\omega_\cS\otimes\omega_{\rm in}$. Here, $H_\cP$ is the Hamiltonian of a single probe. The quantity $\overline{\! E}_m(\tau)$ is the average of the operator $E_m$ over an interaction period $\tau$. By the invariance of $\omega_\cS\otimes\omega_{\rm in}$ under the free dynamics generated by $H_0$ we have, for any observable $A$,
$$
\left. \frac{\d}{\d t}\right|_{t=0} \omega_\cS\otimes\omega_{\rm in}(\e^{\i tH_0}A\e^{-\i tH_0}) = \omega_\cS\otimes\omega_{\rm in}(\i [H_0,A])=0.
$$
Therefore,
$$
\omega_\cS\otimes\omega_{\rm in}\left( \i \lambda\tau [V,\overline{\! E}_m(\tau)]\right)=\tau \,\omega_\cS\otimes\omega_{\rm in}\left( \i [H,\overline{\! E}_m(\tau)]\right),
$$
and we conclude that the linear term in $\lambda$ is given by
$$
f_m'(0) = \tau \frac{\d}{\d t}\Big|_{t=0}\omega_\cS\otimes\omega_{\rm in}\left( \e^{\i tH} \overline{\! E}_m(\tau)\e^{-\i tH}\right),
$$
which is ($\tau$ times) the change in the quantity $\overline{\! E}_m(\tau)$ per unit time, i.e., its flux.

\bigskip

The next result examines the average of the measurement process. Let 
\begin{equation}
\overline{\! X}_{\! n}=\frac1n (X_1+\cdots+X_n)
\label{xbar}
\end{equation}
be the empirical average of the process $\{X_n\}$. 

\begin{thm}[Mean]
\label{prop01'}
Assume that condition {\em{\bf (E)}} holds and denote the resulting asymptotic state of the scatterer without measurements by $\omega_+$. Then we have a law of large numbers, 
$$
\lim_{n\rightarrow\infty} \overline{\! X}_{\! n}=\mu_\infty:= \omega_+\otimes\omega_{\rm in}(\e^{\i\tau H}\M\e^{-\i\tau H}),
$$
where the limit is in the almost everywhere sense. Note that $\mu_\infty=\sum_m mf_m$.
\end{thm}

As for the frequencies, if the interaction is of the form $\lambda V$, then $\mu_\infty = \mu_\infty(\lambda)$ is a holomorphic function in $\lambda$ at the origin, and 
$$
\mu_\infty(\lambda) = \omega_{\rm in}(\M) +\lambda \,\omega_\cS\otimes\omega_{\rm in}\big(\i\tau [V,\overline{\M}(\tau)]\big) +O(\lambda^2).
$$ 
The lowest order correction term is the flux of the averaged measurement operator $$
\overline{\M}(\tau) = \frac1\tau\int_0^\tau \e^{\i s H_\cP}\M\e^{-\i sH_\cP}\d s.
$$

{\bf Relation to other work.} Our approach is based on the repeated interaction setup developed in \cite{BJM1,BJM2,BJM3,BJM4}. In the references \cite{BB,BBB}, a measurement process is considered as well. The analysis  in the latter references is based on the theory of classical stochastic processes (law of large numbers, martingale convergence, large deviation principle). The two mathematical approaches are completely different. The main difference in the models is that we consider {\em energy exchange} scattering processes (our assumption {\bf (E)}), in contrast to the {\em non-demolition models} treated in \cite{BB,BBB}. The non-demolition assumption assumes that there is a preferred basis of system states, called pointer states,  which is preserved by the interacting system-probe dynamics. The pointer states remain unchanged under the successive measurements and evolution. As a consequence of this assumption, the system-probe interaction operator commutes with the system Hamiltonian, so there is no energy transfer between the system and the probes. Assuming this manifold of invariant states, it is shown in \cite{BB} that any initial state of the system converges, under the repeated measurement evolution, to one of the pointer states. The measurement outcome determines which pointer state is chosen. The results are derived for a homogeneous model (same setting for each interaction cycle) and under a `non-degeneracy condition'. They are generalized in \cite{BBB} to non-homogeneous settings and without the non-degeneracy condition. Our assumption {\bf (E)} is in some sense exactly the `opposite' of a non-demolition assumption. Namely, it forces the dynamics (without measurement) to have {\it a single stationary state} (as opposed to an entire basis of stationary states). Both assumptions are reasonable, but they describe different physical situations. As pointed out in \cite{BB,BBB}, the non-demolition setting is realized in some experiments in quantum optics. Our setting describes scattering processes where energy is exchanged. A typical example is that of a `one atom maser', where atoms (probes) interact with modes of the electromagnetic field in a cavity (system) by exciting the field modes, leading to subsequent photon emission \cite{MWM}. These processes necessitate energy exchange. A famous model describing this situation is the Jaynes-Cummings model, in wich energy is not conserved. We discuss in some detail a truncated Jaynes-Cummings, or `spin-spin' model in Section \ref{example}. We mention that in \cite{BB} the incoming states are taken to be {\em pure} (they may be selected randomly from a set of pure states in \cite{BBB}), while in our work, incoming states may be mixed. As we have explained in the intoduction above, the statistical uncertainty in the mixed incoming states contributes to the fluctuation of the measurement process.

\subsection{RI system setup}
\label{qdssetup}

\subsubsection{Multitime measurement process}

We use the Liouvillian formalism introduced in Section \ref{sec:leaky}. In the present setup, both the Hilbert space of the system, ${\cal H}_\cS={\mathfrak h}_\cS\otimes{\mathfrak h}_\cS$, and that of a single probe, ${\cal H}_\cP={\mathfrak h}_\cP\otimes{\mathfrak h}_\cP$ (called `element' in the above-mentioned section), is a finite-dimensional GNS Hilbert space. The doubling of the space is explained in Section \ref{ssec:riliouv}, see \fer{m4}. We consider an initial system-probes state of the form
\begin{equation}
\Psi_0 = \Psi_\cS\otimes_{n\geq 1} B\Psi_\cP,
\label{m2}
\end{equation}
where $\Psi_\cS$ and $\Psi_\cP$ are reference states which are cyclic and separating for 
$$
\fm_\cS={\cal B}({\frak h}_\cS)\otimes\bbbone_\cS\qquad\mbox{and}\qquad\fm_\cP={\cal B}({\frak h}_\cP)\otimes\bbbone_\cP,
$$
respectively.  In \fer{m2}, $B$ belongs to the commutant algebra $\fm'_\cP$. It is not necessary to consider an infinite tensor product as in \fer{m2}, as at any finite time, only finitely many probes have to be described, see also the remarks before and after \fer{eq:systevol2}. The Hilbert space containing the vector $\Psi_0$ is built using the stabilizing sequence $\otimes_{n\geq 1} B\Psi_\cP$, and it is not the same as the one obtained from $\otimes_{n\geq 1} \Psi_\cP$. However, none of our results depend on this distinction. The repeated interaction {\em Schr\"odinger} dynamics is given by
\begin{equation}
\Psi_n = U_n\cdots U_2 U_1\Psi_0,
\label{m1}
\end{equation}
with unitaries $U_k = \e^{-\i\tau \widetilde L_k}$, see also \fer{m3}, \fer{eq:liouvpropagator}. So far, the measurements have not been introduced. For general information about quantum measurements, we refer to \cite{NC,C}. Let $\M\in{\mathcal B}({\mathfrak h}_\cP)$ be a selfadjoint ``measurement'' operator with spectrum
$$
{\rm spec}(\M)=\{m_1,\ldots, m_\mu\},
$$ 
where $1\leq \mu\leq \dim{\mathfrak h}_\cP$ (distinct eigenvalues). Let $S$ be any subset of ${\rm spec}(\M)$ and denote by $E_S$ the spectral projection of $\M$ associated to $S$. We will simply write $E_S$ for $E_S\otimes\bbbone_\cP$, i.e., $E_S\in\fm_\cP$.

The entire system is in the state $\Psi_0$ initially and the following experiment is performed: the system evolves according to $U_1$ and then a measurement of the observable $\M$ is made on the outcoming probe, yielding a value in $S_1\subset {\rm spec}(\M)$. Then the system evolves according to $U_2$ and after this evolution a measurement of $\M$ is made on the outcoming probe and yields a result lying in $S_2\subset {\rm spec}(\M)$. This procedure is repeated $n$ times. According to the principles of quantum mechanics, the probability for obtaining the multitime measurement result lying in $S_1,\ldots, S_n$ is given by
\begin{equation}
P(S_1,\ldots,S_n) = \left\| E_{S_n}U_n\cdots E_{S_2}U_2 \, E_{S_1}U_1\Psi_0\right\|^2.
\label{mm1}
\end{equation}
Furthermore, if this probability is nonzero (so that the outcome of the specific experiment is actually realizable), then the state of the system immediately after the $n$-th measurement is given by the normalized vector
\begin{equation}
\Psi_n = \frac{ E_{S_n}U_n\cdots E_{S_2}U_2 \, E_{S_1}U_1\Psi_0}{\sqrt{P(S_1,\ldots, S_n)}  }.
\label{mm2}
\end{equation}
We have $E_{{\rm spec}(\M)}=\bbbone$, which corresponds to the situation where at the given time step no measurement is performed. The stochastic process associated to the measurements is constructed as follows. Let 
\begin{equation*}
\Omega=\Sigma^{\mathbb N}=\{ \omega=(\omega_1,\omega_2,\ldots)\ : \ \omega_j\in{\rm spec}(\M)\}
\end{equation*}
and let $\cal F$ be the $\sigma$-algebra of subsets of $\Omega$ generated by all cylinder sets of the form
$$
\{\omega\in\Omega\ :\ \omega_1\in S_1, \ldots, \omega_n\in S_n, \ n\in{\mathbb N}, \ S_j\subset {\rm spec}(\M)\}.
$$
On $(\Omega,{\cal F})$ we define the random variables $X_n:\Omega\rightarrow {\rm spec}(\M)$ by $X_n(\omega)=\omega_n$, for $n=1,2,\ldots$ The random variable $X_n$ represents the outcome of the measurement at time-step $n$. The finite-dimensional distribution of the process $\{X_n\}_{n\geq 1}$ is given by 
\begin{equation}
P(X_1\in S_1,\ldots, X_n\in S_n) = P(S_1,\ldots,S_n),
\label{mm3}
\end{equation}
for any $n\in \mathbb N$, any subsets $S_1,\ldots,S_n$ of ${\rm spec}(\M)$, and where the right hand side is defined in \fer{mm1}. $P$ extends uniquely to a probability measure on $(\Omega,{\cal F})$ by the Kolmogorov extension theorem.

\subsubsection{Representation of joint probabilities}
\label{andyn}

Let 
$$
\widetilde\Psi_n = E_{S_n} U_n\cdots E_{S_1}U_1\Psi_0.
$$
Taking into account that the initial probe state is invariant under the dynamics generated by $L_\cP$, and proceeding as in \fer{evol}-\fer{eq:rdoreduction}, one shows that 
\begin{equation}
\|\widetilde\Psi_n\|^2 =  \scalprod{\Psi_{\rm ref}}{[P_1B^*B\e^{\i\tau K_1}E_{S_1}P_1]\cdots [P_nB^*B \e^{\i\tau K_n}E_{S_n}P_n] \Psi_{\rm ref}}, 
\label{m10}
\end{equation}
where $P_j$ is the projection acting trivially on all factors of 
$$
{\cal H}={\cal H}_\cS\otimes{\cal H}_\cP\otimes{\cal H}_\cP\otimes\cdots
$$ 
except on the $j$-th ${\cal H}_\cP$, on which it acts as the rank-one orthogonal projection onto $\Psi_\cP$. The reference vector is given by 
$$
\Psi_{\rm ref} = \Psi_\cS\otimes_{j=1}^n\Psi_\cP.
$$
The operator $K$ is given as in \fer{eq:cliouv}, \fer{def:cliouv}. We leave the details of the derivation of \fer{m10} as an exercise. This yields the following representation for the measurement probabilities \fer{mm1},
\begin{equation}
P(S_1,\ldots,S_n) = \scalprod{\Psi_\cS}{M_{S_1}\cdots M_{S_n}\Psi_\cS},
\label{m11}
\end{equation}
where 
\begin{equation}
M_S = P B^*B\e^{\i\tau K} E_SP
\label{m12}
\end{equation}
for $S\subseteq{\rm spec}(\M)$. Compare also with \fer{eq:rdoreduction}. Here, we view $M_S$ as an operator acting on ${\cal H}_\cS$ only and we write $P=|\Psi_\cP\rangle\langle\Psi_\cP|$ (see also \fer{mP}). We will write simply 
$$
M=PB^*B\e^{\i\tau K}P
$$ 
for $M_{{\rm spec}(\M)}$. Formulas \fer{m11} and \fer{m12} are the basis for the further analysis of the measurement probabilities. 

The following is an easy perturbative result.
\begin{prop}
\label{propx2} 
{}Let $A_j\in\sigma(X_j)$, $j\geq 1$. For any $k\geq 1$, there is a constant $C_k$ such that 
$$
\sup_{n\geq 1}\left|P(A_n,\ldots, A_{n+k})-P(A_n)\cdots P(A_{n+k})\right| \leq C_k \|V\|.
$$ 
\end{prop}

\begin{exo} Prove the above Proposition. 
\end{exo}

\subsubsection{Analysis of joint probabilities}

\label{sectanalproba}

In analogy with Proposition \ref{prop:Mproperties}, we have the following result.
\begin{lem}
\label{lemma1}
The spectrum of $M_S$, \fer{m12}, lies in the closed unit disk centered at the origin of the complex plane. For $S={\rm spec}(\M)$, i.e., $E_S=\bbbone$, we have in addition $M\Psi_\cS=\Psi_\cS$.
\end{lem}

We consider the probability $P(X_n\in S \mbox{\ eventually})$, for $S\subset{\rm spec}(\M)$. This quantity can be expressed using the Riesz spectral projections $\Pi$ and $\Pi_S$ of the operators $M$ and $M_S$ associated to the eigenvalue $1$. They are defined by
\begin{equation}
\Pi_S = \frac{1}{2\pi \i}\oint (z-M_S)^{-1} {\rm d}z,\qquad \Pi=\Pi_{{\rm spec}(\M)},
\label{projections}
\end{equation}
where the integral is over a simple closed contour in the complex plane encircling no spectrum of $M_S$ except the point $1$. If $1$ is not an eigenvalue then $\Pi_S=0$. For the next result, we recall the following definition,
$$
\{ X_n\in S \mbox{\  eventually\ }\}=\{ \omega :  \mbox{ there exists a $k$ s.t. $X_n(\omega)\in S$ for all $n\geq k$}\}.
$$

\begin{lem}
\label{lem3}
We have $P(X_n\in S \mbox{\ eventually}) = \scalprod{\Psi_\cS}{\Pi\,\Pi_S\,\Psi_\cS}$.
\end{lem}

{\it Outline of proof.\ } $\{X_n\in S \mbox{\ eventually}\}$ is the increasing union (in $k$) of $\{X_n\in S \ \forall n\geq k\}$, so 
$$
P(X_n\in S \mbox{\ eventually}) =\lim_{k\rightarrow\infty} P(X_n\in S\ \forall n\geq k).
$$
Next, $\{X_n\in S\ \forall n\geq k\}$ is the intersection of the decreasing  sequence (in $l$) $\{X_n\in S,\ n=k,\ldots,k+l\}$, so 
$$
P(X_n\in S \mbox{\ eventually}) = \lim_{k\rightarrow\infty}\lim_{l\rightarrow\infty} P(X_n\in S,\ n=k,\ldots,k+l).
$$
By using the representation \fer{m11} and the fact that $M^k$, $M_S^k$ converge to their Riesz projections associated to the eigenvalue one, as $k\rightarrow\infty$, one reaches the expression given in Lemma \ref{lem3}.

\medskip
Given a measurement path $X_1=m_1,\ldots,X_n=m_n$, the system state is
$$
\omega_n(A) = \frac{\scalprod{\Psi_\cS}{M_1\cdots M_n A\Psi_\cS}}{\scalprod{\Psi_\cS}{M_1\cdots M_n\Psi_\cS}},
$$
where $A$ is any system observable and $M_j= M_{\{m_j\}}$. The randomness of the measurement paths makes the system state $\omega_n$ a random variable.

\begin{lem}[Evolution of averaged system state]
\label{lemma102}
The expectation of the system state, ${\mathbb E}[\omega_n]$, equals the state obtained by evolving the initial condition according to the dynamics {\em without} measurement.
\end{lem}

{\em Proof of Lemma \ref{lemma102}.\ } Since $P(X_1=m_1,\ldots, X_n=m_n)=\scalprod{\Psi_\cS}{M_1\cdots M_n\Psi_\cS}$ we have
\begin{eqnarray}
{\mathbb E}[\omega_n(A)] &=& \sum_{m_1,\ldots,m_n} P(X_1=m_1,\ldots, X_n=m_n)\frac{\scalprod{\Psi_\cS}{M_1\cdots M_n A\Psi_\cS}}{\scalprod{\Psi_\cS}{M_1\cdots M_n\Psi_\cS}} \nonumber\\
&=&  \sum_{m_1,\ldots,m_n}\scalprod{\Psi_\cS}{M_1\cdots M_n A\Psi_\cS}\nonumber\\
&=& \scalprod{\Psi_\cS}{M^n A\Psi_\cS}.
\label{sstar}
\end{eqnarray}
In the last step, we have used that 
$$
\sum_m M_{\{m\}}=\sum_m PB^*B\e^{\i\tau K}E_{\{m\}}P=PB^*B\e^{\i\tau K}P=M.
$$
The right hand side of \fer{sstar} is the single-step dynamics operator of the system without probe measurements. \hfill $\square$

\subsubsection{Outline of proof of Theorem \ref{corrdeclemma'}} 
\label{proof1+5}

We shall only prove
\begin{equation}
\left| P(A\cap B)-P(A)P(B)\right| \leq c \ \e^{-\gamma(m-l)},
\label{010'simple} 
\end{equation}
and refer to \cite{Merkli-Penney} to explain the extra factor $P(A)$ on the right-hand side of \fer{010'}.

We consider the simplified situation where $A=\{\omega : X_l\in S_l\} \in \sigma(X_l)$ and $B=\{\omega : X_m\in S_m\}\in \sigma(X_m)$, for $S_l, S_m\subset{\rm spec}(M)$. Then
\begin{equation}
P(A\cap B) = \langle \Psi_\cS, M^{l-1} M_{S_l} M^{m-l-1} M_{S_m} \Psi_\cS \rangle.
\label{012.1}
\end{equation}
We now approximate $M^{m-l-1}$ by its value for large $m-l$. Using Assumption {\bf (E)} and since $M\Psi_\cS=\Psi_\cS$, one can write, see also Theorem \ref{thm:idealsmall},
\begin{equation}
\|M^k  - |\Psi_\cS\rangle\langle\Psi_\cS^*| \,\|\leq C\e^{-\gamma k}.
\label{011}
\end{equation}
for some $\gamma>0$ and where $\Psi_\cS^*$ is the unique vector such that $M^*\Psi_\cS^*=\Psi_\cS^*$ and $\scalprod{\Psi^*_\cS}{\Psi_\cS}=1$ ($|\Psi_\cS\rangle\langle\Psi_\cS^*|$ is the Riesz spectral projection associated to the eigenvalue $1$ of $M$).
Inserting \fer{011} in \fer{012.1}, we obtain
\begin{equation}
P(A\cap B) = \scalprod{\Psi_\cS}{M^{l-1}M_{S_l}\Psi_\cS} \scalprod{\Psi_\cS^*}{M_{S_m}\Psi_\cS} +O(\e^{-\gamma(m-l)}).
\label{mame1}
\end{equation}
From \fer{m11} we get
\begin{equation}\label{ng}
\scalprod{\Psi_\cS}{M^{l-1}M_{S_l}\Psi_\cS} = P(A), 
\end{equation}
and, since  $\Psi_\cS$ is normalized, we have 
\begin{eqnarray*}
\scalprod{\Psi_\cS^*}{M_{S_m}\Psi_\cS} & = & \scalprod{\Psi_\cS}{(|\Psi_\cS\rangle\langle\Psi_\cS^*|)  M_{S_m}\Psi_\cS} = \scalprod{\Psi_\cS}{M^{m-1} M_{S_m}\Psi_\cS} +O(\e^{-\gamma m})\\
 & = & P(B)+O(\e^{-\gamma m}),
\end{eqnarray*}
where we have used again \fer{011} in the second step. Combining this with \fer{ng} and \fer{mame1} shows that \fer{010'} holds for this simplified case.

\subsubsection{Outline of proof of Theorems \ref{theorem2} and \ref{prop01'}} 
\label{anotherone}

We have
$$
{\mathbb E}[f_m] = \lim_{n\rightarrow\infty} \frac1n \sum_{m_1,\ldots,m_n} \sum_{j=1}^n \chi(m_j) P(X_1=m_1,\ldots,X_n=m_n),
$$
where $\chi(m_j)$ is the characteristic function, taking the value one if $m_j=m$ and zero otherwise. The double sum equals $\sum_{j=1}^n\sum_{m_k,k\neq j} P(X_1=m_1,\ldots,X_j=m,\ldots X_n=m_n)$, which is $\sum_{j=1}^n\scalprod{\Psi_\cS}{M^{j-1} M_m\Psi_\cS}$. As $\frac1n \sum_{j=0}^n M^j\rightarrow\Pi=|\Psi_\cS\rangle\langle\Psi^*_\cS|$ as $n\rightarrow\infty$, where $\Psi^*_\cS$ is the invariant vector of $M^*$, (see also \fer{011}) we obtain ${\mathbb E}[f_m] = \scalprod{\Psi^*_\cS}{M_m\Psi_\cS}$. Using that $M_m=PB^*B\e^{\i\tau K}E_mP$, see \fer{m12}, we arrive at 
\begin{eqnarray*}
{\mathbb E}[f_m] &=& \scalprod{\Psi^*_\cS\otimes\Psi_\cP}{B^*B \e^{\i\tau K}E_m\Psi_\cS\otimes\Psi_\cP} \\
&=& \scalprod{\Psi^*_\cS\otimes\Psi_\cP}{B^*B \e^{\i\tau L}E_m\e^{-\i\tau L}\Psi_\cS\otimes\Psi_\cP}\\
&=&\omega_+\otimes\omega_{\rm in}(\e^{\i\tau H}E_m\e^{-\i\tau H}).
\end{eqnarray*}
(Here, $L=L_\cS+L_\cP+V_{\cS\cP}$ is the Liouvillian, see also \fer{m2}). This shows convergence of $f_m$ in the mean. To upgrade this to almost-everywhere convergence, one can use a probabilistic `fourth moment method' (see \cite{Merkli-Penney}). 

To prove Theorem \ref{prop01'}, we note that 
\begin{eqnarray*}
{\mathbb E}[\,\overline{\!X}_n] &=& \frac1n\sum_{m_1,\ldots,m_n} (m_1+\cdots +m_n) P(X_1=m_1,\ldots,X_n=m_n) \\
&=&\frac1n \sum_{j=1}^n \scalprod{\Psi_\cS}{M^{j-1} PB^*B\e^{\i\tau K}\M P\Psi_\cS}.
\end{eqnarray*}
Proceeding as above in this proof, one sees that the limit $n\rightarrow\infty$ of the right side is $\omega_+\otimes\omega_{\rm in}(\e^{\i\tau H}\M\e^{-\i\tau H})$.

\subsection{The spin-spin model}
\label{example}

In this model, both the scatterer and the probes have only two degrees of freedom participating in the scattering process. The pure state space of $\cS$ and $\cP$ is ${\mathbb C}^2$, and the Hamiltonians are given by the Pauli $\sigma_z$ operator,
\begin{eqnarray}
H_\cS=H_\cP &=&
 \left(
\begin{array}{cc}
1 & 0\\
0 & -1
\end{array}
\right).
\label{009}
\end{eqnarray}
These Hamiltonians are the same as those introduced before \fer{def:fermicreation} (modulo an additive constant, and where $E_0=E=2$). The interaction between $\cS$ and $\cP$ is determined by the operator 
\begin{equation}
\lambda V= \lambda\left( a^*_\cS\otimes a_\cP + a_\cS\otimes a_\cP^*\right),
\label{0010}
\end{equation}
with coupling constant $\lambda\in {\mathbb R}$, and where  
\begin{equation}
a_{\cS,\cP}= \left(
\begin{array}{cc}
0 & 0\\
1 & 0
\end{array}
\right),\qquad
a^*_{\cS,\cP}= \left(
\begin{array}{cc}
0 & 1\\
0 & 0
\end{array}
\right)
\label{0011}
\end{equation}
are the annihilation and creation operators. The interaction is the same as the one taken after \fer{def:fermicreation}. In the true Jaynes-Cummings model, the system $\cS$ has infinitely many levels (harmonic oscillator), see Section \ref{ssec:qedcavity} and also e.g. \cite{NC}. The total Hamiltonian $H=H_\cS+H_\cP+\lambda V$ describes exchange of energy between $\cS$ and $\cP$, while the total number of excitations, $N=a_\cS^*a_\cS+a_\cP^*a_\cP$, is conserved (commutes with $H$). This allows for a treatment of the system separately in the invariant sectors $N=0,1,2$.

{}For an arbitrary probe observable $X\in{\cal B}({\mathbb C}^2)$ we write $X_{ij}=\scalprod{\varphi_i}{X\varphi_j}$, where $\varphi_1$, $\varphi_2$ are the orthonormal eigenvectors of $H_\cP$ (with $H_\cP\varphi_1=\varphi_1$). Incoming states are invariant, determined by $p\in[0,1]$ via 
\begin{equation}
\omega_{\rm in}(X)=p X_{11} +(1-p) X_{22},
\label{nm2}
\end{equation}
where $X\in{\cal B}({\mathbb C}^2)$ is an arbitrary probe observable.

We will use the notation and definitions of Section \ref{qdssetup} in what follows. In particular, the single time step operator $M_S$ is defined in \fer{m12}. For the following explicit formula, we take the reference state $\Psi_\cS$ to be the trace state,
$$
\Psi_\cS =\frac{1}{\sqrt 2}\big( \varphi_1\otimes\varphi_1 +\varphi_2\otimes\varphi_2\big).
$$

\begin{thm}[Explicit reduced dynamics operator]
\label{thmexplicit}
Set $\varphi_{ij}=\varphi_i\otimes\varphi_j$ and let $X\in\cB(\fh_\cE)$ so that $X\otimes \one_\cE\in \fm_\cE$. In the ordered basis $\{\varphi_{11}, \varphi_{12}, \varphi_{21}, \varphi_{22}\}$ we have 
\begin{eqnarray}
\lefteqn{PB^*B \e^{\i \tau K} \, X\otimes \one_\cE \, P = \omega_{\rm in}(X) \,\e^{\i\tau L_\cS}+}\label{nm1}\\
&&\qquad \nonumber\\
&&\left(
\begin{array}{cc}
(1-p) X_{22} a & (1-p)X_{21}b \nonumber\\
-pX_{12}\e^{2\i\tau} \i \sin(\lambda\tau) & \e^{2\i\tau}(\cos(\lambda\tau)-1)\omega_{\rm in}(X) \nonumber\\
pX_{21}\e^{-2\i\tau}\i\sin(\lambda\tau) & 0 \nonumber\\
-pX_{22}a & -pX_{21} b
\end{array}
\right. \nonumber\\
&&\qquad \nonumber\\
&&\qquad
\left.
\begin{array}{cc}
-(1-p)X_{12} b & -(1-p)X_{11} a \nonumber\\
0 & (1-p)X_{12}\e^{2\i\tau}\i\sin(\lambda\tau)\nonumber\\
\e^{-2\i\tau}(\cos(\lambda\tau)-1)\omega_{\rm in}(X)& -(1-p)X_{21}\e^{-2\i\tau}\i\sin(\lambda\tau)\nonumber\\
pX_{12} b & pX_{11} a\nonumber
\end{array}
\right),
\end{eqnarray}
where $a=-\sin^2(\lambda\tau)$, $b=-\i\sin(\lambda\tau)\cos(\lambda\tau)$. (On the right hand side, we have a matrix with four columns.)
\end{thm}
The proof of Theorem \ref{thmexplicit} is a quite straightforward calculation. 

\medskip

{\em Remarks.\ } 1. The quantity \fer{nm1} is the single-step dynamics operator as a function of the incoming probe state (determined by $B$, or, $p$), and a general operator $X$ on the outgoing probe. If $X$ is the spectral projection associated to a measurement operator then one obtains the measurement process, for $X=\bbbone$ one obtains the single-step dynamics operator without measurements.

2. The vector $[p, 0, 0, 1-p]^t$ is an eigenvector of the adjoint of \fer{nm1} with eigenvalue $\omega_{\rm in}(X)$.

\bigskip

{\bf Resonant and non-resonant system.\ } If $\lambda\tau$ is a multiple of $\pi$ then \fer{nm1} reduces to $PB^*B\e^{\i\tau K}XP= \omega_{\rm in}(X)\, {\rm diag}(1,\pm 1,\pm 1,1)$ with plus and minus signs if the multiple is even and odd, respectively. Then, by using the expression for $P(X_1\in S_1,\ldots,X_n\in S_n)$ given in \fer{m11}, it is readily seen that the random variables $X_j$ are independent, and $P(X_j\in S)=\omega_{\rm in}(E_S)$. When  $\lambda\tau\in \pi{\mathbb Z}$, the system is called the {\it resonant}, otherwise it is called {\it non-resonant}. This is the same terminology as used in Section \ref{ssec:qedcavity}, Definition \ref{def:resonant} and also \cite{BP}. One can understand the resonant regime as follows: consider the dynamics on $\cS$ and a single probe $\cP$, generated by the Hamiltonian $H=H_\cS+ H_\cP + \lambda V$. The probability of transition from the initial state $\varphi_2^\cS\otimes\varphi_1^\cP$, where the $\cS$ is in the ground state and $\cP$ in the excited state, to the opposite state $\varphi_1^\cS\otimes\varphi_2^\cP$, at time $t$, is given by $P_t = \left|\scalprod{\varphi_1^\cS\otimes\varphi_2^\cP}{\e^{-\i tH}\varphi_2^\cS\otimes\varphi_1^\cP}\right|^2= \sin^2(\lambda t)$. For  $\lambda t\in \pi{\mathbb Z}$ this probability vanishes. If the interaction time $\tau$ in the repeated interaction system is a multiple of $\pi/\lambda$, then interaction effects are suppressed. It is not hard to see that in this case, the system does not feel the interaction with the probes in the sense that $\omega_n(A)=\omega_0(\alpha^\cS_n(A))$ for all $n\geq 1$, where $\alpha^\cS_n(A)$ is the reduced dynamics of $\cS$ alone. We focus now on the non-resonant situation.

\medskip

{\bf Asymptotics.\ }  We show that 

\begin{itemize}
\item[{}]
{\it For $\omega_{\rm in}=|\!\!\uparrow\rangle\langle\uparrow\!\! |$, the process $X_n$ converges to $|\!\!\uparrow\rangle\langle\uparrow\!\! |$, i.e., 
$$
P(X_n=|\!\uparrow\rangle \mbox{\ eventually})=1.
$$
Moreover, the incoming state of the probes is copied onto the scatterer, in the limit of large times.}
\end{itemize}
This result is non-perturbative and holds for all $\lambda\in\mathbb R$. Intuitively, the result is explained as follows: since the incoming probes are in the state up, the effect of the interaction \fer{0010} is to de-excite them into the state down, while the scatterer does the opposite, passing from the ground to the excited state. After some scattering interactions, the scatterer thus tends to be in the state up. However, the state where both the scatterer and the probe are in the state up is invariant under the dynamics generated by $H=H_\cS+H_\cP+\lambda V$. To show the result, let $\omega_{\rm in}=|\!\!\uparrow\rangle\langle\uparrow\!\! |$, which corresponds to $p=1$ in \fer{nm2}.  The matrix in \fer{nm1} becomes lower triangular and we obtain 
\begin{equation}
{\rm spec}(PB^*B\e^{\i\tau K}XP) = X_{11}\ \{ 1, \e^{2\i\tau}\cos(\lambda\tau), \e^{-2\i\tau}\cos(\lambda\tau), \cos^2(\lambda\tau)\}.
\label{starstar}
\end{equation}
Condition {\bf (E)}  is always satisfied in the non-resonant case. $\Psi_\cS$ and $\varphi_{11}$ are the eigenvectors associated to the eigenvalue $1$ of $PB^*B\e^{\i\tau K}P$ and its adjoint, respectively. So the Riesz projection associated with the eigenvalue $1$ of $T$ is 
$$
\Pi = \sqrt{2} |\Psi_\cS\rangle\langle\varphi_{11}|
$$
(see \fer{projections}). One finds the following results for the asymptotic mean  and the frequencies, and for an arbitrary measurement operator $\M$:
\begin{equation*}
\mu_\infty=\omega_{\rm in}(\M),\quad f_m=\omega_{\rm in}(E_m).
\end{equation*} 
This suggests that the cavity becomes `transparent' for large times (no effect on incoming probes). And indeed, the Riesz projection $\Pi_S$ associated to the value 1 of $M_S$ (with $S\subset{\rm spec}(\M)$) vanishes unless $(E_S)_{11}=1$, see \fer{starstar}. Thus we have $P(X_n\in S \mbox{\ eventually})=0$ if $(E_S)_{11}\neq 1$ (see Lemma \ref{lem3}). If $(E_S)_{11}=1$ then $E_S$ is the projection $|\varphi_1\rangle\langle\varphi_1|$, and one finds easily that $\Pi_S = |\varphi_{11}\rangle\langle\varphi_{11}|$ and (using Lemma \ref{lem3}), that $P(X_n=|\!\!\uparrow\rangle \mbox{\ eventually})=1$. In the situation where the measurement outcomes stabilize, $X_n\rightarrow |\!\uparrow\rangle$, we can determine the asymptotic state of the scatterer $\cS$ \cite{Merkli-Penney}. It is given by $\omega_\infty(A) =\omega_{\rm in}(A)$, $A\in{\cal B}({\mathbb C}^2)$. The above `asymptotic transparency' statement follows.

\medskip

{\bf Large deviations for the mean.\ }The logarithmic moment generating function \cite{DZ} is defined by 
\begin{equation}
\Lambda(\alpha) = \lim_{n\rightarrow\infty}\frac1n \log\E[\e^{n\alpha\, \overline{\!X}_{\! n}}],
\label{001}
\end{equation}
for all $\alpha\in\mathbb R$ s.t. the limit exists as an extended real number. The existence of the logarithmic generating functional can be analyzed for general repeated measurement systems \cite{Merkli-Penney}. We give here only a discussion for the spin-spin example at hand.

Using Theorem \ref{thmexplicit} (with  $p=1$) we find that 
$$
\Lambda(\alpha)=\log\omega_{\rm in}(\e^{\alpha \M}),
$$
for $\alpha\in\mathbb R$. The Legendre transformation of $\Lambda(\alpha)$,
\begin{equation}
\Lambda^*(x) = \sup_{\alpha\in{\mathbb R}} \ \alpha x-\Lambda(\alpha),
\label{b1}
\end{equation}
{}for $x\in\mathbb R$, is called the {\it rate function}. Its usefulness in the present context is due to the G\"artner-Ellis theorem \cite{DZ}, which asserts that for any closed set $F\subset\mathbb R$ and any open set $G\subset\mathbb R$, we have
\begin{eqnarray}
\limsup_{n\rightarrow\infty} \frac1n \log P\left(\,\overline{\! X}_{\! n} \in F \right)&\leq& -\inf_{x\in F}\Lambda^*(x)\nonumber \\
\liminf_{n\rightarrow\infty} \frac1n \log P\left(\, \overline{\! X}_{\! n}\in G\right)&\geq &-\inf_{x\in G\cap{\cal F}}\Lambda^*(x).
\label{b10}
\end{eqnarray}
Here, $\cal F$ denotes the set of `exposed points of $\Lambda^*$' (see \cite{DZ} for the definition). We now evaluate the Legendre transform locally. Note that $\Lambda$ is twice differentiable, and the second derivative w.r.t. $\alpha$ of the argument of the supremum in \fer{b1} is less than or equal to zero. Therefore, for fixed $x$, the supremum is taken at $\alpha\in\mathbb R$ satisfying 
\begin{equation}
x = \Lambda'(\alpha)=\frac{\omega_{\rm in}(\M\e^{\alpha \M})}{\omega_{\rm in}(\e^{\alpha \M})}.
\label{b2}
\end{equation}
For $\alpha=0$ we have $x=\omega_{\rm in}(\M)$. If $\Lambda''(0)= {\rm Var}(\M):=\omega_{\rm in}(\M^2)-\omega_{\rm in}(\M)^2 \neq 0$, then equation \fer{b2} has an implicit solution $\alpha=\alpha(x)$, locally around $x=\omega_{\rm in}(\M)$.  Since $\Lambda'(\alpha)$ is holomorphic at $\alpha=0$, the implicit solution  is holomorphic at $x=\omega_{\rm in}(\M)$. The Taylor expansion of \fer{b2} is 
\begin{equation}
x=\omega_{\rm in}(\M) +\alpha{\rm Var}(\M) +c\alpha^2 +O(\alpha^3),
\label{b3}
\end{equation}
where 
$$
c=\textstyle\frac12\{ \omega_{\rm in}(\M^3)-3\omega_{\rm in}(\M^2)\omega_{\rm in}(\M)+2\omega_{\rm in}(\M)^3\}.
$$ 
We solve equation \fer{b3} implicitly for $\alpha=\alpha(x)$, which is the point where the supremum in \fer{b1} is taken, i.e.,
\begin{equation}
\Lambda^*(x) = \frac{\left( x-\omega_{\rm in}(\M)\right)^2}{2{\rm Var}(\M)} +O\!\left( (x-\omega_{\rm in}(\M))^4\right).
\label{localLT}
\end{equation}

As an application we consider a {\em measurement of the outgoing spin angle}. 
Since $\omega_{\rm in}$ is the state `spin up', we have $\omega_{\rm in}(\M)=\M_{11}$ and ${\rm Var}(\M)= |\M_{12}|^2$. Imagine an experiment where we measure the angle of the spins as they exit the scattering process. Let $\theta\in[0,\pi)$ be the angle measuring the altitude ($\theta=0$ is spin up). The measurement operator ``spin in direction $\theta$'' is given by
$$
\M=\left[
\begin{array}{cc}
\cos\theta &  \sin\theta\\
 \sin\theta & -\cos\theta
\end{array}
\right],
$$
see e.g. \cite[Chapitre IV, (A-19)]{CTDL}. The eigenvectors of $\M$ associated to the eigenvalues $\pm 1$ of $\M$ are 
\begin{eqnarray*}
\chi_+ &=& \cos(\theta/2) \varphi_1 + \sin(\theta/2) \varphi_2 \\
\chi_- &=& -\sin(\theta/2) \varphi_1 + \cos(\theta/2) \varphi_2.
\end{eqnarray*}
The eigenprojection $E_+$ measures the spin in the positive direction $\theta$. By using Lemma \ref{lem3} is easy to see that
$$
P\left( \mbox{$X_n$ is in direction $\theta$ eventually}\right) = \left\{
\begin{array}{ll}
1 & \mbox{if $\theta =0$}\\
0 & \mbox{if $\theta\neq 0$.}
\end{array}
\right.
$$
This is another manifestation of the asymptotic transparency of the cavity.

We obtain from theorem \ref{prop01'} (with $\mu_\infty=\cos\theta$) that for any $\epsilon>0$,  
$$
\lim_{n\rightarrow\infty} P(|\,\overline{\! X}_{\! n}-\cos\theta|\geq \epsilon)=0.
$$
The speed of convergence can be estimated using \fer{b10} and \fer{localLT}. It is easy to see that the logarithmic generating function and the rate function associated to the shifted random variable $\overline{\!X}_{\! n}-\cos\theta$ are given by $\Lambda_{\rm shift}(\alpha) = \Lambda(\alpha)-\alpha\cos\theta$ and  $\Lambda_{\rm shift}^*(x)=\Lambda^*(x+\cos\theta)$, respectively. Next, we note that all points in the vicinity of zero belong to ${\cal F}_{\rm shift}$, the set of exposed points of $\Lambda^*_{\rm shift}$. Indeed, if $x=\Lambda_{\rm shift}'(\alpha)$ for some $\alpha\in{\mathbb R}$, then $x\in{\cal F}_{\rm shift}$ (\cite{DZ}, Lemma 2.3.9). But $x=0=\Lambda_{\rm shift}'(0)$, and $\Lambda'_{\rm shift}$ is invertible around zero (as $\Lambda''_{\rm shift}(0)\neq 0$). This shows that ${\cal F}_{\rm shift}$ contains a neighbourhood of the origin.

Take $0<\epsilon<\epsilon'<\!\!<1$, set $G=(-\epsilon',-\epsilon)\cup (\epsilon,\epsilon')$, and let $F$ be the closure of $G$. Then (use \fer{localLT})
$$
\inf_{x\in F}\Lambda_{\rm shift}^*(x) = \inf_{x\in G\cap{\cal F}_{\rm shift}}\Lambda_{\rm shift}^*(x) = \frac{\epsilon^2}{2{\rm Var}(\M)} +O((\epsilon')^4).
$$
Combining this with the two bounds \fer{b10} (for the shifted random variable), we obtain
$$
P\big(\epsilon \leq |\,\overline{\! X}_{\! n}-\cos\theta|\leq \epsilon')\sim \exp\left[-n\Big\{\frac{\epsilon^2}{2\sin^2\theta} + O((\epsilon')^4)\Big\}\right],\quad n\rightarrow\infty,
$$
which is a large deviation statement for the average $\,\overline{\! X}_{\! n}$.

\bigskip
{\bf Acknowledgements.\ } The research of L.B. and A.J. was partially supported by the Agence Nationale de la Recherche (grant ANR-09-BLAN-0098). The research of M.M. was partially supported by NSERC through an NSERC Discovery Grant. The authors warmly thank V. Jak\v si\'c, C.-A. Pillet and R. Seiringer for the invitation to the summer school ``Non-equilibrium Statistical Mechanics'' which was held in Montreal and is at the origin of this review, as well as CRM and McGill University for financial support.



\end{document}